\documentclass[twocolumn,amstext,amssymb,amsmath,superscriptaddress,showpacs,nofootinbib,aps,prd,preprintnumbers,floatfix,natbib]{revtex4-1}
\usepackage{lmodern}
\usepackage[T1]{fontenc}
\usepackage[utf8]{inputenc}
\usepackage{amstext,amssymb,amsmath}
\usepackage{hyperref}
\usepackage[table]{xcolor}
\usepackage{multirow}
\usepackage{graphicx}
\usepackage{slashed}
\usepackage{enumitem}
\usepackage[normalem]{ulem}
\usepackage{comment}
\usepackage{cases}

\definecolor{LightGray}{gray}{0.91}
\definecolor{LightBlue}{rgb}{0.87, 0.94, 1}

\newcommand{\eq}[1]{\eqref{eq:#1}}
\newcommand{\fig}[1]{Fig.~\ref{fig:#1}}
\newcommand{\tab}[1]{Tab.~\ref{tab:#1}}

\newcommand{\hc}{\text{h.c.}}

\newcommand{\Sec}[1]{Sec.~\ref{sec:#1}}

\newcommand{\MPl}{M_\text{Pl}}
\newcommand{\kt}{\tilde{\kappa}}

\newcolumntype{G}{>{\columncolor{LightGray}}c}
\newcolumntype{C}{>{$}c<{$}}

\makeatletter
    \def\CT@@do@color{%
      \global\let\CT@do@color\relax
            \@tempdima\wd\z@
            \advance\@tempdima\@tempdimb
            \advance\@tempdima\@tempdimc
    \advance\@tempdimb\tabcolsep
    \advance\@tempdimc\tabcolsep
    \advance\@tempdima2\tabcolsep
            \kern-\@tempdimb
            \leaders\vrule
                    \hskip\@tempdima\@plus  1fill
            \kern-\@tempdimc
            \hskip-\wd\z@ \@plus -1fill }
    \makeatother

\begin{document}
\preprint{DO-TH 21/08}

\title{Portals into Higgs vacuum stability}

\author{Gudrun~Hiller}
\author{Tim~H\"ohne}
\affiliation{TU Dortmund University, Department of Physics, Otto-Hahn-Str.4, D-44221 Dortmund, Germany}
\author{Daniel~F.~Litim}
\affiliation{Department of Physics and Astronomy, University of Sussex, Brighton, BN1 9QH, U.K.}
\author{Tom~Steudtner}
\affiliation{TU Dortmund University, Department of Physics, Otto-Hahn-Str.4, D-44221 Dortmund, Germany}
\affiliation{Department of Physics and Astronomy, University of Sussex, Brighton, BN1 9QH, U.K.}
\begin{abstract}
We address the notorious metastability  of the standard model (SM) and promote it to  a model building task: 
What are the new ingredients required to  stabilize the SM up to the Planck scale without  encountering subplanckian Landau poles?  
Using the SM extended by vector-like fermions (VLFs), we chart out the  corresponding landscape of Higgs stability. We find that  the gauge portal mechanism, triggered by new SM charge carriers, opens up  sizeable room for stability in a minimally invasive manner. We also find  models with Higgs criticality, and Yukawa portals  opening up at stronger coupling. 
Several models allow for  VLFs  in the TeV-range, which can be searched for at the LHC.
For nontrivial flavor structure severe flavor-changing neutral current constraints arise which complement those from stability, and push
  lower  fermion masses up to $\mathcal{O}(10^3\,\text{TeV})$.
   \end{abstract}
\maketitle
\tableofcontents
\flushbottom

\section{Introduction}

The discovery of the Higgs  particle \cite{CMS:2012qbp,ATLAS:2012yve} a decade ago, together with theoretical precision calculations \cite{Degrassi:2012ry,Buttazzo:2013uya} evidenced the instability of the   standard model (SM) vacuum.  While  a theory of nature with a decaying ground state  would be  unacceptable, vacuum metastability due to a lifetime sufficiently large compared to the age of the universe has become a widely accepted narrative.
Moreover, it is expected that new physics prior to the scale of  metastability ($\mu_{\rm inst} \approx 10^{10}$ GeV), and  possibly  not too far above the TeV scale, may  stabilize the ground state.
This view has been confirmed in many scenarios for physics beyond the SM, for instance~\cite{Elias-Miro:2012eoi,
Gonderinger:2012rd,
Anchordoqui:2012fq,
Arkani-Hamed:2012dcq,
Joglekar:2012vc,
Fairbairn:2013xaa,
Altmannshofer:2013zba,
Gabrielli:2013hma,
BhupalDev:2013xol,
Datta:2013mta,
Khan:2014kba,
Costa:2014qga,
Khoze:2014xha,
Xiao:2014kba,
Coriano:2014mpa,
DiChiara:2014wha,
Haba:2014oxa,
Lalak:2014qua,
Altmannshofer:2014vra,
Belanger:2014bga,
Salvio:2015cja,
Chakrabarty:2015kmt,
Blum:2015rpa,
Bagnaschi:2015pwa,
Duch:2015jta,
Dhuria:2015ufo,
Salvio:2015jgu,
Son:2015vfl,
Hamada:2015bra,
Ng:2015eia,
Falkowski:2015iwa,
Han:2015hda,
Lindner:2015qva,
Lindner:2015qva,
DelleRose:2015bms,
Latosinski:2015pba,Chowdhury:2015yja,
Khan:2015ipa,
Ferreira:2015rha,
Ferreira:2015pfi,
Swiezewska:2015paa,
Coriano:2015sea,
Oda:2015gna,
Das:2015nwk,
Haba:2015rha,
Haba:2015nwl,
Das:2016zue,
Chakrabarty:2016smc,
Bandyopadhyay:2016oif,
Khan:2016sxm,
Haba:2016zbu,
Mambrini:2016dca,
Chang:2016pya,
Hamada:2016vwk,
Ghosh:2017pxl,
Oda:2017kwl,
Garg:2017iva,
Ghosh:2017fmr,
Goswami:2018jar,
DuttaBanik:2018emv,
Schuh:2018hig,
Marzo:2018nov,
Ellis:2018khn,
Boucenna:2018wjc,
Gopalakrishna:2018uxn,
Wang:2018lhk,
Okada:2019bqa,
Hiller:2019mou,
Bhattacharya:2019fgs,
Mandal:2019ndp,
Hiller:2020fbu,
Borah:2020nsz,
Chakrabarty:2020jro,
Mandal:2020lhl,Fabbrichesi:2020svm,DuttaBanik:2020jrj,Bandyopadhyay:2021kue,Bause:2021prv}.
On the other hand, much less attention has been given to vacuum stability in its own right.

In this work, we propose a change of perspective by bringing the SM instability into the fore,  and promoting it into a primary model building task. 
There are  several reasons for doing so now:
\begin{itemize}

\item
The continuous success of the SM in the LHC era  
with only a few anomalies, and the absence of  clear new physics signatures  at colliders 
or elsewhere, calls for new theory  ideas and directions in model building. 
 
 \item The onset of the SM instability is a high energy effect and out of reach for future colliders. However, the existence of the instability is not directly pointing at a mass scale for new physics. Its remedy, therefore, may arise from new phenomena  at any scale below the Planck scale,  and possibly  from scales as low as  a few TeV. The latter   opens a door for indirect tests of stability at present and future colliders. 
  \item Understanding the RG running of couplings and masses between TeV and the Planck scale with high precision is key for testing vacuum stability beyond the SM.
 Powerful and versatile   
computer-algebraic tools  such as \texttt{ARGES} \cite{Litim:2020jvl} have  become available, which offer easy access to advanced perturbative beta functions for any 4d QFT and SM extension. 

 \end{itemize}

In this spirit, we  investigate  ingredients and mechanisms  required to reinstate stability.
We will be satisfied if this happens at and possibly  the whole way up to the Planck scale where quantum gravity is expected to kick in. 
In addition, we   demand that the renormalization group (RG) running of couplings up to the Planck scale remains finite and well-defined, without encountering  any singularities or Landau poles along the way. We focus on  SM extensions by vector-like fermions and the study of gauge portals -- which arise because new SM gauge charge carriers are available, and Yukawa portals -- which arise as soon as the the new fermions admit new Yukawa couplings with SM matter.
 Our main interests are the constraints dictated by stability on BSM matter fields, their masses, multiplicities, or interactions. We are particularly interested in settings where stability arises through effects at the TeV scale, which can be probed at colliders.

\begin{figure*}
    \centering
    \includegraphics[trim={2cm 0 2cm 0},clip,width=2.1\columnwidth]{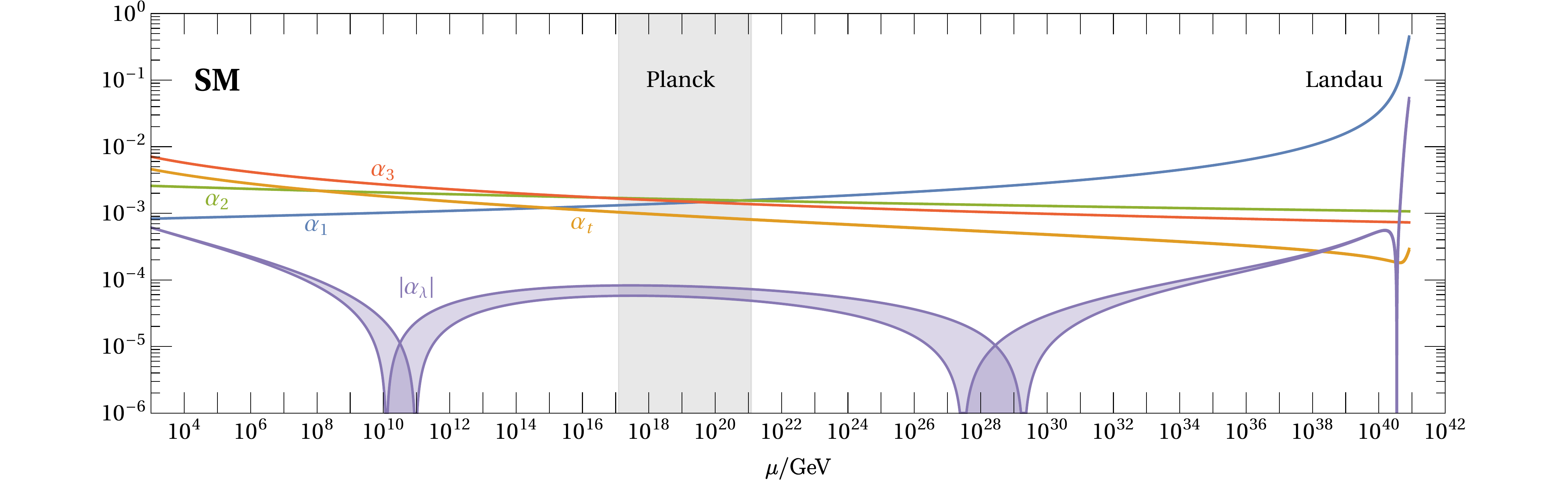}
    \caption{Shown is the running of the Higgs, top Yukawa, and gauge couplings
    in the Standard Model between the TeV and the Planck scale  (gray band), and further up to the Landau pole of the hypercharge coupling where the running breaks down. The SM  vacuum becomes unstable around $\mu \approx 10^{10}$ GeV and remains so up to the Planck scale, with the largest source of uncertainty arising from the  top pole mass
    (\ref{eq:mtop}), 
     indicated by   $1\sigma$-band widths for all couplings (see main text).} 
    \label{fig:SM-run}
\end{figure*}

The paper is organized as follows: We  recall the state of affairs in the SM and  basic strategies towards stability 
(\Sec{approach}).
We then explain why and how the gauge portal mechanism provides stability in a minimally invasive manner (\Sec{VLF}). 
We further include Yukawa interactions and contrast the availability of gauge vs Yukawa stability portals, 
and discuss phenomenology (\Sec{SingleYuk}).
  We conclude in \Sec{con}.

\section{Vacuum Stability
\label{sec:approach}}

\subsection{Standard Model}
We begin by taking stock of Higgs stability in the SM. 
To that end, we plot the 3-loop  running of SM couplings up to the Planck scale and beyond (\fig{SM-run}).
We also introduce   the $U(1)_Y \times SU(2)_L \times SU(3)_c$ gauge couplings $g_{\ell}\ (\ell=1,2,3)$, Yukawa interactions $Y$, and the Higgs quartic $\lambda$, all normalized in units of  loop factors, and write them as
 \begin{equation} \label{eq:alpha}
  \alpha_\ell = \frac{g_\ell^2}{(4\pi)^2}, \quad \alpha_Y = \frac{Y^2}{(4\pi)^2}, \quad \alpha_\lambda = \frac{\lambda}{(4\pi)^2}\,.
\end{equation}
 Due to their smallness, SM Yukawas other than those of the  top ($y_t$) and bottom ($y_b$) are immaterial for the RG evolution and can be safely neglected,
 corresponding to the approximation
 \begin{equation} \label{eq:YukApprox}
     Y_{ij}^{u}=y_t \,\delta_{i3}\,\delta_{j3}, \quad Y_{ij}^{d}=y_b \,\delta_{i3}\,\delta_{j3}, \quad Y_{ij}^{\ell}=0
 \end{equation}
for the up- and down-quark as well as the charged lepton Yukawa matrices, respectively.
SM initial conditions (central values) at the reference scale $\mu_0 =$1~TeV are determined as 
\begin{equation}\label{eq:Initial}
\begin{aligned}
\alpha_1(\mu_0)  &= 8.30 \cdot 10^{-4} ,\\
    \alpha_2(\mu_0) &= 2.58 \cdot 10^{-3} ,\\
     \alpha_3(\mu_0)& = 7.08 \cdot 10^{-3} ,\\
    \alpha_\lambda(\mu_0) &= 6.09 \cdot 10^{-4} ,\\
    \alpha_t(\mu_0) &= 4.61 \cdot 10^{-3} ,\\
   \alpha_b(\mu_0) &= 1.22 \cdot 10^{-6}\,.
\end{aligned}
\end{equation}
 The uncertainties  \cite{Buttazzo:2013uya} in the initial values \eq{Initial} due to the strong gauge coupling, Higgs and  $W$ mass are quantitatively irrelevant. 
The dominant source of uncertainty  originates from the determination of the top mass~\cite{Zyla:2020zbs}   
\begin{equation}\label{eq:mtop}
m_t = 172.76  \pm 0.30~{\rm GeV}\,.
\end{equation} 
We then have integrated the SM renormalization group equations (RGEs) from the TeV scale up to the hypercharge Landau pole (\fig{SM-run}).    
Due to its smallness, the bottom Yukawa $\alpha_b(\mu)$ is  not displayed but included in the numerics.
  We observe  that most SM couplings run slowly. We have also indicated the  $1\sigma$~uncertainty band \eq{mtop}  for all couplings.
In practice the band width is visible only for  the Higgs quartic (violet band), which takes higher (lower) values for a smaller (larger) top mass. 
The Higgs quartic invariably changes sign ($\mu\approx 10^{10}$~GeV) and the vacuum becomes unstable,  signaled by  the downward spike prior to the Planck scale (gray band). 
Stability at the Planck scale would require that the top mass deviates by more than $3\sigma$ from its presently determined central value. 
This suggests that a negative value for the Higgs quartic at the Planck scale and the possibility of an unstable ``great desert'' should be taken for real.

Taking the liberty of ignoring quantum gravity effects,  we  extend the SM flow into the transplanckian regime. Perhaps unexpectedly, we find that the Higgs becomes stable again  
($\mu_{\rm stab}\approx 10^{10} M_{\rm Pl}$), with the second sign change largely triggered by the mild but continued growth of the hypercharge coupling. 
The vacuum becomes  fully unstable at higher scales ($\mu\approx 10^{23} M_{\rm Pl}$),  shortly before the hypercharge Landau pole 
is reached which finally brings  down the SM as we know it.

We conclude that new effects  are required to stabilize the vacuum, either at the Planck scale such as from higher dimensional operators \cite{Branchina:2014rva}, or fully-fledged quantum gravity, or  from below the Planck scale such as from new particles or interactions. 

\subsection{Portals}
Next, we are interested how vacuum stability can be  improved using low-scale BSM mechanisms, some of  which  can  be tested at colliders.

The perhaps simplest such mechanism consists in the addition of  BSM fermions $\psi$  in a non-trivial representation under the $U(1)_Y \times SU(2)_L \times SU(3)_c$ SM gauge group. The main effect of these ``gauge portals''
\begin{align} \label{eq:Gauge}
\mathcal{L} \supset \bar \psi i   \!  \! \not \! \! D \psi \, 
\end{align}
arises through modifications of the RG-running of SM interactions. The exploration of  gauge portals, a main novelty of this work, is given in Sec.~\ref{sec:VLF}.

Alternatively,  ``Yukawa portals'' arise when  the Higgs couples directly to a BSM fermion $\psi$ and a SM fermion $f_{\text{SM}}$,
\begin{align} \label{eq:Yukawa}
\mathcal{L} \supset - \kappa \bar \psi H f_{\text{SM}} \,.
\end{align}
Unlike the gauge portals which solely involve new SM charge carriers, Yukawa portals are additionally controlled by new Yukawa interactions $\kappa$.  In general, Yukawa portals are flavorful with three SM generations  $j=1,2,3$ and  a BSM flavor index $i=1,.., N_F$.
These interactions allow the $\psi$ to decay to SM particles and avoid potential issues with otherwise too long-lived VLFs. 
To leading loop order, Yukawa couplings contribute to the running of the Higgs quartic as
\begin{equation}\label{eq:YukawaHiggs}
\beta_\lambda = \beta_\lambda^{\rm SM} +I_{\kappa\lambda}\, \alpha_\kappa\, \alpha_\lambda - I_{\kappa\kappa}\,\alpha_\kappa^2 
+{\cal{O}}(\text{2-loop})
\end{equation}
where both $I_{\kappa\lambda}, I_{\kappa\kappa} >0$~\cite{Machacek:1984zw,Hiller:2020fbu}. 
Sufficiently small  $\alpha_\kappa$ does not spoil   stability  as it  induces a positive one-loop contribution to the beta-function of the quartic. However, once $\alpha_\kappa\gtrsim \alpha_\lambda$, which is bound to happen if the Higgs potential becomes very flat, the second term takes over and starts to destabilize the vacuum.  It has also been observed that the Yukawa portal can stabilize the Higgs at larger coupling through a walking regime.\footnote{\label{walking}
Walking regimes refer to a parametrically slowed-down running of gauge couplings due to a near-zero of their beta functions, e.g.~\cite{Cohen:1988sq,Cacciapaglia:2020kgq}. More recently, walking regimes have also been observed in Yukawa and scalar sectors~\cite{Hiller:2019mou,Hiller:2020fbu,Bause:2021prv}.}
In addition, the Yukawa portal  to the Higgs always  implies a gauge portal \eq{Gauge}, and their interplay is analyzed in  \Sec{SingleYuk}.  We also note that  several  BSM Yukawas with $H$ can arise if several suitable representations of BSM fermion exist, say $\psi$ and $\chi$,  allowing for additional Yukawa portals such as $\bar \psi  H \chi $.

Finally, in the presence of BSM scalars,  ``Higgs portals'' arise as soon as the Higgs $H$ couples to the BSM scalar $S$
through a portal coupling  $\delta$ 
\begin{align} \label{eq:Higgs}
\mathcal{L} \supset - \delta\, S^\dagger S\, H^\dagger H \,.
\end{align}
This new portal interaction  improves vacuum stability as it provides a positive one-loop contribution to the running of the Higgs quartic
$\beta_\lambda \approx \beta_\lambda^{\rm SM} +I_\delta \alpha_\delta^2$ and $ I_\delta >0$~\cite{Machacek:1984zw,Hiller:2020fbu}. Also, Higgs portals come along with gauge portals \eq{Gauge} 
provided the BSM scalar is charged under the SM; see \cite{Hiller:2020fbu} for  models with gauge, Higgs and Yukawa portals simultaneously.

Previous works have  studied vacuum stability from the viewpoint of concrete BSM models.
These include SM extensions by
single uncharged real~\cite{Falkowski:2015iwa,Khan:2014kba,Han:2015hda,Garg:2017iva}, complex~\cite{Gabrielli:2013hma,Elias-Miro:2012eoi,Gonderinger:2012rd,Costa:2014qga,Khoze:2014xha,Anchordoqui:2012fq}, or charged scalars 
such as leptoquarks \cite{Bandyopadhyay:2016oif,Bandyopadhyay:2021kue} or other~\cite{Chakrabarty:2020jro,Hamada:2015bra}.
Many works feature both scalar and fermionic BSM fields, such as
type~I~\cite{Ng:2015eia,Mandal:2019ndp,Lindner:2015qva}, type~II~\cite{BhupalDev:2013xol,Khan:2016sxm,Haba:2016zbu,Ghosh:2017pxl}
  type~III~\cite{Goswami:2018jar,Lindner:2015qva} and inverse seesaw models \cite{Mandal:2020lhl,DelleRose:2015bms}
 as well as theories with matrix scalar fields~\cite{Latosinski:2015pba,Hiller:2019mou,Hiller:2020fbu,Bause:2021prv}, or models 
with uncharged  scalars coupling to charged BSM fermions~\cite{Xiao:2014kba,Dhuria:2015ufo,Salvio:2015jgu,Son:2015vfl,DuttaBanik:2018emv,Borah:2020nsz}.
More complex extensions include
Two-Higgs-doublet models, e.g.~\cite{Chowdhury:2015yja,Khan:2015ipa,Ferreira:2015rha,Ferreira:2015pfi,Bhattacharya:2019fgs,Swiezewska:2015paa,Chakrabarty:2016smc,Schuh:2018hig,Bagnaschi:2015pwa},  $U(1)'$ extensions~\cite{Duch:2015jta,Coriano:2014mpa,Coriano:2015sea,Oda:2015gna,Das:2015nwk,Das:2016zue,Marzo:2018nov,Haba:2015rha,Oda:2017kwl,DiChiara:2014wha,Datta:2013mta,Bause:2021prv}, or extended dark gauge sectors~\cite{Altmannshofer:2014vra},
GUTs~\cite{Mambrini:2016dca,Ellis:2018khn,Haba:2015nwl,Boucenna:2018wjc,Haba:2014oxa,Okada:2019bqa,Fabbrichesi:2020svm} and many others~\cite{Salvio:2015cja,Belanger:2014bga,Ghosh:2017fmr,Chakrabarty:2015kmt,Chang:2016pya,DuttaBanik:2020jrj,Lalak:2014qua,Wang:2018lhk,Hamada:2016vwk}.
Extensions that rely solely on BSM fermions are less common, notable exceptions include \cite{Arkani-Hamed:2012dcq,Joglekar:2012vc,Fairbairn:2013xaa,Blum:2015rpa,Altmannshofer:2013zba,Gopalakrishna:2018uxn}.

In this work, we systematically explore SM extensions with $N_F$ vector-like fermions (VLFs) of mass $M_F$ 
and charged under the $U(1)_Y \times SU(2)_L \times SU(3)_c$ SM gauge group. 
By design, all models are free of gauge anomalies, and allow for direct Dirac mass terms.
Our setup gives  access to gauge portals \eq{Gauge} which we discuss in   \Sec{VLF}, and to gauge-Yukawa portals \eq{Yukawa} which we discuss  \Sec{SingleYuk}. 
To test vacuum stability quantitatively, we match SM extensions onto the SM at the mass scale of the BSM fermions which we take to be in the TeV range or above,
\begin{equation}\label{eq:matching-scale}
    \mu_0 = M_F \gtrsim 1 \, \text{TeV}\,.
\end{equation}
SM couplings are matched to their BSM counterparts via
  \begin{equation}\label{eq:matching}
  \alpha_{1,2,3,t,b,\lambda}^\text{BSM} (\mu_0) = \alpha_{1,2,3,t,b,\lambda}^\text{SM} (\mu_0)\,.
  \end{equation}
  We briefly comment on threshold corrections, which are neglected in this work. 
 In principle, threshold corrections arise at loop level  and replace the dependence on the unphysical matching scale $\mu_0$ by the physical mass  $M_F$ (see e.g.~\cite{Manohar:2018aog} for a pedagogical introduction). 
     One-loop thresholds to the gauge couplings are well known to be $\propto \ln\tfrac{\mu_0}{M_F}$~\cite{Weinberg:1980wa} and therefore vanish as we  match at $\mu_0=M_F$, see \eq{matching-scale}.
  Moreover, for BSM fermions which solely interact with the SM gauge sector,  threshold corrections to the SM Yukawas and Higgs self-interaction only start at two loops.
  Consequently, our matching procedure \eq{matching} only neglects threshold corrections that are suppressed by two loop orders and higher.

 We employ the tool \texttt{ARGES} \cite{Litim:2020jvl} to generate the  two-loop $\beta$-functions for all running couplings, which are then integrated numerically between the scale of new physics $\mu_0$ and the Planck scale. The goal is to identify the set of BSM parameters such as the mass scale $M_F$  and the field multiplicities  where the RG flow remains well-defined without Landau poles and the vacuum becomes stable at the Planck scale. We refer to this set of parameters as the ``BSM critical surface''. 

As a side aspect, we also watch out for settings where the Higgs coupling remains stable all the way up to the Planck scale, as opposed to settings where it may have changed its sign (say, twice) along the trajectory up to the Planck scale. The former stipulates that loop corrections, resummed by the RG evolution, may not overturn the tree-level condition at any scale. The latter seeks to minimize radiative corrections to the resummed potential by imposing stability at the highest scale.\footnote{In our models,  if a quartic changes sign twice prior to the Planck scale, it still remains  larger than in  the SM ($\alpha_\lambda\simeq - 10^{-4})$, see Fig.~\eqref{fig:SM-run}. Hence, models are always less unstable than the SM, or even outright stable due to positive quartics at the Planck scale (see  Fig.~\eqref{fig:pulldown} below for examples). Stability can further be confirmed by  computing  the full Higgs effective potential.}

\section{Gauge Portals} \label{sec:VLF}

In this section we  introduce and explain the mechanism of  gauge portals for vacuum stability. 
We then illustrate the different gauge portals quantitatively by looking into  concrete extensions of the SM and  their main features and characteristics.

\subsection{Gauge Portals at Work
 \label{sec:inter}}

We begin by explaining why gauge portal mechanisms can stabilize the Higgs potential in a minimally-invasive manner.
The essence of the mechanism is the addition of $N_F$ vector-like fermions of mass $M_F$,  charged under the SM gauge group as  
\begin{equation}\label{eq:VLF}
(Y_F,d_2,d_3)\,,
\end{equation}
and without adding any  new interactions. The latter can be achieved either by specific choices of representations which forbid Yukawa interactions with SM fields, or by noting that Yukawa couplings are natural and cannot be switched on through quantum fluctuations.
The net primary effect of the new matter fields is a modification of the running gauge couplings. 
Modifications to the running of the Higgs quartic, which is the ultimate goal of the mechanism, are much milder in that they only arise   indirectly, namely  through a modified running of the top Yukawa and the gauge  couplings and without changing the beta function itself. 

To understand the gauge portal mechanism more quantitatively, 
we start with  the running of the gauge couplings 
at one-loop
\begin{equation}\label{eq:beta-gauge}
    \beta_i \approx - B_i\,\alpha_i^2\,.
\end{equation} 
Integrating the flow between the matching scale $\mu_0$ and some higher scale $\Lambda$ we find 
\begin{equation} \label{eq:1loop}
    \alpha_i(\Lambda) = \frac{\alpha_i(\mu_0)}{1 + \alpha_i(\mu_0) B_i \ln\left(\frac{\Lambda}{\mu_0}\right)}.
\end{equation}
The one-loop gauge coefficients
\begin{equation}
     \begin{aligned}
     B_1 &= -\tfrac{41}3 - \delta B_1, \\ 
     B_2 &= \ \ \tfrac{19}3 - \delta B_2, \\ 
     B_3 &= \ \ 14 - \delta B_3\,,
    \end{aligned}
\end{equation}
contain the SM contributions, while the  BSM contributions  \eq{VLF} are hidden in the coefficients
 $\delta B_{i} \geq 0$ which  read
\begin{equation} \label{eq:deltaB}
    \begin{aligned}
      \delta B_1 &= \tfrac83 N_F \, d_2\, d_3 \,Y_F^2,\\
      \delta B_{2,3} &=\tfrac83 N_F \, d_{3,2} \, S_2(d_{2,3})\,.
    \end{aligned}
\end{equation}
Here, $S_2(d_{2,3})$ denote the respective Dynkin indices of the $SU(2)_L$ and $SU(3)_c$ representations. 
If BSM fields are integrated out at the matching scale $\mu_0$, the gauge couplings generically take larger values than the  SM running at a high scale $\Lambda > \mu_0$. Using \eq{1loop}, we have
\begin{equation}\label{eq:deviation-gauge}
    \frac{1}{\alpha_i(\Lambda)} - \frac{1}{\alpha_i^\text{SM}(\Lambda)} = - \delta B_i \ln\left(\frac{\Lambda}{\mu_0}\right) <0 
\end{equation}
in comparison with the running in the SM. 

Next, we turn to the top Yukawa coupling. Owing to its magnitude in the SM,  it plays a significant role for the stability of the Higgs potential. 
Its  one-loop flow is given by  
\begin{equation}\label{eq:beta-top}
    \beta_t \approx \alpha_t \left[ 9\,\alpha_t - \tfrac{17}6\, \alpha_1 - \tfrac92\, \alpha_2  - 16 \, \alpha_3\right]\,.
\end{equation}
Together with \eq{deviation-gauge} it implies that $\alpha_t$ decreases faster than in the SM.  Integrating the flow to leading order in $\delta B_i$ and to leading logarithmic accuracy, we find
\begin{equation}\label{eq:deviation-top}
\begin{aligned}
      \alpha_t(\Lambda) &-  \alpha_t^\text{SM}(\Lambda)  \approx - \alpha_t(\mu_0) \, \Delta_t (\mu_0)\, \ln^2 \left(\frac{\Lambda}{\mu_0}\right) <0 ,  \\
     \Delta_t & =  \tfrac{17}{12} \delta B_1 \,\alpha_1^2 
       + \tfrac94 \delta B_2 \,\alpha_2^2 + 8 \,\delta B_3 \, \alpha_3^2 .
\end{aligned}
\end{equation}
Note that the result is a resummation effect as \eq{beta-gauge} and \eq{beta-top} are one loop.
One-loop contributions to \eq{deviation-top} are absent because the leading-order top running \eq{beta-top} is insensitive to charged new fields, but there is a modification of the one-loop running of gauge couplings contained within \eq{beta-top}, which manifests itself at two-loop for $\alpha_t(\mu)$.

Turning  to the Higgs quartic interaction $\alpha_\lambda$, we note that its value  is much smaller than the gauge  and  top Yukawa couplings. Hence, its RGE is primarily driven by the inhomogeneous one-loop terms
\begin{equation}\label{eq:quartic-beta}
    \beta_\lambda \approx \tfrac{3}{8}\left[\alpha_1^2 + 2\, \alpha_1 \alpha_2 + 3\, \alpha_2^2\right] - 6\, \alpha_t^2 \,.
\end{equation}
We observe that the gauge interactions increase the quartic coupling while  the top contribution reduces it. Hence, larger $\alpha_{1,2}$ couplings \eq{deviation-gauge} and  a smaller $\alpha_t$ coupling \eq{deviation-top} are  beneficial for stability.
Integrating the flow and subtracting the SM result, and expanding  to leading orders in $\delta B_{1,2,3}$ and leading logarithms, we find
\begin{equation}\label{eq:quartic-enhance}
    \begin{aligned}
        &\alpha_\lambda(\Lambda) - \alpha^\text{SM}_\lambda(\Lambda) \approx \\
        &\qquad + \tfrac38 \alpha_1^2(\mu_0)\left[\alpha_1(\mu_0) + \alpha_2(\mu_0)\right] \delta B_1 \, \ln^2\left(\frac{\Lambda}{\mu_0}\right) \\
        &\qquad + \tfrac38 \alpha_2^2(\mu_0)\left[\alpha_1(\mu_0) + 3\alpha_2(\mu_0)\right] \delta B_2 \, \ln^2\left(\frac{\Lambda}{\mu_0}\right) \\
        &\qquad + 32\,\alpha_t^2(\mu_0)\,\alpha_3^2(\mu_0)\,\delta B_3 \,\ln^3\left(\frac{\Lambda}{\mu_0}\right)\,, 
    \end{aligned}
\end{equation}
which is the central result of this section. A number of comments are in order:
\begin{itemize}

\item
From the result \eq{quartic-enhance} we observe that the enhancement of $\alpha_{1}$ and $\alpha_{2}$ over SM values  \eq{deviation-gauge} also yields a direct uplift of $\alpha_\lambda$ at two-loop level and leading-log, characterized by the terms $\propto \delta B_{1,2}$. There is no such contribution $\propto \delta B_{3}$ because the Higgs is not charged under 
$SU(3)_c$.

\item
The leading impact from the modified running of the strong gauge coupling is channeled through the last term in \eq{quartic-beta} and
therefore suppressed by another loop compared to weak isospin and hypercharge effects. 
The detour via the top Yukawa implies that the leading contributions $\propto \delta B_{3}$ to \eq{quartic-enhance} arise
 at  three loop and through leading logs. Quantitatively, the higher-loop suppression may well be compensated by the larger numerical 
 coefficients in the last term of \eq{quartic-enhance} and the size of $\alpha_{3,t}(\mu_0)$. 
 Similar three-loop enhancements via the top coupling also arise $\propto \delta B_{1,2}$, 
but have not been shown as they are subleading to the terms already in \eq{quartic-enhance}.

\item
Since all terms on the right hand side of \eq{quartic-enhance} are  positive, we  conclude  that
\begin{equation}\label{eq:quartic-enhancement}
        \alpha_\lambda(\Lambda) - \alpha^\text{SM}_\lambda(\Lambda)>0\,,
\end{equation}
leading   to an  uplift of the Higgs quartic $\alpha_\lambda(\Lambda)$, and, potentially, stability of the vacuum.
The result also establishes that the  SM gauge charges   \eq{VLF} of the newly-added matter fields  all pull  into the  direction of stability, even though the Higgs does not carry color. We conclude that vector-like fermions  are natural candidates for stability.

\item
Owing to the fact that the leading loop coefficients of the scalar and top Yukawa beta functions have not changed, the accelerated running of gauge couplings \eq{deviation-gauge} implies that $\alpha_\lambda$  runs along the SM trajectory but with an accelerated speed, $\alpha_\lambda(\mu)\approx \alpha^{\rm SM}_\lambda(\mu_{\rm SM}(\mu))$ with $\mu_{\rm SM}(\mu)>\mu$. As such, the new $\alpha_\lambda$ trajectory comes out as a squeezed version of the SM one (\fig{SM-run}), plus a small uplift \eq{quartic-enhancement}.
 
\item
A consequence of  \eq{quartic-enhancement}   is that the Higgs quartic, for any gauge portal extension of the SM, is naturally bounded from below by its most negative value achieved in the SM, reading  $\alpha_\lambda \gtrsim - 10^{-4}$ if we discard the region of the third sign change close to the hypercharge Landau pole.
\end{itemize}
It is  interesting to explore the space of SM extensions with $\delta B_i>0$ quantitatively. 
In the remainder, we focus on the strong gauge portal $(\delta B_3>0)$, the weak gauge portal $(\delta B_2>0)$, and the hypercharge portal $(\delta B_1>0)$ separately. We also point out that the strong gauge portal as a mechanism for stability has previously been noted in \cite{Gopalakrishna:2018uxn}.

\begin{figure}
  \centering
  \renewcommand*{\arraystretch}{0}
    \includegraphics[width=.75\columnwidth]{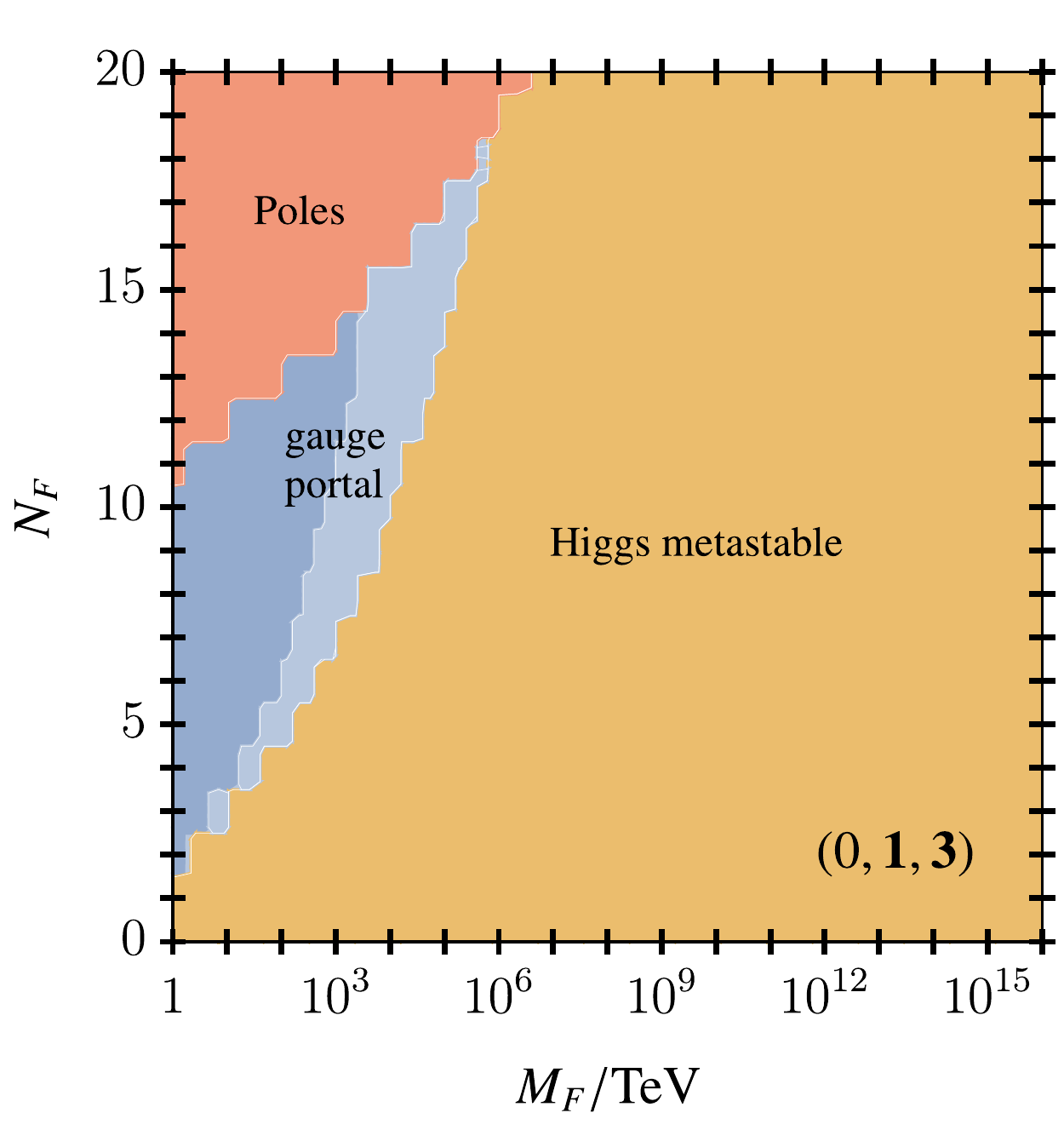}
  \caption{Illustration of the strong gauge portal for  SM extensions with $N_F$ generations of VLQs  of mass $M_F$ and in the representation $(0,\mathbf{1},\mathbf{3})$. Shown is the  critical surface in the $(N_F,M_F)$  plane indicating parameter  regions where the Higgs potential at the Planck scale remains either metastable such as in the SM (yellow), is plagued by a  subplanckian Landau pole (red),  or has become stable (blue). We further indicate whether the Higgs quartic remains positive along the entire trajectory (dark blue) or whether it has become negative in some range  prior to the Planck scale (light blue). 
  We observe that the strong gauge portal opens a  wedge  which extends moderately into the high mass / high multiplicity region.
  }
  \label{fig:StrongPortal}
\end{figure}

\subsection{Strong Gauge Portal \label{sec:strong}}

We study SM extensions with  $N_F$ new VLFs  in the representation $(0,\mathbf{1},d_3)$ and of mass $M_F$, with  parameters $\delta B_3>0$ and $\delta B_{1,2}=0$. We refer to the strong gauge portal as  the  set of parameters $N_F$, $M_F$, and $d_3$ which ensure stability of the vacuum at the Planck scale.  

To illustrate the gauge portal mechanism we consider vector-like quarks (VLQs) with $d_3={\bf 3}$. We integrate the RG flow up to the Planck scale  and take $N_F$ and $M_F$ as free parameters. The UV critical surface of parameters is shown in  \fig{StrongPortal},
indicating parameter regions where the vacuum at the Planck scale remains metastable (yellow), has become stable (blue), and regions where  the RG flow has blown up prior to the Planck scale due to Landau poles (red).
We observe that if $M_F$ is too large and the multiplicity $N_F$  too small,  there is not enough ``RG time'' between $M_F$ and $M_{\rm Pl}$ to lift the potential into stability. The impact   is ``too little too late'', and we are left with metastability of the vacuum just as in the SM (yellow area).  
On the other hand, if $N_F$ is too large, i.e.~$N_F > \frac{21}{2}$, asymptotic freedom in $\alpha_3$ is lost.  If additionally $M_F$ is too small, this may lead to a breakdown of predictivity due to a Landau pole prior to the Planck scale (red area). In particular, this excludes  higher representations $d_3 \geq {\bf 10}$ as a single one of these would cause subplanckian Landau poles.

The sweet spot of SM extensions with stable vacua are situated in the wedge (blue regions) between the regions of metastability and Landau poles, characterized by  upper and lower bounds on $N_F$, and an upper bound on $M_F$. The latter can be understood as follows: For too high values of $M_F$,  SM running implies that $\alpha_{3,t}$ has become too small at the matching scale $\mu_0$ for the uplift \eq{quartic-enhance} to be sufficient. In fact, even if asymptotic freedom is lost, the BSM growth of $\alpha_3$   and the thereby induced decrease of $\alpha_t$ are insufficient to generate  enough uplift \eq{quartic-enhance}. We conclude that once $\alpha_t(M_F)$ is too small, the strong gauge portal mechanism  is insufficient. 

\begin{figure}
    \centering
    \includegraphics[width=.9\columnwidth]{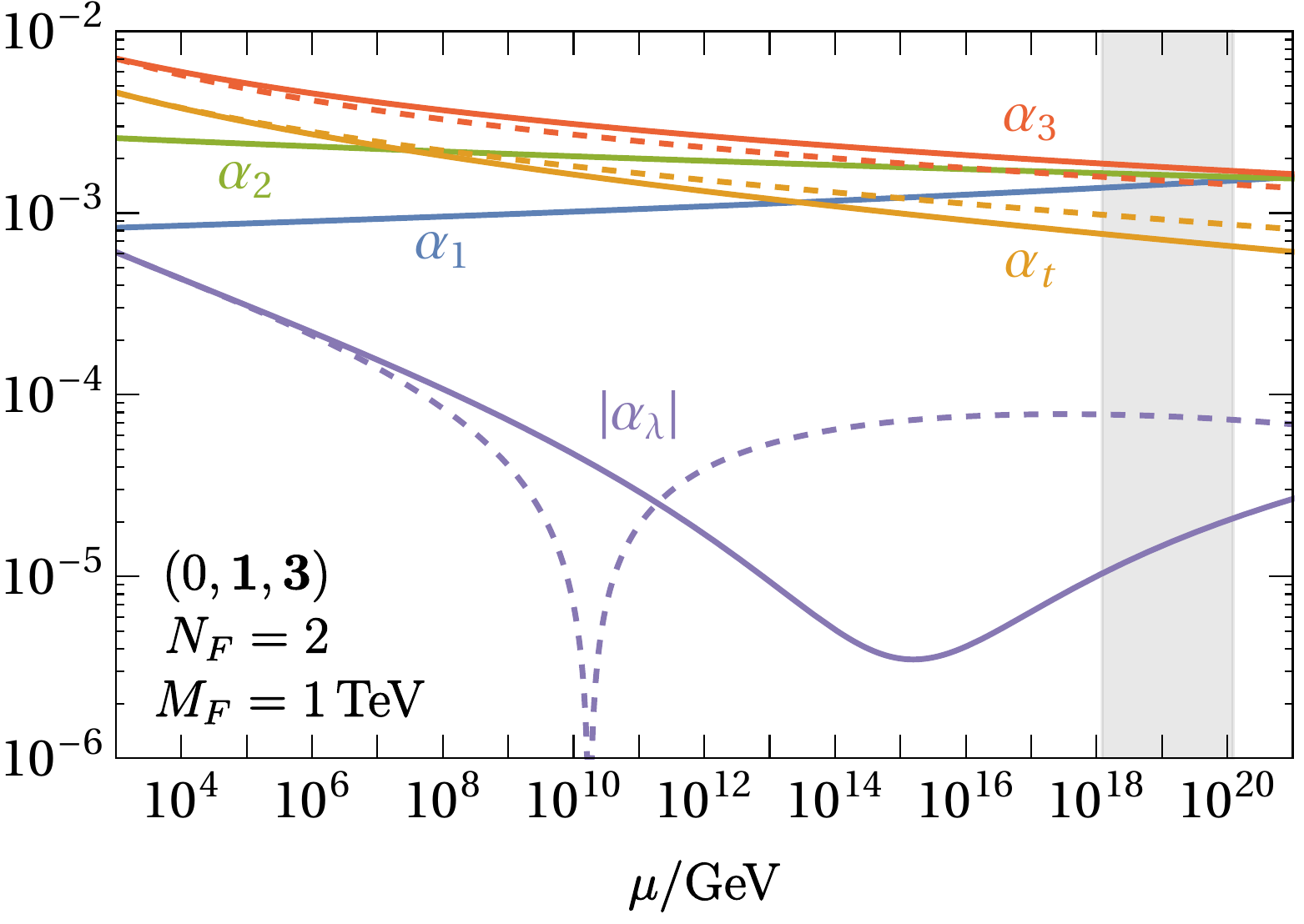}
    \caption{Example for a strong gauge portal, showing the two-loop running of couplings for a SM extension featuring
    $N_F = 2$ Dirac fermions [solid lines] in the representation
     $(0, \mathbf{1}, \mathbf{3})$, in comparison with  the SM [dashed lines]. 
     The mild enhancement of  $\alpha_3$ together with  the mild decrease of $\alpha_t$ ensure that $\alpha_\lambda$ remains positive throughout.}
     \label{fig:VLQ-run}
\end{figure}

Regions of stability (blue)  arise in  two manners, either by uplifting the first sign change of $\alpha_\lambda$ leading to $\alpha_\lambda\ge 0$ for all scales below $M_{\rm Pl}$ (dark blue region), or by pulling the second sign change down to subplanckian energies (light blue region). 
The first case is illustrated in  \fig{VLQ-run}  for  $N_F = 2$ and $M_F=1$~TeV, where the two-loop  RG trajectories (solid lines) are  contrasted with  the SM (dashed lines). We observe that the  mild enhancement of $\alpha_{3}$ (red)  and the mild reduction of the top Yukawa (yellow) over SM running is sufficient to fully stabilize the Higgs quartic (violet), in accord with the analytical estimate  in \eq{quartic-enhance}.

The second case are settings where   the uplift \eq{quartic-enhance} is not enough, but where the squeezing of the $\alpha_\lambda$ trajectory has pulled down  the second sign change to below planckian energies. 
Consequently, at the boundary between metastability (yellow) and stability (blue) we  find a 
vanishing quartic at the Planck scale,
\begin{equation}\label{eq:flat}
\lambda|_{\mu=M_{\rm Pl}} = 0\,.
\end{equation}
Concrete examples and more details  are discussed  in \Sec{hyp} (\fig{pulldown} and \fig{VLL-run-YF}) below.

Altogether, we observe from \fig{StrongPortal}  that stability at the Planck scale leads to the  parameter constraints
\begin{equation}
    \begin{aligned} \label{eq:PureVLQBounds}
    M_F \lesssim 10^3\,\text{TeV},& \qquad 2 \leq N_F \leq 14, \\
   (M_F \lesssim 10^6\,\text{TeV},& \qquad 2 \leq N_F \leq 18).
    \end{aligned}
\end{equation}
provided we demand stability with $\alpha_\lambda\ge 0$ all the way up to (at least at) the  Planck scale. Similar bounds are found for other representations $d_3<{\bf 10}$. 

The result \eq{PureVLQBounds} also demonstrates  the constraining power of  Planck scale stability.
Further, TeV-mass VLQs without or with super-feeble Yukawas to SM fields can be experimentally probed in $R$-hadron or dijet resonance searches, or  measurement of $\alpha_3$-running \cite{ATLAS:2020mee,Bond:2017wut}.

\begin{figure}
    \centering
    \includegraphics[width=.9\columnwidth]{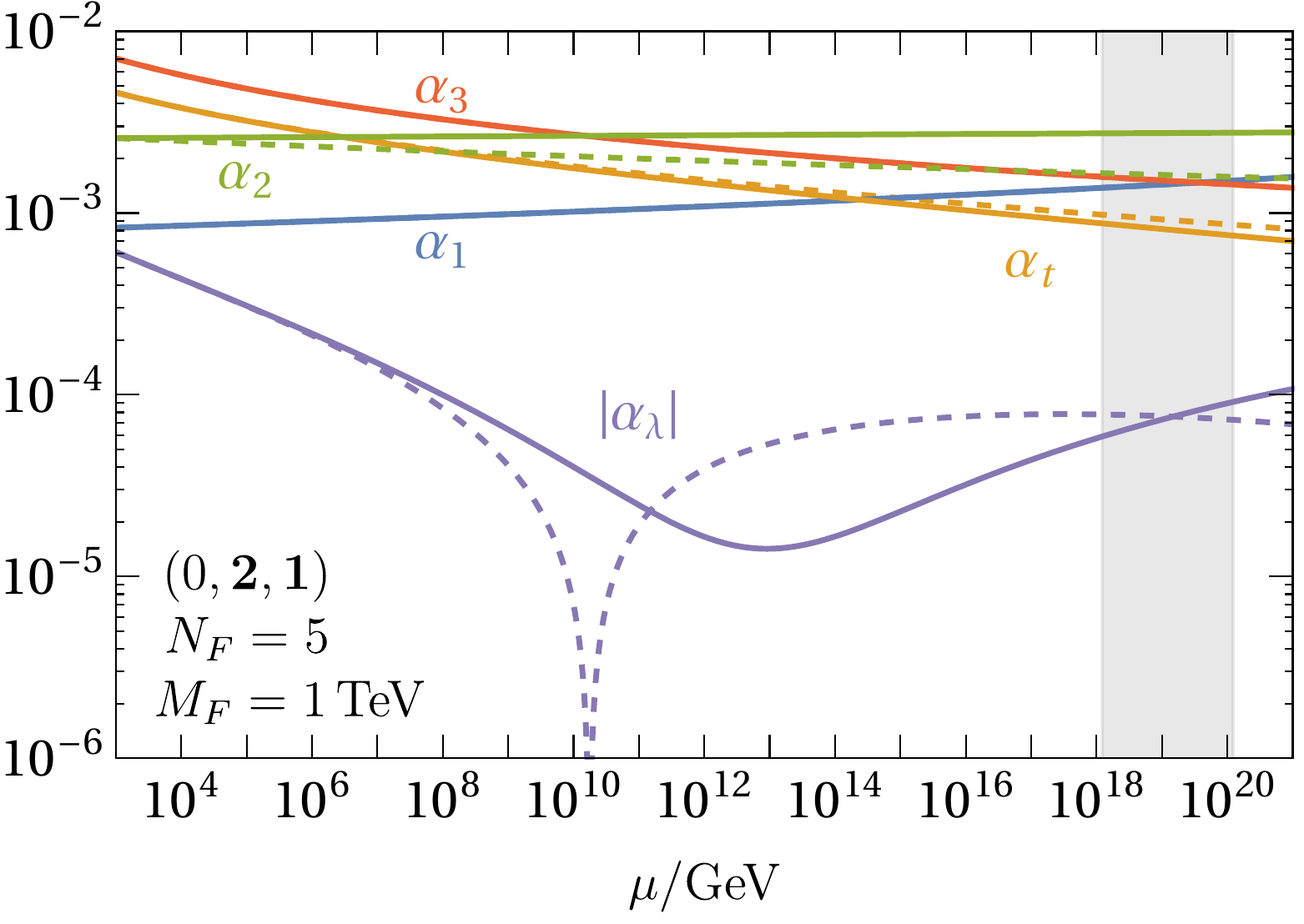}
    \caption{Example for a weak gauge portal, showing the two-loop running of couplings for a SM extension featuring
    $N_F = 5$ Dirac fermions [solid lines] in the representation $(0, \mathbf{2}, \mathbf{1})$, in comparison with  the SM [dashed lines]. 
    The mild enhancement of $\alpha_2$ ensures that  $\alpha_\lambda$ remains positive throughout.}
    \label{fig:VLL-run}
\end{figure}

\begin{figure}
  \centering
  \renewcommand*{\arraystretch}{0}
  \includegraphics[width=.75\columnwidth]{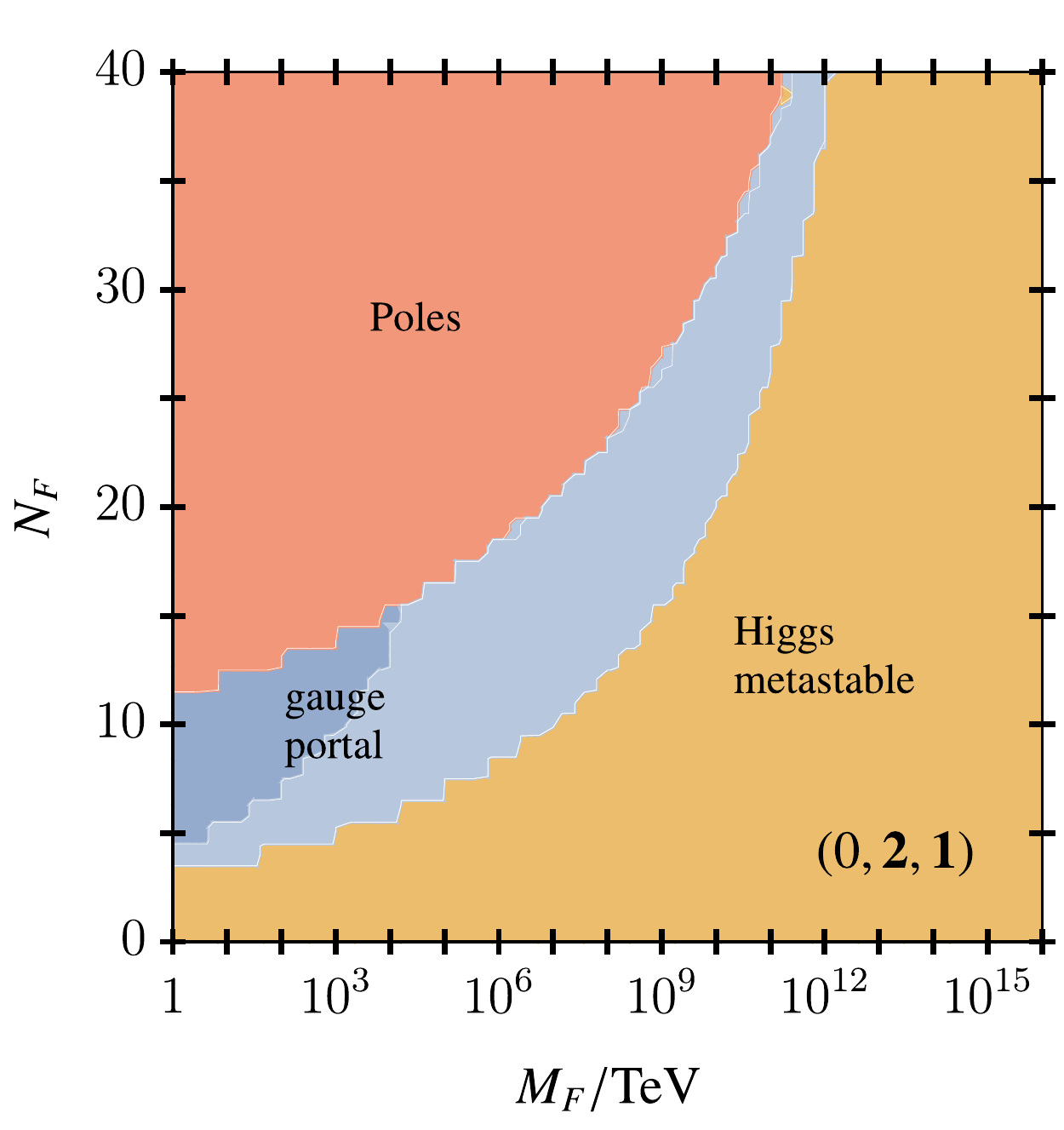}
  \caption{Illustration of the weak gauge portal  for  SM extensions with $N_F$ generations of VLLs  of mass $M_F$ and in the representation $(0,\mathbf{2},\mathbf{1})$, showing  the  critical surface of parameters in the $(N_F,M_F)$ plane with color coding as in \fig{StrongPortal}. Poles relate to   subplanckian Landau poles in $\alpha_2$. We observe that the weak gauge portal extends substantially into the high mass / high multiplicity region.}
  \label{fig:WeakPortal}
\end{figure}

\subsection{Weak Gauge Portal \label{sec:weak}}

Next, we consider the weak gauge portal  characterized by $N_F$ new vectorlike leptons (VLLs) of mass $M_F$ in the representation $(0,d_2,\mathbf{1})$, corresponding to    parameters $\delta B_2>0$ and $\delta B_{1,3}=0$. We search for  parameter windows where the VLL contributions  are sufficiently large to ensure stability of the quartic, but not too large to generate Landau poles in the electroweak couplings below $\MPl$. 

To illustrate the weak gauge portal at work, a concrete example  is depicted in \fig{VLL-run} featuring $N_F=5$ VLLs of mass $M_F=1$~TeV in the  $(0, \mathbf{2}, \mathbf{1})$ representation.  We observe that the new VLLs induce a slight enhancement of the weak gauge coupling  $\alpha_2$ (solid, green),  which proves sufficient to stabilize the Higgs quartic (solid violet curve) along the entire trajectory up to the Planck scale, and  in accord with the analysis leading up to  \eq{quartic-enhance}. 

The BSM critical surface for  SM extensions with $N_F$ VLLs of mass $M_F$ and in the $(0, \mathbf{2}, \mathbf{1})$ representation
  is shown   in \fig{WeakPortal}. We again observe that the wedge of stability is situated between the metastability and pole regions. Unlike in the strong gauge portal,  no upper bound on $N_F$ arises. Consequently, there is no upper limit on $M_F$  either,
except at their natural cut-off values around $10^{10}$ GeV (to stop the sign flip of $\alpha_\lambda$) or up to  $M_{\rm Pl}$ (elsewise). 
Larger $M_F$ implies that gauge couplings are much smaller at the matching scale. However, as long as subplanckian Landau poles remain absent, this is countered by larger $N_F$, which  allows $\alpha_2$
to grow fast enough to stabilize the Higgs. 
We emphasize that this mechanism is not operative for the strong gauge portal.

From \fig{WeakPortal} we note  that the weak gauge portal  starts out at  $N_F \geq 5 (4)$, depending on whether we demand $\alpha_\lambda\ge 0$ up to (at) the Planck scale.
Choosing triplet VLLs $(0, \mathbf{3}, \mathbf{1})$, stability starts for smaller $N_F \geq 2 (1)$,  and the window is  narrower.
We also find that  $d_2 \geq {\bf 5}$ leads to subplanckian poles already for  a single generation.

\begin{figure}
  \centering
  \renewcommand*{\arraystretch}{0}
  \includegraphics[width=.75\columnwidth]{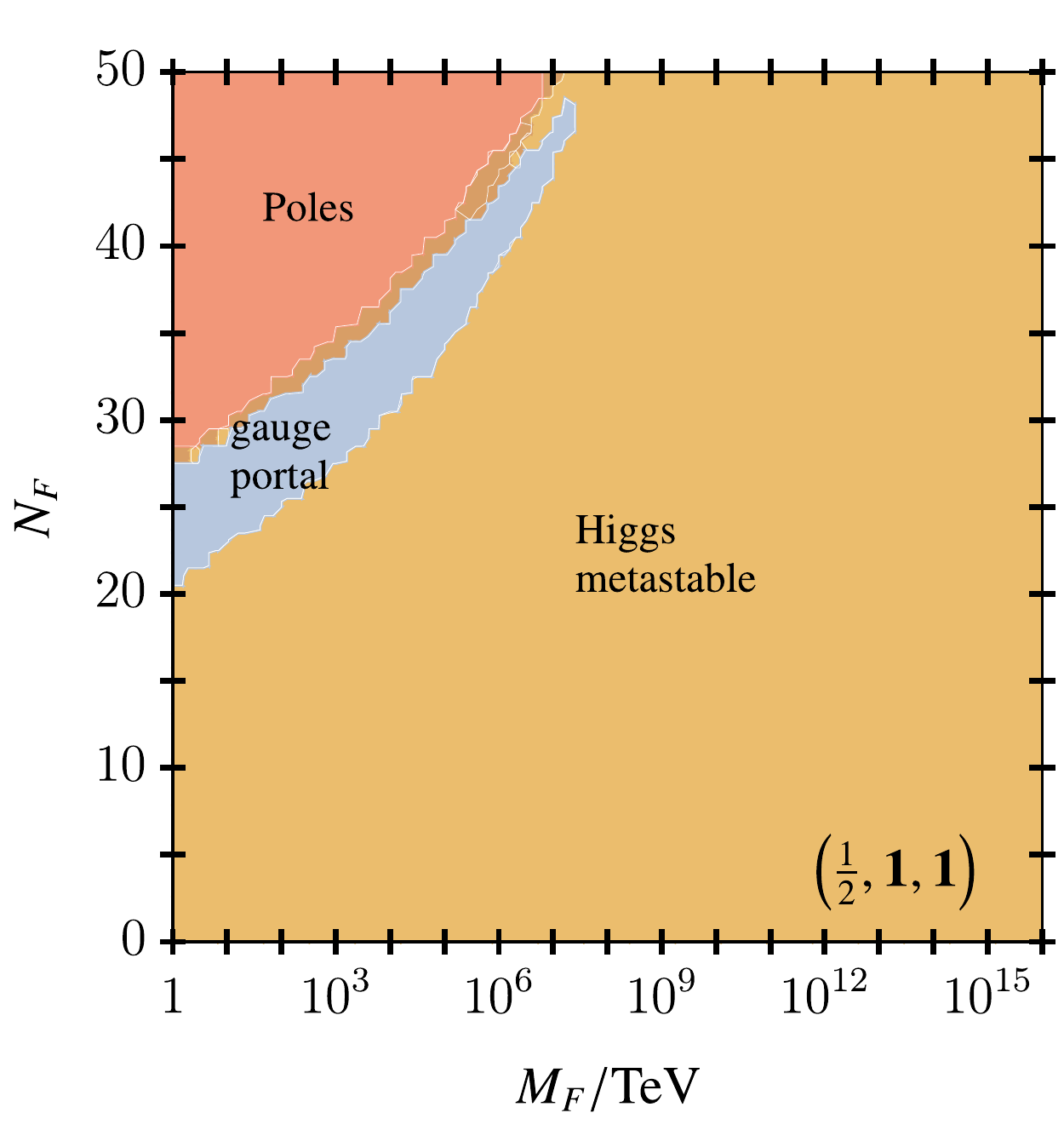}
  \caption{Illustration of the hypercharge portal  for  SM extensions with $N_F$ generations of VLLs  of mass $M_F$ and in the representation $(\frac{1}{2},\mathbf{1},\mathbf{1})$, showing  the  critical surface of parameters in the $N_F$, $M_F$ plane with color coding as in \fig{StrongPortal}. Here, poles relate to   subplanckian Landau poles in $\alpha_1$. We observe that the hypercharge portal extends  into the high mass / high multiplicity region. Much unlike  the strong and weak gauge portals, we find no regions where the Higgs quartic is positive along the entire trajectory.}
  \label{fig:HyperchargePortal}
\end{figure}

\subsection{Hypercharge Portal \label{sec:hyp}}

Finally, we consider the hypercharge  portal  characterized by $N_F$ new VLLs of mass $M_F$ in the representation $(Y_F,\mathbf{1},\mathbf{1})$, corresponding to    parameters $\delta B_1>0$ and $\delta B_{2,3}=0$. 
Interestingly, also the hypercharge portal is open, despite the looming Landau pole.
The BSM critical surface for $Y_F=\frac{1}{2}$ is shown  in \fig{HyperchargePortal}, where we observe that the stability wedge has a boundary with the  metastability  region. 
Most notably, we do not find any region where $\alpha_\lambda>0$ all the way up to the Planck scale. This states that the uplift in  \eq{quartic-enhance} due to  hypercharge alone  is insufficient. 

\begin{figure}
    \centering
\vskip-.3cm
\hspace*{-.7cm}    
\includegraphics[width=1.15\columnwidth]{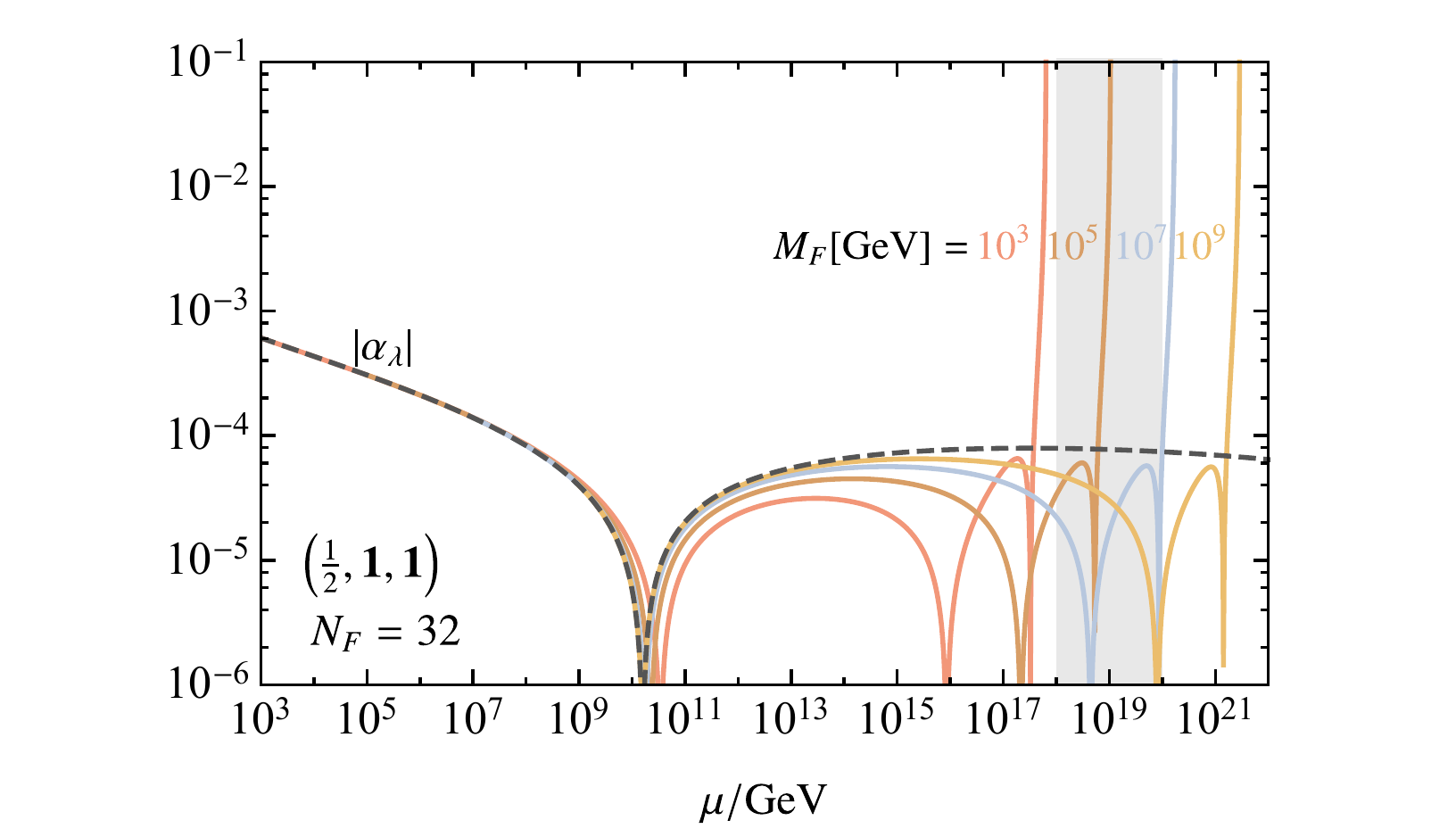}
    \caption{Example for the ``pull-down'' of sign flips, showing  the two-loop running of the Higgs quartic  
    for a SM extension with $N_F = 32$ Dirac fermions in the representation $(\frac{1}{2}, \mathbf{1}, \mathbf{1})$  for different masses $M_F$ [full lines] in comparison with the SM [dashed line].  While the location of the first sign flip is only mildly affected, the location of the second and third ones have moved down significantly from much above the Planck scale (\fig{SM-run})  to  the Planckian region and below. 
    The effect is more pronounced for smaller $M_F$.}        \label{fig:pulldown}
\end{figure}

Instead, stability solely arises through the pull-down of the second sign change,
which is further  illustrated  in \fig{pulldown}  for $N_F=32$ and $Y_F=\frac{1}{2}$. Varying the VLL masses $M_F$, which corresponds to a horizontal cut across \fig{HyperchargePortal}, we observe that the  sign changes of $\alpha_\lambda$ located at transplanckian energies in the SM (\fig{SM-run}), are pulled-down to below planckian energies. As such, this  illustrates the ``squeezing'' of the Higgs quartic SM trajectory, see~\Sec{inter}. Evidently, the effect is more pronounced for smaller $M_F$ as this triggers an earlier start of the accelerated $\alpha_1$ growth.  Moreover, if the pull-down is too substantial, even the third sign change may arise prior to the Planck scale, typically around $\alpha_1 \gtrsim \text{few} \times 10^{-2}$, followed by a Landau pole shortly thereafter.
In consequence, the stability region in \fig{HyperchargePortal} develops a border with  a new  (narrow) region where the vacuum is unstable, and just before  Landau poles take over. For \fig{HyperchargePortal},
we conclude that stability 
is achievable for $M_F \lesssim 3 \cdot 10^7$ TeV, and $N_F \lesssim 48$.

The hypercharge portal  disappears either by increasing $M_F$  thus leaving insufficient RG time for the pull-down to be operative, or by increasing $N_F$ leading to a third sign change or a Landau pole.
Similarly, increasing the hypercharge $|Y_F|$ causes the $N_F$-window to become narrower and to move towards lower $N_F$, and vice versa,
reflecting that new terms in the gauge beta functions   scale as $\propto N_F Y_F^2$ and $N_F Y_F^4$. Maximal hypercharges are achieved for smallest number of flavors.
We find that for fixed $M_F=1$ TeV, $N_F=6$ flavors achieve stability with $Y_F=1$. For a single flavor $N_F=1$, one would need $2.28 < Y_F \leq 2.54$.
Increasing the mass for fixed $N_F$ slowly enhances the range of viable hypercharges.
For $M_F \simeq 10^{10}$ GeV,  the maximal values of $Y_F$  correspond to  a narrow  window around $Y_F \sim 3.2-3.3$, after which stability can no longer be achieved.

\begin{figure}
    \centering
    \includegraphics[width=.9\columnwidth]{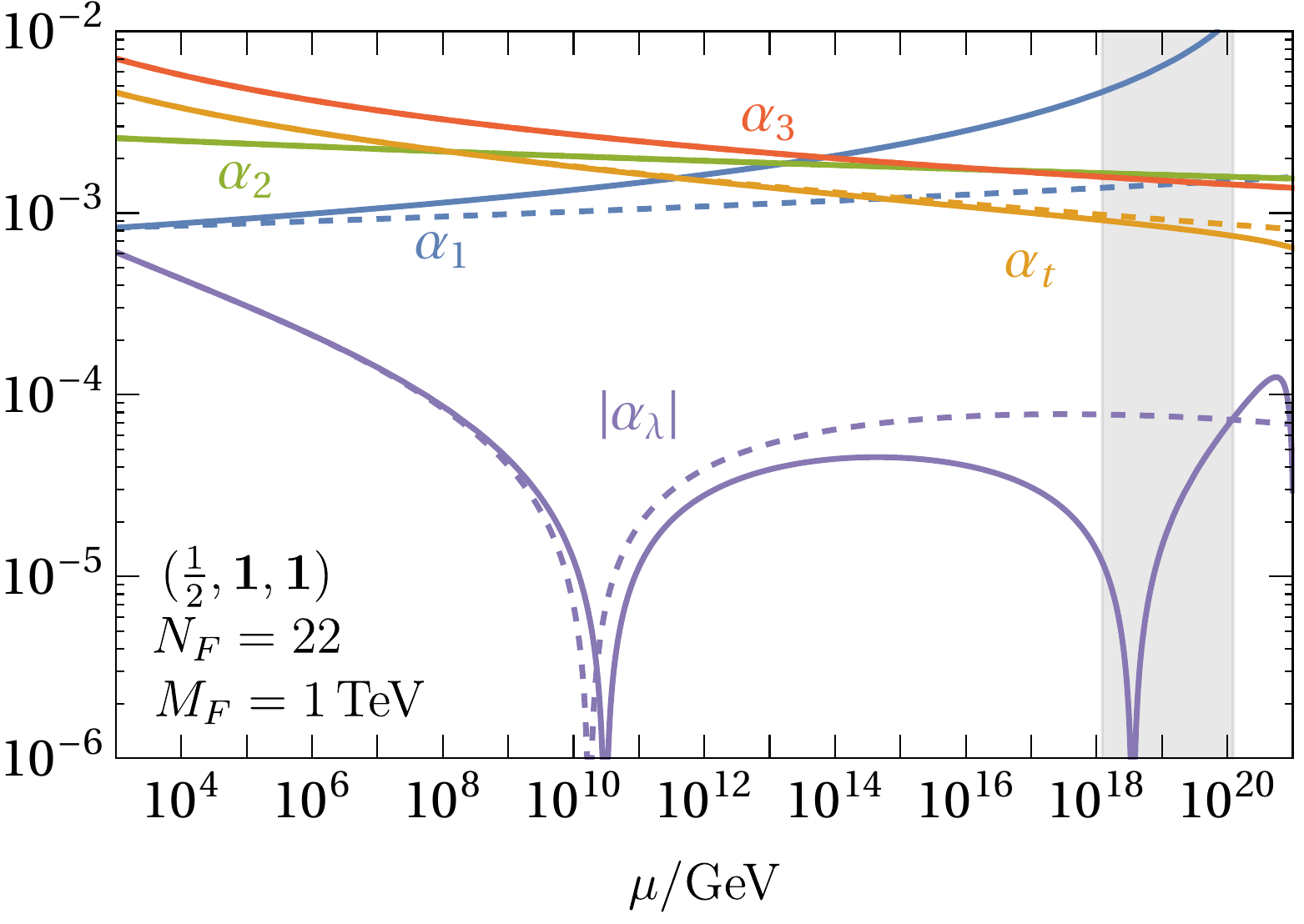}
    \caption{Example for a hypercharge portal, showing  the two-loop running of couplings 
    for a SM extension with $N_F = 22$ Dirac fermions [solid lines] in the representation $(\frac{1}{2}, \mathbf{1}, \mathbf{1})$  in comparison to the SM [dashed lines].  Notice that the enhancement of  $\alpha_1$ has effectively pulled down the second sign change of $\alpha_\lambda$ in the SM from much above the Planck scale (\fig{SM-run})  to  the Planckian region.}
            \label{fig:VLL-run-YF}
\end{figure}

We now turn to a concrete example where the Higgs quartic vanishes identically at Planckian energies. To that end, we exploit the hypercharge portal  and tune parameters ($N_F=22$, $M_F=1$~TeV, $Y_F=\frac{1}{2}$) appropriately, with results given in \fig{VLL-run-YF}. We observe that $\alpha_\lambda$ changes sign twice, with the second sign flip happening precisely at the Planck scale. We conclude that a vanishing quartic \eq{flat} can be achieved at the Planck scale, with the third sign flip still located beyond $M_{\rm Pl}$.

Theory constraints from demanding stability at the Planck scale  can be confronted with experimental measurements of the electroweak precision parameters $Y,W$ \cite{Farina:2016rws}, which depend directly on $M_F$ \cite{Alves:2014cda} as well as on $N_F$ via $\delta B_{1,2}$ given in \eq{deltaB}. We find that existing constraints for $M_F \gtrsim 1$ TeV  are not affecting
 the weak or hypercharge stability windows.
 In the absence of Yukawa portals and corresponding decays,  LHC searches for  long lived charged  particles \cite{Altakach:2022hgn} that leave ionization tracks or $\bar \psi \psi $ resonances decaying to diphotons are particularly important. They are available for larger values of electric charge $ \geq 1$ and $N_F=1$, and do not exceed $\sim 1$ TeV presently.
 A dedicated analysis of the data and reach of the LHC and future experiments with $N_F$ fermions is desirable but beyond the scope of this work.
 
\subsection{Higgs Criticality}\label{sec:crit}

It is interesting to discuss our findings from the viewpoint of Higgs criticality.
It has been noticed previously that the SM Higgs quartic approximately obeys
$\beta_\lambda|_{\mu=M_{\rm Pl}}\approx 0$ together with $\lambda|_{\mu=M_{\rm Pl}}\approx 0$ to within an accuracy of about ${\cal O}(10^{-4})$ \cite{Buttazzo:2013uya}.
The vanishing of the quartic and its beta function  is reminiscent of a free RG fixed point for the quartic.

Given that gauge portals offer mild modifications to the running of the Higgs quartic, it is   natural to investigate whether SM extensions can be found where the quartic and its beta function vanish identically. We can answer this question to the affirmative. Specifically, for any gauge portal,  $\alpha_\lambda$ does not change sign  (changes sign twice) in the dark blue (light blue) regions of the corresponding critical surface. Therefore, the dark/light blue boundary line corresponds to models where $\alpha_\lambda$ achieves a double-zero at some intermediate scale $M_F\le \mu_{\rm crit}\le M_{\rm Pl}$, 
\begin{equation}\label{eq:crit}
\lambda|_{\mu_{\rm crit}}= 0 \quad {\rm and}\quad \beta_\lambda|_{\mu_{\rm crit}}= 0\,,
\end{equation}
corresponding to Higgs criticality at the scale $\mu_{\rm crit}$. 

\begin{figure}
  \centering
\vskip-.4cm
\hspace*{-.5cm}
  \includegraphics[width=1.1\columnwidth]{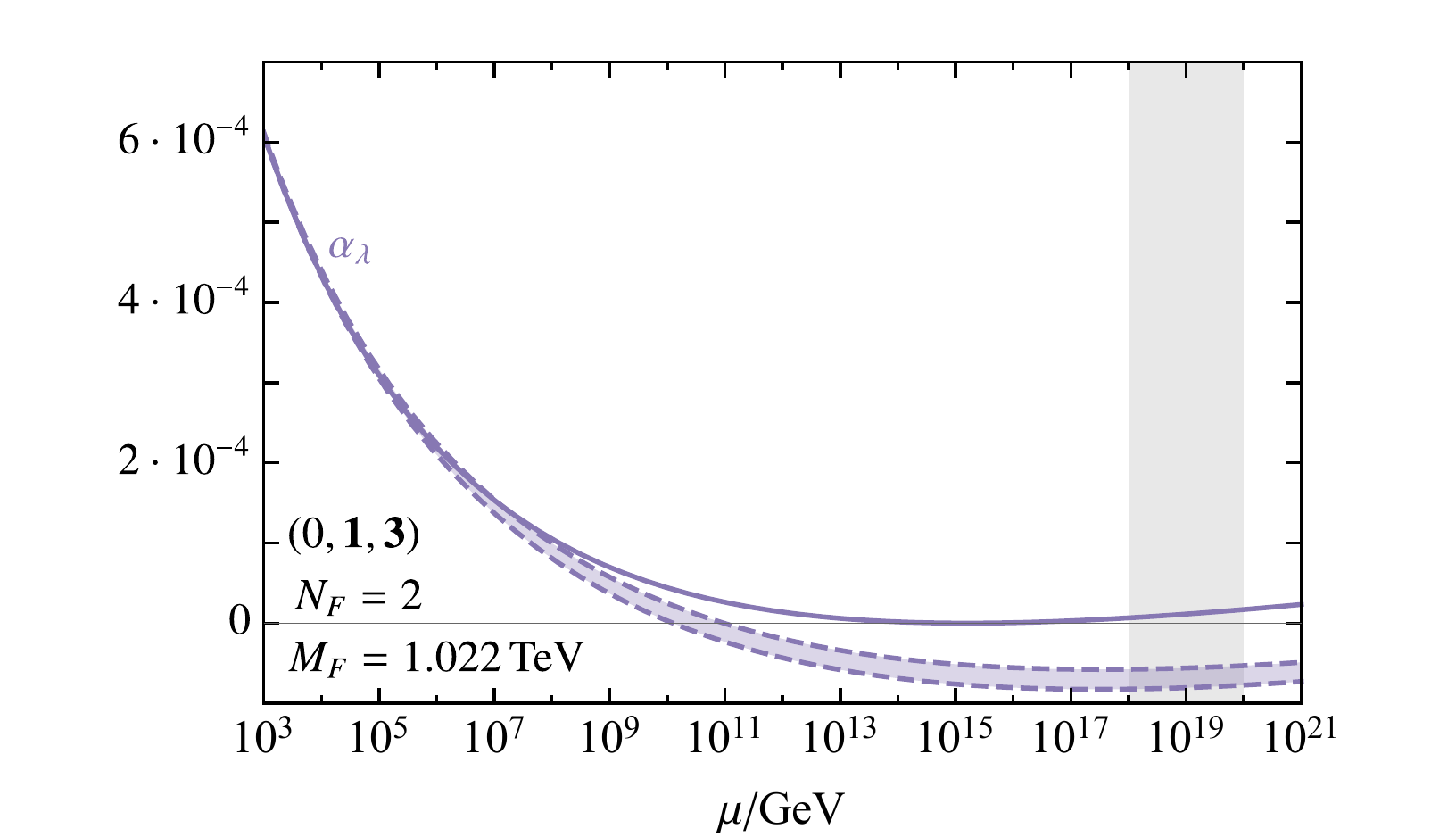}
  \caption{Higgs criticality   in  a SM extension with two vector-like quarks 
 of mass $M_F=1.022$~TeV.
  Shown are the running of the BSM Higgs coupling  [solid] in comparison to the SM including uncertainties [dashed]. We observe that the Higgs quartic develops a double-zero at $\mu_{\rm crit}\approx 10^{15}$~GeV and remains positive throughout  before  settling around $\alpha_\lambda\approx 10^{-5}$ at the Planck scale.
  }
  \label{fig:Criticality}
\end{figure}

An example for Higgs criticality is shown in \fig{Criticality}, where the strong gauge portal is exploited with two vector-like quarks in the representation $(0, \mathbf{1}, \mathbf{3})$. For $m_t = 172.76\,{\rm GeV},
$ we find that their mass  must read $M_F=1.022$~TeV  to achieve \eq{crit}. In particular, the BSM Higgs quartic  remains $\alpha_\lambda\ge 0$  throughout and achieves   a double-zero $\alpha_\lambda= 0$  just above $\mu_{\rm crit}\approx 10^{15}$~GeV before  settling around $\alpha_\lambda\approx 10^{-5}$ at the Planck scale. 

Including the $1\sigma$ uncertainty band for the top mass \eq{mtop}  translates into the range  $M_F\approx  0.872\, -\,3.767\, \text{TeV}$  for the critical mass, where larger $M_F$ relates to  lower $m_t$. As discussed in \Sec{inter}, the reason why a  small uncertainty range in $m_t$ translates into a   wide  range in $M_F$ is that the value of top Yukawa has a direct impact on the running of $\alpha_\lambda$, \eq{quartic-beta}, while the BSM fermions modify $\alpha_\lambda$  only very indirectly \eq{quartic-enhance}.

One may also ask whether the scale for Higgs criticality can be pushed up to  the Planck scale,
\begin{equation}\label{eq:HiggsCrit}
\beta_\lambda|_{\mu=M_{\rm Pl}} =0\,,\quad \lambda|_{\mu=M_{\rm Pl}}=0\,.
\end{equation}
If so, on the critical surface this would have to happen on the intersection of  the dark blue/light blue boundary line (where \eq{crit} holds true) with the yellow/blue  boundary line (where $\alpha_\lambda|_{M_{\rm Pl}}=0$ holds true). This is where \eq{flat} and \eq{crit} are valid simultaneously, and the scale $\mu_{\rm crit}$ in \eq{flat} then coincides with the Planck scale. For this to happen we need that the two sign changes of $\alpha_\lambda$ in the SM coincide at the Planck scale. From our model studies, we observe that the strong and weak gauge portals can achieve sufficient uplift for that,    \fig{StrongPortal} and \fig{WeakPortal}, but the scale of criticality  comes out in the region 
\begin{equation}\label{eq:mucrit}
\mu_{\rm crit} \approx 10^{11}-10^{15}~\text{GeV}\,,
\end{equation}
a few orders of magnitude below $M_{\rm Pl}$. The hypercharge portal on its own cannot achieve enough uplift, \fig{HyperchargePortal}, and criticality  is out of reach. 
 
 Finally, we  emphasize that models with  Higgs criticality around \eq{mucrit} lead to $\beta_\lambda|_{\mu=M_{\rm Pl}}\approx 0$ together with $\lambda|_{\mu=M_{\rm Pl}}\approx 0$, and to an accuracy similar or lower than observed in the SM (see \fig{Criticality}).
 Future work should  clarify whether more sophisticated gauge portals can push the scale of criticality fully up to the Planck scale.

\section{Gauge-Yukawa Portals} \label{sec:SingleYuk}

In this section, we consider SM extensions with vector-like fermions $\psi_{L,R}$ in  representations that allow for renormalizable Yukawa interactions with  SM and BSM fermions and the Higgs.  Thirteen different types of SM extensions exist, given in \tab{VLF-models}, featuring vector-like leptons (Models A--F) or vector-like quarks (Models G--M).

\begin{table}[t]
  \centering
  \renewcommand*{\arraystretch}{1.2}
  \setlength\arrayrulewidth{1.3pt}
  \rowcolors{2}{LightGray}{}
  \begin{tabular}{|cc|c|}
  \hline
  \rowcolor{LightBlue} \bf Model & $(Y_F, d_2, d_3)$  & \bf Yukawa Interactions 
  \\ \hline
    A & $(-1,\mathbf1,\mathbf1)$ & $\kappa_{ij}\, \overline{L}_i H \psi_{Rj} + \hc$  \\ 
    B & $(-1,\mathbf3,\mathbf1)$ & $\kappa_{ij}\, \overline{L}_i \psi_{Rj} H + \hc$ \\
    C & $(-\tfrac12,\mathbf2,\mathbf1)$ & $\kappa_{ij}\, \overline{\psi}_{Li} H E_j + \hc$ \\
    D & $(-\tfrac32,\mathbf2,\mathbf1)$ & $\kappa_{ij}\, \overline{\psi}_{Li} H^c E_j + \hc$\\
    E & $(\ 0, \ \mathbf1,\mathbf1)$ & $\kappa_{ij}\, \overline{L}_i H^c \psi_{Rj} + \hc$ \\
    F & $(\ 0, \ \mathbf3,\mathbf1)$ & $\kappa_{ij}\, \overline{L}_i \psi_{Rj} H^c + \hc$ \\
    G & $(-\tfrac13,\mathbf1,\mathbf3)$ & $\kappa_{ij}\, \overline{Q}_i H \psi_{Rj} + \hc$ \\ 
    H & $(+\tfrac23,\mathbf1,\mathbf3)$ & $\kappa_{ij}\, \overline{Q}_i H^c \psi_{Rj} + \hc$\\
    I & $(-\tfrac13,\mathbf3,\mathbf3)$ & $\kappa_{ij}\, \overline{Q}_i \psi_{Rj} H + \hc$ \\
    J & $(+\tfrac23,\mathbf3,\mathbf3)$ & $\kappa_{ij}\,\overline{Q}_i \psi_{Rj} H^c + \hc$ \\
    K & $(+\tfrac16, \mathbf2, \mathbf3)$ & $\kappa^u_{ij}\, \overline{\psi}_{Li} H^c U_j  + \kappa^d_{ij} \,\overline{\psi}_{Li} H D_j + \hc$ \\
    L & $(+\tfrac76, \mathbf2, \mathbf3)$ & $\kappa_{ij}\, \overline{\psi}_{Li} H U_j + \hc$ \\
    M & $(-\tfrac56, \mathbf2, \mathbf3)$ & $\kappa_{ij}\, \overline{\psi}_{Li} H^c D_j + \hc$ \\ \hline
  \end{tabular}
  \caption{
 Complete list of vector-like fermion extensions of the SM  with Yukawa portals to  the Higgs and SM fermions, also showing  the respective  gauge charges   and  interactions;   $H^c = i \sigma_2 H^*$. Note that Model K offers two Yukawa portals.
 }
  
  \label{tab:VLF-models}
\end{table}

The presence of Yukawa interactions modifies the RG running and may impact both the position of Landau poles and the stability of the Higgs potential.
Generically, Landau poles in the gauge sector are moved towards higher scales, but the extent of the shift depends on fermion representations as well as the structure and strength of the Yukawa portal. 
Similarly, estimating the effects on vacuum stability requires a dedicated study for each of the models in \tab{VLF-models}. 

We begin investigating the models in the limit of feebly small Yukawa portals in Sec.~\ref{sec:feeble}. Next, we conduct a stability  analysis for Yukawas
 to the third generation of SM fermions in Sec.~\ref{sec:3}. In Sec.~\ref{sec:walk} we detail the running and walking windows in VLQ models,  before turning to flavorful models in Sec.~\ref{sec:FullYuk}.

\subsection{Feeble Yukawas and Gauge Portals\label{sec:feeble}}

 Yukawa portals from vector-like fermions to the Higgs are technically natural and can hence be made feebly small, or even vanishing.
In this limit, the vacuum stability is controlled by the gauge portal  (Sec.~\ref{sec:VLF}).

\begin{table}[t]
  \centering
  \renewcommand*{\arraystretch}{1.2}
  \setlength\arrayrulewidth{1.3pt}
  \rowcolors{2}{LightGray}{}
  \begin{tabular}{|cc|c|c|c|}
    \hline
    \rowcolor{LightBlue} 
    \multicolumn{2}{|c|}{ \bf Model}
    & {\rm \bf Poles}
    &\multicolumn{2}{c|}{\bf Stability}\\ 
     \rowcolor{LightBlue} & $(Y_F, d_2, d_3)$
     &$N_F^\text{pole}$ \ $\alpha^\text{pole}$
     & \multicolumn{1}{c}{$\alpha_\lambda|_{M_{\rm Pl}}\ge 0$}
    & 
    $\alpha_\lambda\ge 0$ \\ \hline
    A & $(-1,\mathbf1,\mathbf1)$ & 7 \ \ $\alpha_1$
    & $N_F = 6$ & $\mathsf{\color{red}X}$\\ 
    B & $(-1,\mathbf3,\mathbf1)$ & 3  \ \  $\alpha_1$
    & $1 \leq N_F \leq 2$ & $1 \leq N_F \leq 2$ \\
    C & $(-\tfrac12,\mathbf2,\mathbf1)$ & 12  \  $\alpha_2$ 
    & $3 \leq N_F \leq 11$ & $5 \leq N_F \leq 11$\\
    D & $(-\tfrac32,\mathbf2,\mathbf1)$ & 2 \ \  $\alpha_1$ 
    & $N_F=1$ & $\mathsf{\color{red}X}$ \\ 
    E & $(\ 0 \ ,\mathbf1,\mathbf1)$ & as in SM
    & $\mathsf{\color{red}X}$ & $\mathsf{\color{red}X}$ \\
    F & $(\ 0\ ,\mathbf3,\mathbf1)$ & 3 \ \  $\alpha_2$ 
    & $1 \leq N_F \leq \tfrac52$ & $\tfrac32 \leq N_F \leq \tfrac52$ \\ 
    G & $(-\tfrac13,\mathbf1,\mathbf3)$ & 11  \  $\alpha_3$ 
    &  $2 \leq N_F \leq 10$ & $2 \leq N_F \leq 10$ \\ 
    H & $(+\tfrac23,\mathbf1,\mathbf3)$ & 6 \ \  $\alpha_1$
    & $2 \leq N_F \leq 5$ & $2 \leq N_F \leq 5$ \\
    I & $(-\tfrac13,\mathbf3,\mathbf3)$ & 1 \ \  $\alpha_2$
    & $\mathsf{\color{red}X}$ & $\mathsf{\color{red}X}$ \\
    J & $(+\tfrac23,\mathbf3,\mathbf3)$ & 1 \ \   $\alpha_2$
    & $\mathsf{\color{red}X}$ & $\mathsf{\color{red}X}$ \\ 
    K & $(+\tfrac16, \mathbf2, \mathbf3)$ & 4 \ \  $\alpha_2$ 
    & $1 \leq N_F \leq 3$ & $1 \leq N_F \leq 3$ \\
    L & $(+\tfrac76, \mathbf2, \mathbf3)$ & 1 \ \  $\alpha_1$
    & $\mathsf{\color{red}X}$ & $\mathsf{\color{red}X}$ \\ 
    M & $(-\tfrac56, \mathbf2, \mathbf3)$ & 2 \ \  $\alpha_1$
    & $N_F = 1$ & $N_F = 1$ \\ \hline
  \end{tabular}
  \caption{
  Planck scale stability  of SM extensions  as in \tab{VLF-models} for $M_F=1$~TeV and 
 vanishing Yukawa portal couplings $\alpha_\kappa|_\text{1 TeV} \simeq 0$.  Parameter regions indicate  stability at the Planck scale ($\alpha_\lambda|_{M_{\rm Pl}}\ge 0$),  stability  for all scales up  to $\MPl$ ($\alpha_\lambda\ge 0$), and the type of singularity above the upper end of stability. Model E remains metastable as the SM, and Models  I, J, and L display Landau poles prior to $\MPl$. }
  \label{tab:SM+VLFs} 
\end{table}

For $M_F \simeq 1$~TeV, stability constraints  on $N_F$  for  the various models A--M are collected in \tab{SM+VLFs}. 
We observe that models E, I, J, and L  do not achieve a stable vacuum at the Planck scale, not even with a single BSM generation. For  model E this is trivially so because it does not exhibit a gauge portal as the BSM fermions are singlets, hence SM metastability prevails.  
Models I, J (and L ) are afflicted by subplanckian Landau poles in $\alpha_2$ ($\alpha_1$), which  can be remedied  provided the new fields have larger masses $(M_F \gg$ TeV), whereby viable gauge portals open up   (we demonstrate this explicitly for model L in \fig{SingleYukawaSurface} below).
 Vector-like lepton models A--D and F have a finite range of stability  in $N_F$. 
Models are constrained by Landau poles in $\alpha_{1,2}$. Note that since the representations in E and F are real, an odd number of BSM Weyl fermions is compatible with gauge anomaly cancellation, corresponding to half-integer values of $N_F$.
Finally, for suitable values of $N_F$, for vector-like quark models G, H, K, M, and $M_F=1$~TeV, we find that $\alpha_\lambda\ge 0$ all the way up to the Planck scale. 

\begin{figure}
  \centering
\vskip-.2cm
  \renewcommand*{\arraystretch}{0}
  \includegraphics[width=.75\columnwidth]{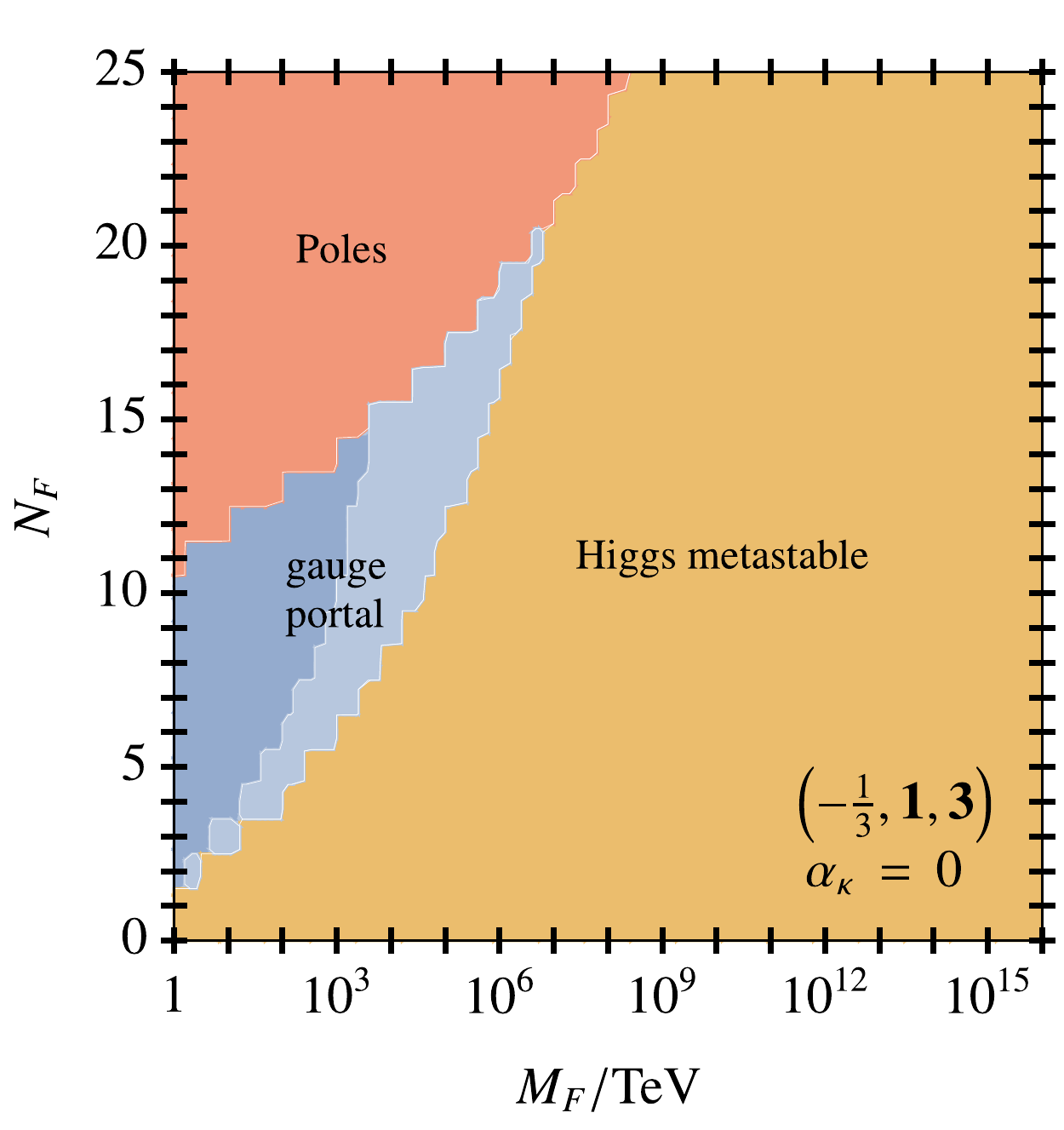}
  \caption{Illustration of a gauge portal for stability from strong gauge and hypercharge contributions (Model G with $\alpha_\kappa=0$), showing  the  critical surface of parameters in the $(N_F,M_F)$ plane with color coding as in \fig{StrongPortal}. The pole region relates to   subplanckian Landau poles in $\alpha_3$.}
  \label{fig:MixedPortal}
\end{figure}

Keeping both $N_F$ and $M_F$ as free parameters,  the critical surface is exemplarily shown for model G in \fig{MixedPortal}. 
We observe that a wedge of stability arises between regions of metastability and Landau poles, bounded by upper and lower limits on $N_F$, and an upper limit on $M_F$. Similar results are found for models B, C, F, H and K (not shown).

\subsection{Third Generation Yukawa Portals \label{sec:3}}

In this section we study the impact of Yukawa portals. We consider the SM extensions given in \tab{VLF-models}  (Model A--M) taking $M_F = 1$~TeV and a minimal number of BSM fermions, $N_F=1$. For simplicity, we only retain a single Yukawa  $\kappa$ (there are two in Model K), coupling to third-generation SM fermions. This is in accord with the flavor symmetry implications of \eq{YukApprox}.

We then integrate the running couplings up to the Planck scale and perform a scan by using the BSM Yukawa coupling $\alpha_\kappa|_{M_F}$ as a free parameter to find conditions for  vacuum stability. Our main findings are compiled  in  \tab{SM+VLF+Yukawa}. 
For feeble Yukawas, models display  diverse phenomena such as stability with $\alpha_\lambda\ge 0$  (models B, K, M) or with $\alpha_\lambda|_{M_{\rm Pl}}\ge 0$ (models D, F), Higgs metastablity at the Planck scale (models A, C, E, G, H) or subplanckian Landau poles (models I, J, L). 
We further observe that all models which achieve stability through a gauge portal for vanishing Yukawa interactions, \tab{SM+VLFs}, continue to do so up to a small upper limit for the Yukawa coupling $\alpha_\kappa$ (models B, D, F, K and M). 
This is in accord with expectations from perturbation theory which showed that  small Yukawa interactions counteract stability, \eq{YukawaHiggs}.

\begin{table*}
  \centering
  \renewcommand*{\arraystretch}{1.2}
  \setlength\arrayrulewidth{1.3pt}
  \rowcolors{2}{LightGray}{}
  \begin{tabular}{|cc|c|c|c|}
  \hline
  \rowcolor{LightBlue} {\bf Model }& $(Y_F, d_2, d_3)$  & \bf Interactions 
  & \bf Gauge Portal & \bf Yukawa Portal\\ \hline
    A & $(-1,\mathbf1,\mathbf1)$ & $\kappa\, \overline{L}_3 H \psi_R$ 
    & $\mathsf{\color{red}X}$ &$\alpha_\kappa|_\text{1 TeV} \gtrsim 0.20\, (6 \cdot 10^{-3})$ \\ 
    B & $(-1,\mathbf3,\mathbf1)$ & $\kappa\, \overline{L}_3 \psi_R H$ 
    & $\alpha_\kappa|_\text{1 TeV} \lesssim 2 \cdot 10^{-4}\, (1.6 \cdot 10^{-3})$ & $\alpha_\kappa|_\text{1 TeV} \gtrsim 0.4 \,(1.6 \cdot 10^{-2})$ \\
    C & $(-\tfrac12,\mathbf2,\mathbf1)$ & $\kappa\, \overline{\psi}_L H E_3$ 
    & $\mathsf{\color{red}X}$&$\alpha_\kappa|_\text{1 TeV} \gtrsim 0.20\, (6 \cdot 10^{-3})$\\
    D & $(-\tfrac32,\mathbf2,\mathbf1)$ & $\kappa\, \overline{\psi}_L H^c E_3$ 
    & ($\alpha_\kappa|_\text{1 TeV} \lesssim 3 \cdot 10^{-5})$ & $\alpha_\kappa|_\text{1 TeV} \gtrsim 0.20\, (8 \cdot 10^{-3})$\\
    E & $(\ 0, \ \mathbf1,\mathbf1)$ & $\kappa\, \overline{L_3} H^c \psi_R$ 
    & $\mathsf{\color{red}X}$&$\alpha_\kappa|_\text{1 TeV} \gtrsim 0.20\, (5 \cdot 10^{-3})$\\
    F & $(\ 0, \ \mathbf3,\mathbf1)$ & $\kappa\, \overline{L_3} \psi_R H^c$ 
    & ($\alpha_\kappa|_\text{1 TeV} \lesssim 10^{-3}$)& $\alpha_\kappa|_\text{1 TeV} \gtrsim 0.4\, (1.6 \cdot 10^{-2})$\\
    G & $(-\tfrac13,\mathbf1,\mathbf3)$ & $\kappa\, \overline{Q}_3 H \psi_R$ 
    &$\mathsf{\color{red}X}$ &$\alpha_\kappa|_\text{1 TeV} \gtrsim 0.20\, (1\cdot 10^{-2})$\\ 
    H & $(+\tfrac23,\mathbf1,\mathbf3)$ & $\kappa\, \overline{Q}_3 H^c \psi_R$ 
    &$\mathsf{\color{red}X}$ &$\alpha_\kappa|_\text{1 TeV} \gtrsim 0.20\, (6 \cdot 10^{-3})$\\
    I & $(-\tfrac13,\mathbf3,\mathbf3)$ & $\kappa\, \overline{Q}_3 \psi_R H$ 
     &$\mathsf{\color{red}X}$ &$\alpha_\kappa|_\text{1 TeV} \gtrsim 0.6\, (0.3)$\\
    J & $(+\tfrac23,\mathbf3,\mathbf3)$ & $\kappa\,\overline{Q}_3 \psi_R H^c$ 
    &$\mathsf{\color{red}X}$& $\alpha_\kappa|_\text{1 TeV} \gtrsim  0.6\, (0.3)$\\
    K & $(+\tfrac16, \mathbf2, \mathbf3)$ & {$\kappa_t\, \overline{\psi}_L H^c U_3 $ $ + \ \kappa_b \,\overline{\psi}_L H D_3$} 
    &  $\alpha_{\kappa_t,\kappa_b(\kappa_t)}|_\text{1 TeV} \lesssim 10^{-5}\, (10^{-4})$ & $\alpha_{\kappa_t,\kappa_b}|_\text{1 TeV} \gtrsim 0.25\, (0.13)$\\
    L & $(+\tfrac76, \mathbf2, \mathbf3)$ & $\kappa\, \overline{\psi}_L H U_3$ 
    & $\mathsf{\color{red}X}$&$\alpha_\kappa|_\text{1 TeV} \gtrsim 0.20\, (10^{-2})$\\
    M & $(-\tfrac56, \mathbf2, \mathbf3)$ & $\kappa\, \overline{\psi}_L H^c D_3$ 
    & 
    $\alpha_\kappa|_\text{1 TeV} \lesssim 8 \cdot 10^{-4}\, (1.4 \cdot 10^{-3})$ & $\alpha_\kappa|_\text{1 TeV} \gtrsim 0.2\, (8 \cdot 10^{-3})$
    \\ \hline
  \end{tabular}
  \caption{
  Vacuum stability at the Planck scale for SM extensions with a single vector-like fermion of mass $M_F=1$~TeV and Yukawa couplings to the Higgs and third-generation SM fermions. Shown are the parameter ranges of Yukawa couplings at the matching  scale ($\mu_0=M_F)$ which ensure  stability via either the gauge  or the Yukawa portal. Values without (with) brackets  refer to settings where $\alpha_\lambda\ge 0$ ($\alpha_\lambda|_{M_{\rm Pl}}\ge 0$).  For small $\alpha_\kappa(\mu_0)$, we observe that a few models offer stability via a gauge portal. For moderate or large $\alpha_\kappa(\mu_0)$, stability is provided by the Yukawa portal and a strongly coupled walking regime. For vanishing Yukawas, the results of \tab{SM+VLFs} are recovered.  }
  \label{tab:SM+VLF+Yukawa}
\end{table*}

\begin{figure}[b]
    \centering
\vskip-.3cm
     \hspace*{-0.3cm}
    \includegraphics[width=\columnwidth]{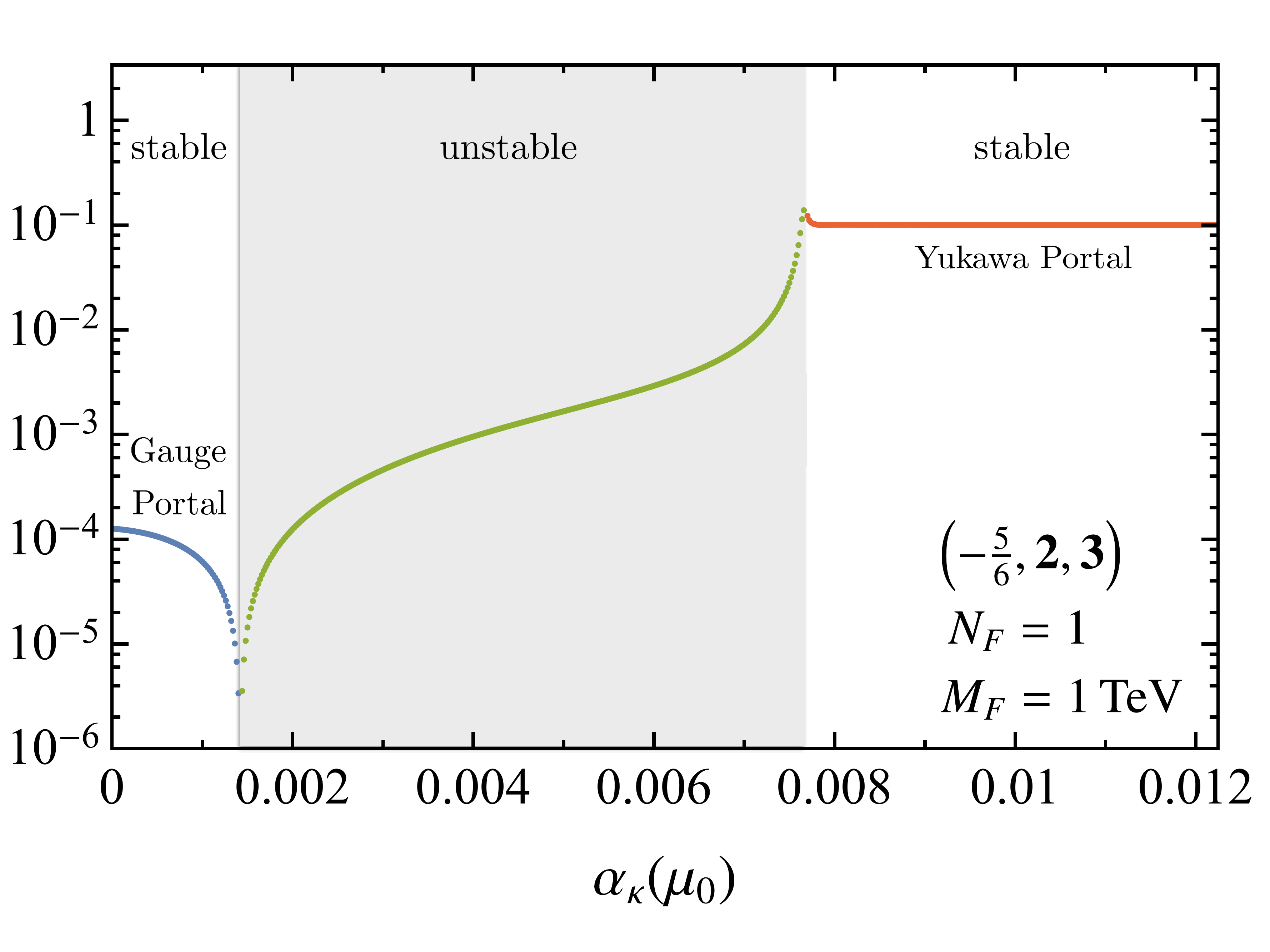}
   \vskip-0.4cm
    \caption{Shown is the Higgs quartic coupling at the Planck scale $|\alpha_\lambda(\MPl)|$ as a function of  the Yukawa portal coupling $\alpha_\kappa$ 
    (Model M, $N_F=1, M_F=1$~TeV). Stability is achieved through the gauge portal (blue) or the Yukawa portal (red).  In between, stability cannot be achieved  (green).}
       \label{fig:lambdaofkappa}
\end{figure}

By the same token, away from feeble or  weak-sized  values, Yukawa portal interactions destabilize the Higgs potential, and we find that stability cannot be achieved for a range of  Yukawa couplings.
However, a new effect sets in for moderate or larger Yukawa couplings, which trigger a  walking regime whereby couplings are 
almost interlocked due to the proximity of a partial fixed point. 
Incidentally, this effect helps stabilizing both the running of the scalar coupling as well as preventing sub-planckian Landau poles \cite{Hiller:2020fbu,Bause:2021prv}. In consequence, a new window into Higgs stability opens up to which we refer as the Yukawa portal. We note that  Yukawa portals are available for all models including for those without viable  gauge portals for small Yukawas. The onset for this can happen for Yukawas as low as $\alpha_\kappa \gtrsim 5\cdot 10^{-3}$, depending on the model, though for some models the onset is borderline non-perturbative ($\alpha_\kappa \gtrsim 0.3$). Upper bounds on stability  are dictated by the loss of perturbative control. 
Regions of stability are illustrated in \fig{lambdaofkappa} (Model M).

We note that stability is largely insensitive to our choice of connecting the Yukawa portal to the third generation, 
as  mixed terms in beta-functions containing both SM and BSM Yukawas are numerically small with very little impact on the running.
The reason we selected the third generation of SM fermions  is that their BSM search window is larger. 
We discuss  phenomenological constraints   in Sec.~\ref{sec:mass} below.

\begin{figure}[b]
    \centering
    \hspace*{-1cm}
    \includegraphics[width=1.2\columnwidth]{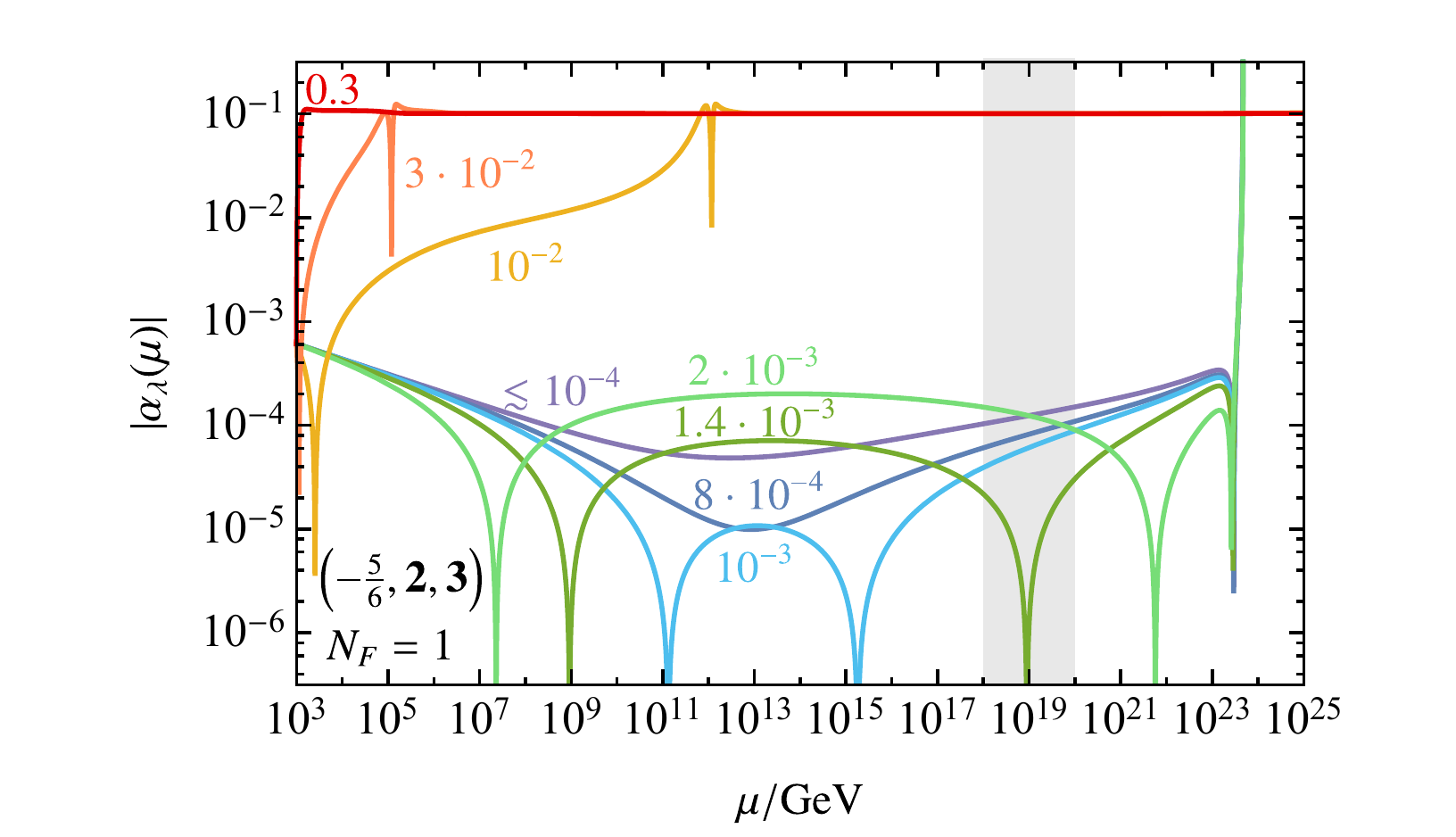}
    \caption{
    Shown is the transition from the gauge to  the Yukawa portal (Model M, $N_F=1, M_F=$1~TeV) by varying the Yukawa coupling 
    $\alpha_\kappa|_{M_F}$ between $10^{-4}$ and $3\cdot 10^{-1}$. For feeble $\alpha_\kappa$ (violet) the Higgs potential is stabilized by the gauge portal. Increasing $\alpha_\kappa$ generates two sign flips for $\alpha_\lambda$ (blue), one of which moves beyond $M_{\rm Pl}$ to destabilize the vacuum (green). For $\alpha_\kappa$ beyond $10^{-2}$  (yellow, orange, red)  the Higgs is stabilized through a strongly coupled Yukawa portal.}

    \label{fig:Running_Lambda}
\end{figure}

\subsection{Running and Walking \label{sec:walk} }

To further appreciate the  different mechanisms for stability as indicated in \tab{SM+VLF+Yukawa}, 
we investigate the RG running of $\alpha_\lambda$ in dependence of $\alpha_\kappa(\mu_0)$  in more detail for Model M (\fig{lambdaofkappa}). 
  
 \fig{Running_Lambda} shows the running of $\alpha_\lambda$ between $\mu_0=1$~TeV and the Planck scale for selected values of $\alpha_\kappa$ at the matching scale $\mu_0$.
 For any  $\alpha_\kappa$, we find a Landau pole in the transplanckian regime $\mu \simeq 5 \cdot 10^{23}$ GeV owing to the hypercharge.
 For  sufficiently small $\alpha_\kappa\lesssim 8 \cdot 10^{-4}$  (dark blue, violet) the gauge portal is operative leading to stability $\alpha_\lambda\ge 0$ along the entire trajectory.
  We  observe that $\alpha_\lambda$ exhibits a minimum around $\mu_{\text min} \approx 10^{13}$~GeV where the value of $\alpha_\lambda(\mu_{\text min})$  decreases with increasing $\alpha_\kappa$. 
 Within the range $ 8 \cdot 10^{-4} \lesssim \alpha_\kappa \lesssim 1.4 \cdot 10^{-3}$ illustrated by the light blue curve,  the running $\alpha_\lambda$ becomes mildly negative for a range of scales, but stability prevails in the end $(\alpha_\lambda(M_{\rm Pl})\ge 0)$.
 The width of the metastability regime quickly increases with growing $ \alpha_\kappa$  until    the second zero  occurs at the Planck scale ($\alpha_\kappa \simeq 1.4 \cdot 10^{-3}$) implying  a flat potential \eq{flat}. Increasing $\alpha_\kappa$ even further pushes this second sign flip beyond the Planck scale and stability is lost (green) for a range of $\alpha_\kappa$.
 
 A new effect sets in when   $\alpha_\kappa\simeq 8 \cdot 10^{-3}$. 
 While we observe a region of instability 
 starting above 1~TeV, the Higgs coupling is subsequently pulled-up to positive values  
 (yellow and orange curves), and achieves stability with $\alpha_\lambda(M_\text{Pl}) \approx 0.1$.
 Since the quartic and Yukawa couplings run very slowly,  we refer to this as the ``walking regime'' (see footnote \ref{walking}). 
 The larger $\alpha_\kappa$ the earlier $\alpha_\lambda$ enters into a walking regime.  
For large $\alpha_\kappa \gtrsim 0.2$   the  walking  regime starts straight after the matching scale and we have $\alpha_\lambda\ge 0$ along the entire trajectory. 
 
\begin{figure}
    \centering
    \hspace*{-1cm}
    \includegraphics[width=1.2\columnwidth]{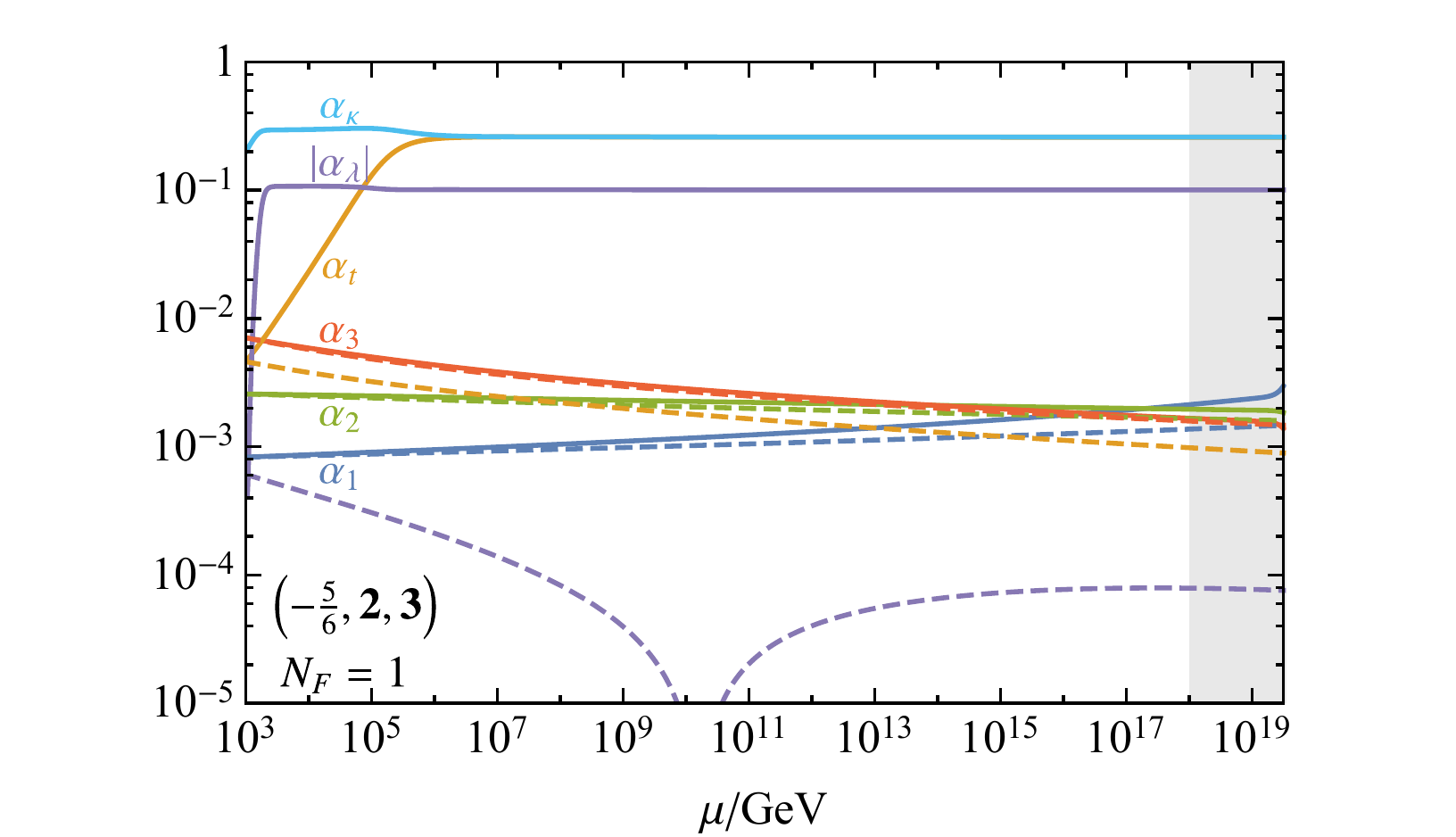}
    \caption{
    Example of a Yukawa portal, showing the two-loop running of couplings in Model M of \tab{SM+VLF+Yukawa} (SM plus a VLF of mass $1$~TeV and BSM Yukawa coupling $\alpha_{\kappa}$).
    The RG flow is depicted for large initial $\alpha_\kappa|_{\mu_0=M_F} = 0.2$. We observe  a walking regime for the Yukawa and quartic couplings $\alpha_{\kappa,t,\lambda}$ at sizeable values, while the gauge couplings continue to evolve within a weakly coupled regime all the way up to the Planck scale, similar to their running in the SM (dashed).}
    \label{fig:Walking}
 \end{figure}

 The latter  result  can also be inferred from \fig{Walking} where the running of couplings (with $\alpha_\kappa = 0.2$) is displayed [full lines] and compared with the SM [dashed lines]. We observe that the Higgs, the top and the BSM Yukawa couplings run very slowly, while gauge couplings continue to evolve as in the SM.\footnote{Neglecting  contributions from $\alpha_{1,b,\lambda}$, we find that the BSM and top Yukawa beta functions are identical after  exchanging $\alpha_\kappa \leftrightarrow  \alpha_t$. This explains why $\alpha_\kappa$ and $\alpha_t$ asymptote to nearby values.}

\begin{figure}
  \centering
  \renewcommand*{\arraystretch}{0}
    \begin{tabular}{c}
    \includegraphics[width=.75\columnwidth]{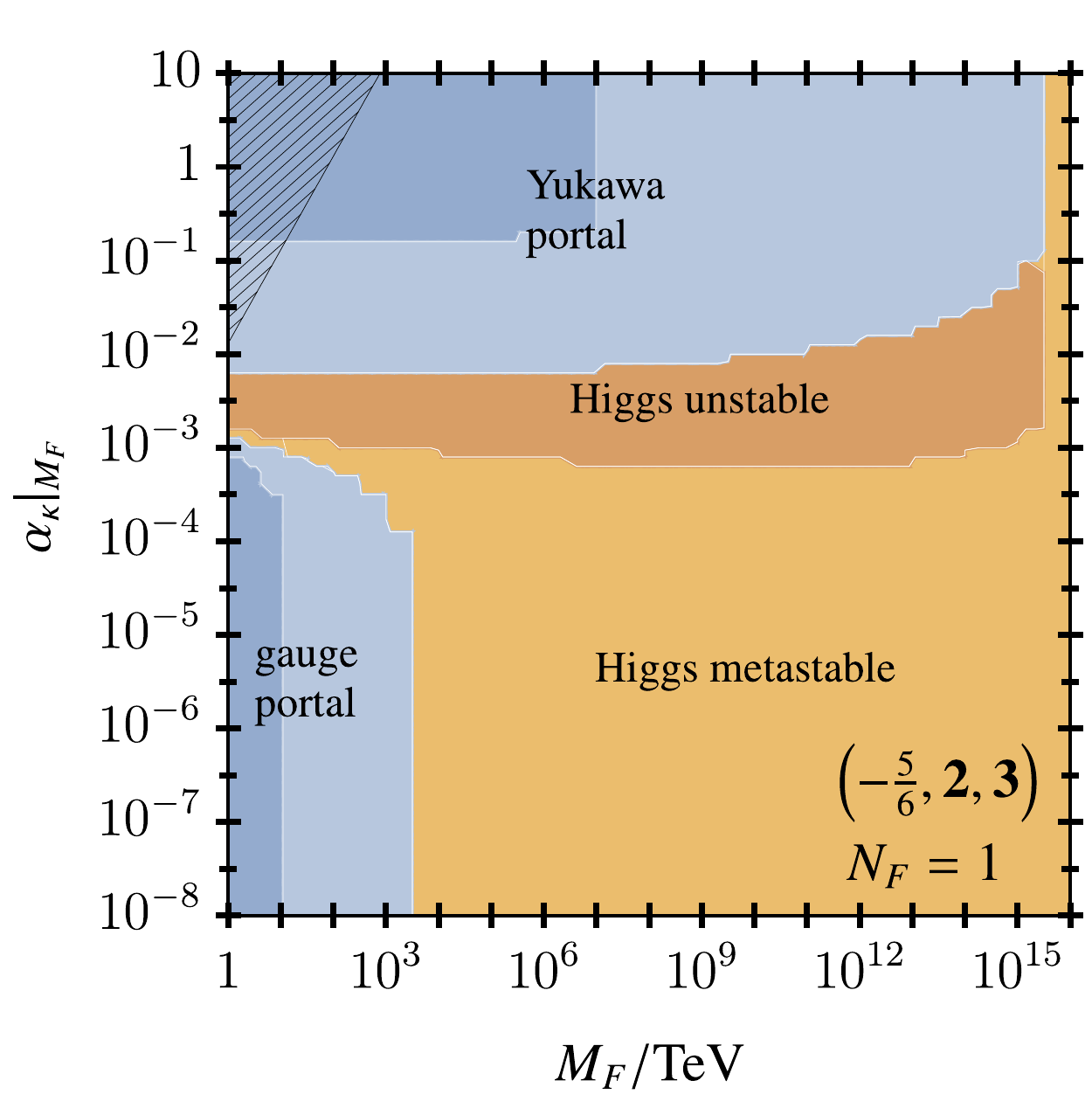}\\
    \includegraphics[width=.75\columnwidth]{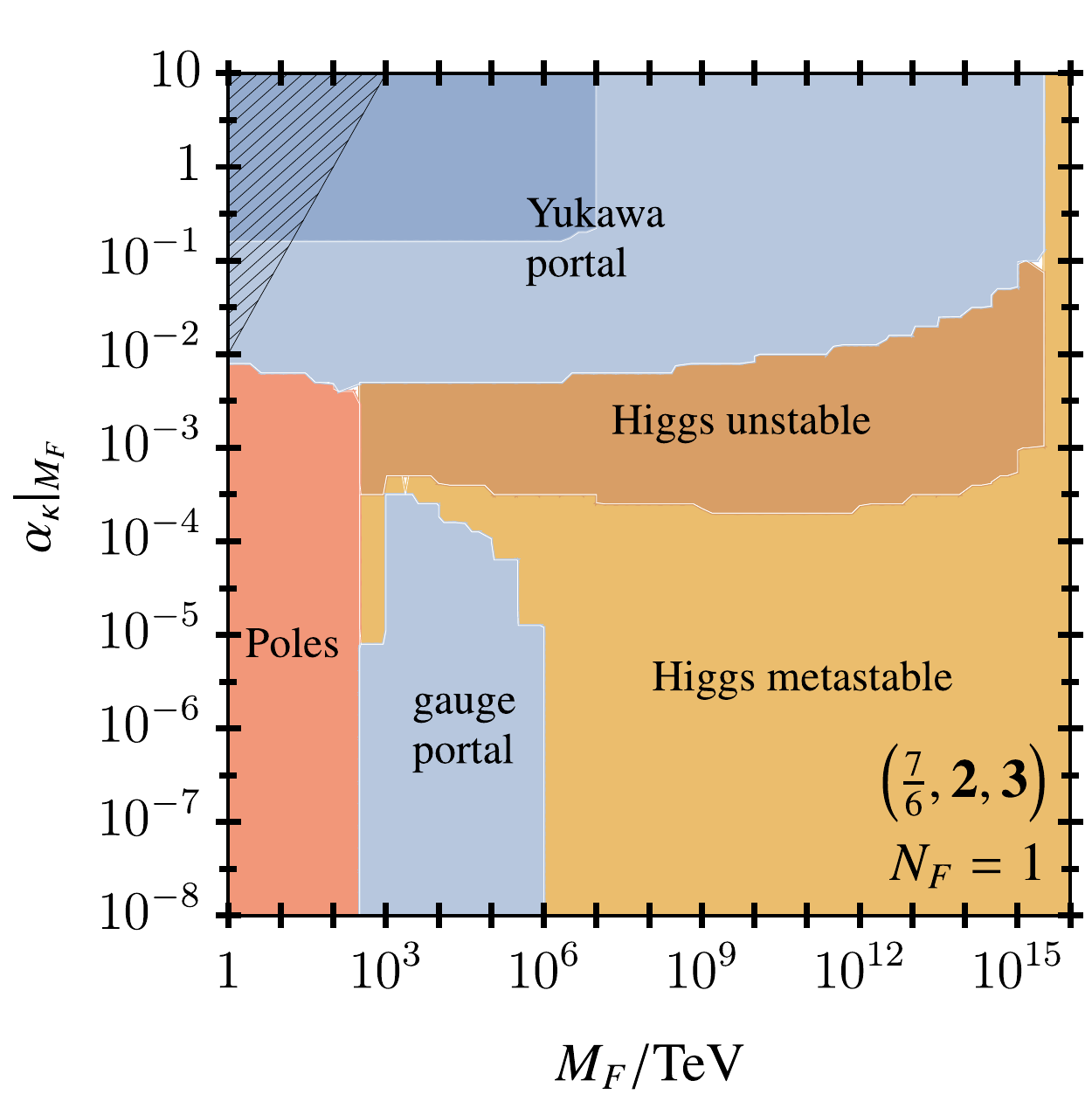}
  \end{tabular}
  \caption{Critical surfaces in the $M_F$ and $\alpha_{\kappa}|_{M_F}$ plane for Model M (top) and L (bottom) for $N_F=1$;
  color coding as in \fig{StrongPortal}, also indicating unstable potentials 
  $ \alpha_\lambda < -10^{-4}$ (brown). SMEFT  90\% C.L. exclusion regions on $\alpha_\kappa/M_F^2$  \cite{Falkowski:2019hvp,Ellis:2020unq}
  are indicated by the black hatched regions.}
  \label{fig:SingleYukawaSurface}
\end{figure}

Next, we study the impact of the BSM mass scale $M_F$ on the RG-fate of two complementary VLQ models:
Model M has stability windows from the gauge and the Yukawa portal,
while  model L only features stability from the latter. 
The RG-anatomy of VLL models  has been the subject of~\cite{Hiller:2020fbu}.
The critical surface spanned by  $M_F$ and $\alpha_{\kappa}(M_F)$  is displayed   in \fig{SingleYukawaSurface} for model M (upper plot) and model L (lower plot).

 For model M, we find that stability is achieved via the gauge portal  for masses up to $M_F\approx 3\cdot 10^3$~TeV, and for Yukawas up to $\alpha_{\kappa}(M_F)\approx 10^{-3}$. Increasing $M_F$  further leads to metastability. Increasing $\alpha_{\kappa}(M_F)$ we find a narrow band of instability, for any $M_F$, after which the Yukawa portal opens up. The latter extends   up to $M_F\simeq 10^7$~TeV if we demand stability along the entire trajectory, $\alpha_\lambda\ge 0$. Interestingly, the upper limit on the VLF mass 
 coincides with the scale where the Higgs quartic flips sign in the SM, which is the natural upper limit.
 Similarly,  if we demand stability at the Planck scale $\alpha_\lambda|_{M_{\rm Pl}}\ge 0$, we find stability in a more substantial range up to $M_F\simeq 10^{16}$~TeV, in accord with \tab{SM+VLF+Yukawa}.

In contrast, model L does not offer a gauge portal for small $M_F$ and small $\alpha_\kappa$. However, a gauge portal opens up in the range $3\cdot 10^2\,\text{TeV} \lesssim M_F \lesssim 10^6\,\text{TeV}$ for small $\alpha_\kappa(M_F)\lesssim 3\cdot 10^{-4}$. The reason for this is that the Landau pole at $\mu\simeq10^{14}$~TeV, which arises for low $M_F\approx 1$~TeV, is pushed beyond the Planck scale  by increasing $M_F$ by two orders of magnitude, and stability is  achieved  $\alpha_\lambda|_{M_{\rm Pl}}\ge 0$. 
 For larger masses the potential becomes metastable, much like in model M. Also, increasing $\alpha_{\kappa}(M_F)$ leads to a narrow band of instability, for any $M_F$, after which the Yukawa portal opens up, which quantitatively is very similar to the one found in model M.

\subsection{Flavorful Yukawas} \label{sec:FullYuk}

\begin{figure}
  \centering
  \renewcommand*{\arraystretch}{0}
  \begin{tabular}{c}
    \includegraphics[width=.75\columnwidth]{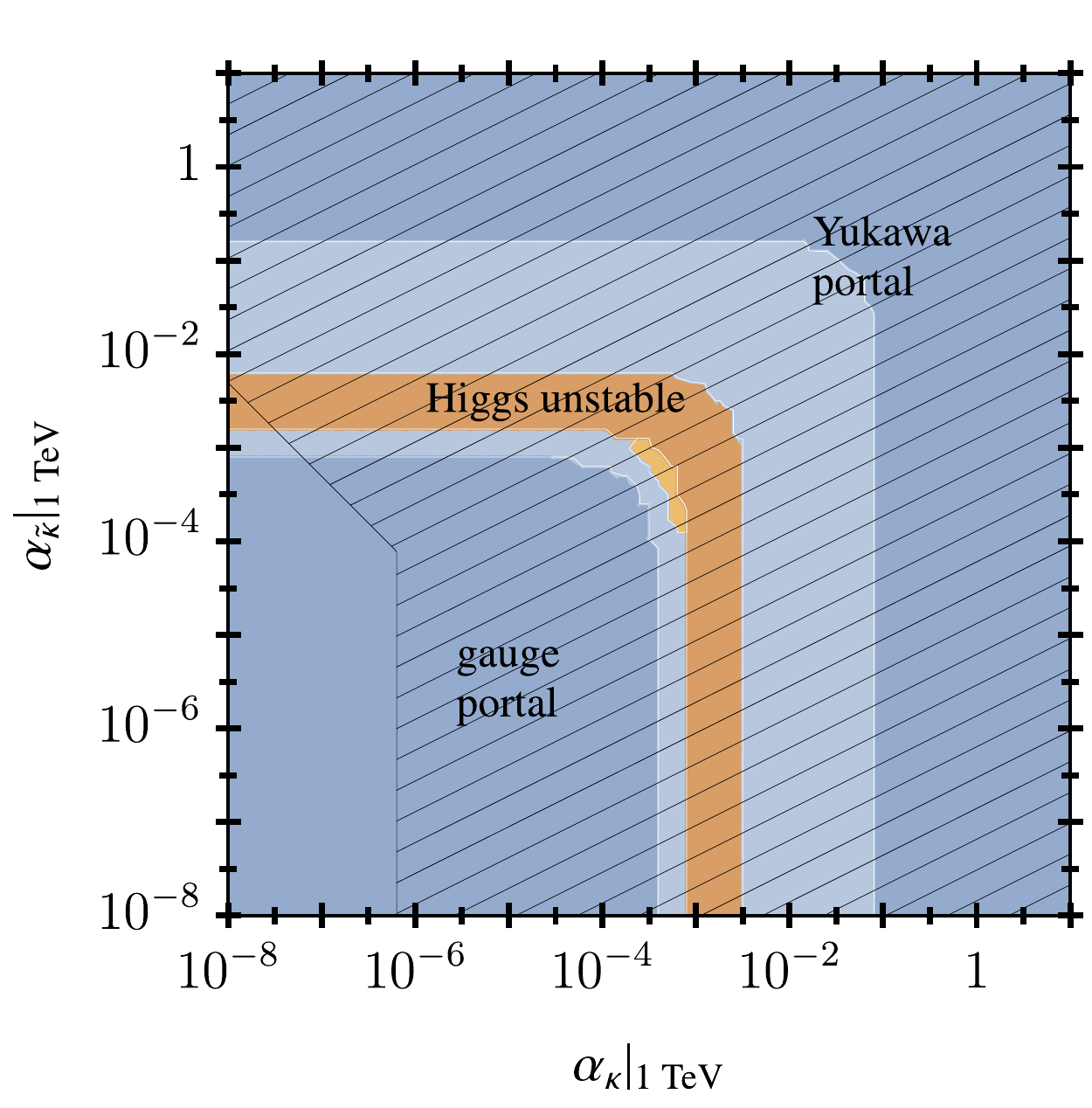}\\
     \includegraphics[width=.75\columnwidth]{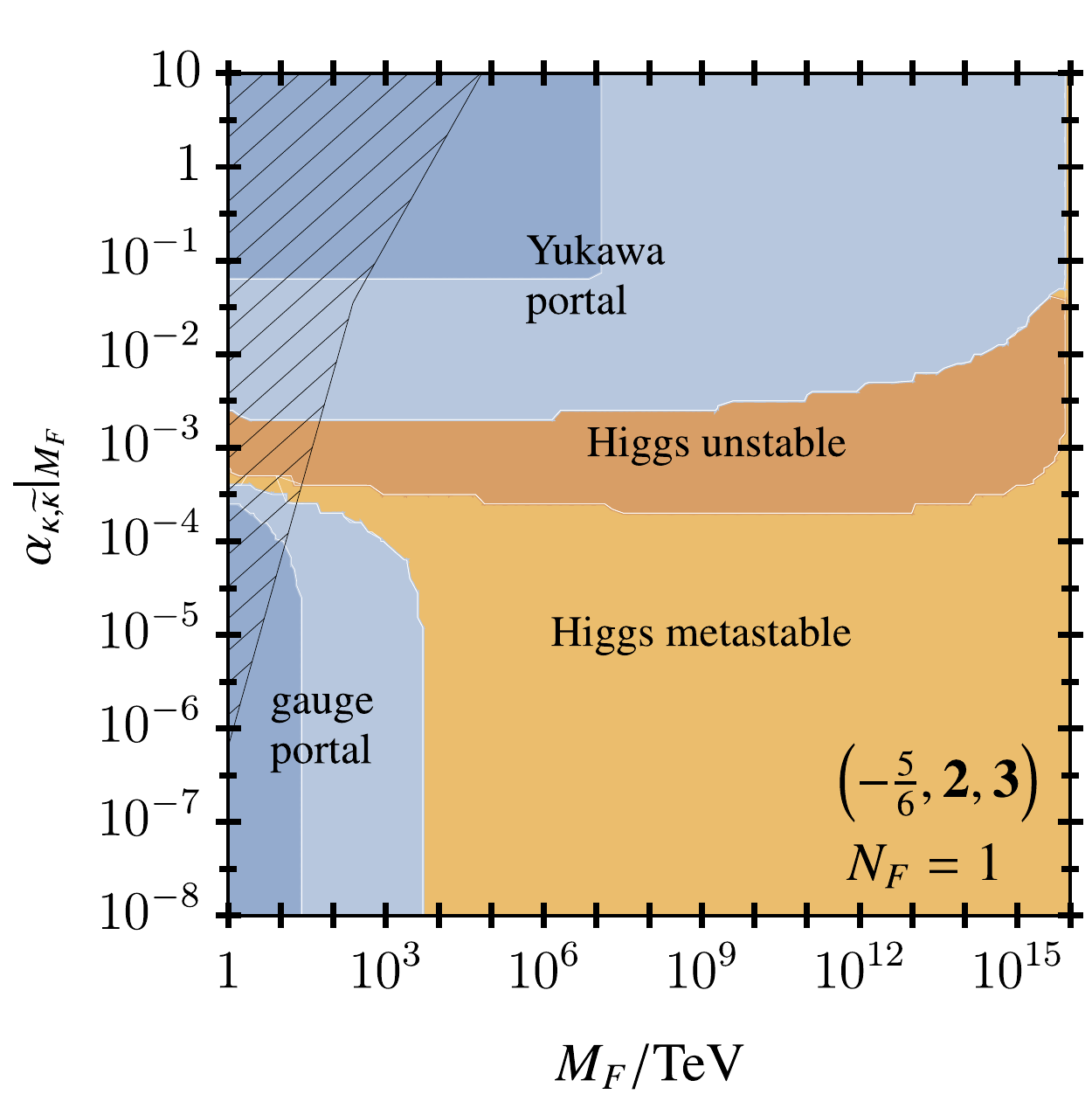}
  \end{tabular}
  \caption{BSM critical surface for Model M with $N_F=1$ and  Yukawa couplings to all  SM fermion generations \eq{FlavorfulYuk}  scanning over  $\alpha_{\kappa,\kt}|_\text{1~TeV}$ (top) as well as $M_F$ and $\alpha_{\kappa}|_{M_F}=\alpha_{\kt}|_{M_F}$ (bottom); color coding as in \fig{StrongPortal} and \fig{SingleYukawaSurface}. 
Regions excluded by $\Delta F =1$ or $\Delta F =2$ FCNC bounds on $\alpha_\kappa/M_F^2$ and $\alpha_\kappa^2/M_F^2$, respectively, are indicated by the black  hatched areas.
  Most stringent constraints are from $K_L \to  \mu \mu$ decays and $K$-mixing \cite{Ishiwata:2015cga}. }
  \label{fig:FlavorfulModelM}
\end{figure}

\begin{figure}
  \centering
  \renewcommand*{\arraystretch}{0}
  \begin{tabular}{c}\textbf{}
      \includegraphics[width=.75\columnwidth]{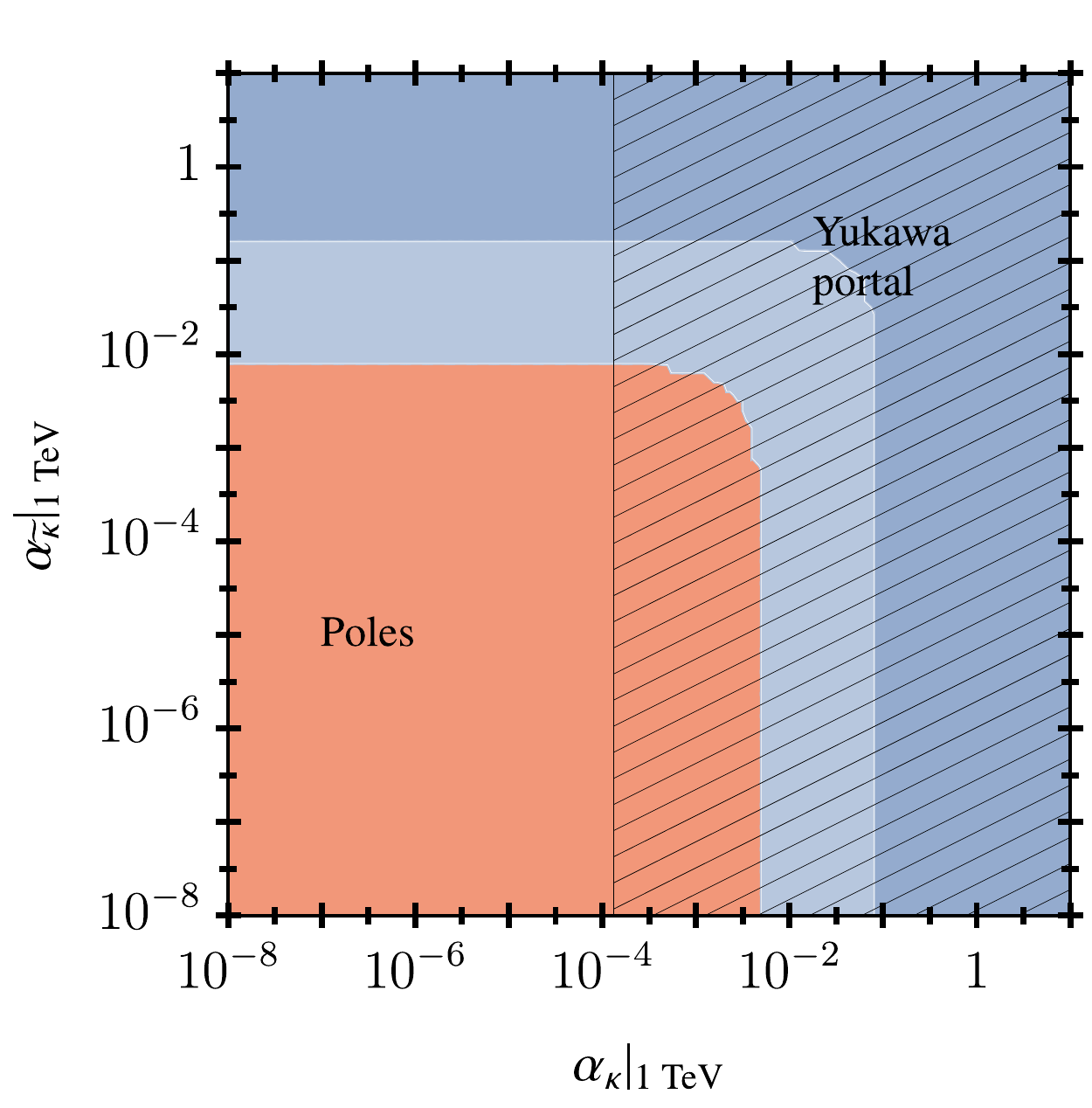}\\
     \includegraphics[width=.75\columnwidth]{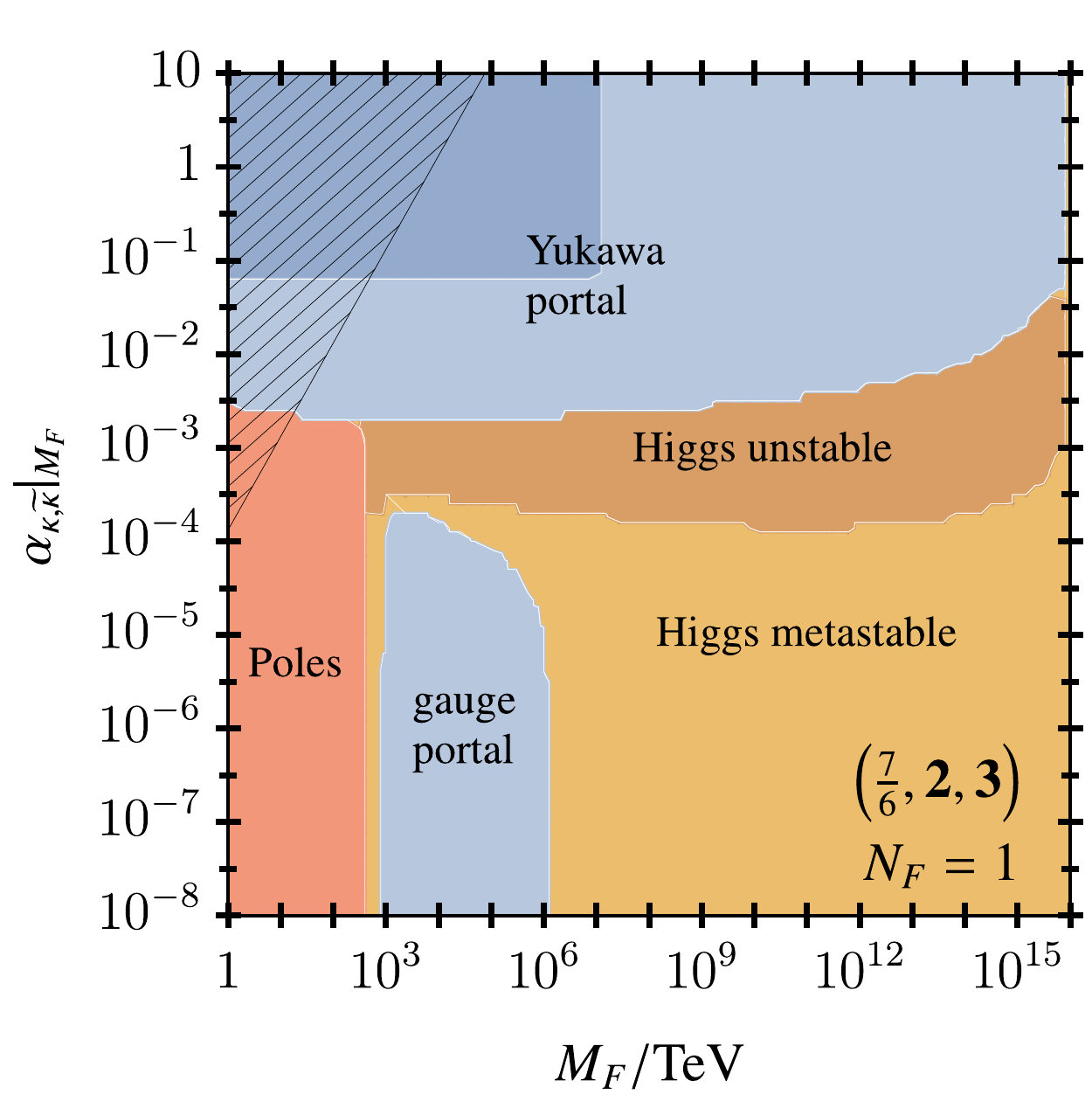}
   \end{tabular}
  \caption{BSM critical surface as in Fig.~\ref{fig:FlavorfulModelM}  for Model L with $N_F=1$ and Yukawa couplings to all three SM fermion generations scanning over $\alpha_{\kappa,\kt}|_\text{1~TeV}$ (top) as well as $M_F$ and $\alpha_{\kappa}|_{M_F}=\alpha_{\kt}|_{M_F}$ (bottom); color coding as in \fig{StrongPortal} and \fig{SingleYukawaSurface}.
   The most stringent FCNC bound is from 1-loop contributions to $D$-mixing  \cite{Ishiwata:2015cga}.}
  \label{fig:FlavorfulModelL}
\end{figure}

In this section, we generalize the SM extensions from \Sec{3}  by allowing for Yukawa couplings to all three generations of SM fermions, while keeping $N_F=1$ BSM fermions.
This accounts for all renormalizable portal interactions compatible with symmetries, rendering portal couplings $\vec\kappa$ a vector in the SM generations.
The  approximation \eq{YukApprox} implies that the non-abelian flavor symmetry
\begin{equation} \label{eq:FlavorSym}
      U(2)_Q \times U(2)_U \times U(2)_D \times U(3)_L \times U(3)_E
\end{equation}
is retained for SM leptons as well as the first and second quark generations. Hence, a suitable ansatz for the portal couplings reads
\begin{equation}\label{eq:FlavorfulYuk}
    \vec\kappa = \begin{cases}
    (\kappa,\,\kappa,\,\kappa)^\intercal & \text{ for VLLs} \\
    (\kappa,\,\kappa,\,\kt)^\intercal & \text{ for VLQs}
    \end{cases} .
\end{equation}
featuring a universal coupling  to parametrize the  portal in  VLL models, and $\kappa$ and $\kt$  for VLQs. 

Changing from single coupling  $\vec\kappa = (0,0, \kappa)$ discussed in \Sec{3}  to triple  \eq{FlavorfulYuk}   has minor qualitative impact on the BSM critical surfaces.
The only notable difference for VLL Models (A--F)  is a shift  by roughly half an order of magnitude towards lower values of $\alpha_{\kappa}(M_F)$ compared to the single Yukawa case. 
The origin for this minor modification is that 
 contributions to the $\beta$-functions from fermion bubble diagrams involving $\alpha_\kappa$ pick up an additional factor of three compared to the single Yukawa case.

For VLQs, on the other hand, with  \eq{FlavorfulYuk} two independent Yukawa couplings $\alpha_{\kappa,\kt}$ are present.
A section of the BSM critical surface with $\alpha_{\kappa}(M_F) = \alpha_{\kt}(M_F)$ versus  $M_F$ is displayed in the bottom panels of \fig{FlavorfulModelM} and \fig{FlavorfulModelL} for  Model M and L, respectively. One notes a strong resemblance to \fig{SingleYukawaSurface} in the single Yukawa case up to a factor $\lesssim 3$ shift towards lower values of $\alpha_{\kappa}(M_F)$.

The upper panels of \fig{FlavorfulModelM} and \fig{FlavorfulModelL} display  $\alpha_\kappa(M_F)$ versus $\alpha_{\kt}(M_F)$ for   $M_F = 1$~TeV.
For sufficiently small $\alpha_{\kappa}(M_F)$ the stability conditions  are consistent with  \tab{SM+VLF+Yukawa}.
Moreover, the surface plot is basically symmetric under interchange $\alpha_\kappa \leftrightarrow 2 \,\alpha_{\kt}$. This reflects the finding from \Sec{3}   that the RG evolution is 
essentially unchanged if the VLF couples to a first- or second- instead of third-generation SM fermion as mixed SM-BSM Yukawa contributions to the RGEs are numerically small and of minor importance.

While the impact of a larger Yukawa portal is qualitatively minor for stability, it  is important for phenomenology, discussed in the next section.

\subsection{Mass Limits \label{sec:mass}}

We briefly comment on experimental probes of  VLF models. In the presence of the Yukawa portal that induces VLF decay, pair and single production searches (e.g. \cite{ATLAS:2018ziw} and \cite{Erdmann:2018axt}) can be used to constrain $M_F$ and $\alpha_\kappa/M_F^2$, respectively. 
In addition, global standard model effective field theory (SMEFT) fits allow to constrain $\alpha_\kappa/M_F^2$, too. A detailed analysis of constraints is given elsewhere \cite{HHLS22plus,Ishiwata:2015cga,Aguilar-Saavedra:2013qpa}.
Here we show that stability requirements and  experimental constraints are complementary and allow to obtain new limits on the BSM masses.
In particular,  collider and flavor measurements are probing the  walking regime with large Yukawas, i.e., the Yukawa portal at work.

We discuss  the theory-collider interplay in the case of model L, beginning with the third generation only scenario: The SMEFT   exclusion region on $\alpha_\kappa/M_F^2 < 0.01 \, \text{TeV}^{-2}$  at 90 \% C.L. from a single operator fit  \cite{Ellis:2020unq}
  is indicated by the black hatched region in the lower plot of \fig{SingleYukawaSurface}.
 There exists   also a lower limit $M_F > 1.37$ TeV  at 95 \% C.L.  from pair production from ATLAS \cite{ATLAS:2018ziw} for this model, not shown for clarity.
Combining the  stability constraints with the experimental limits yields the allowed mass windows
\begin{equation}\label{eq:SMEFTBoundL}
\begin{aligned}
    4\,\text{TeV} \lesssim &M_F \lesssim 10^7 \,\text{TeV},\\
    (1\,\text{TeV} \lesssim &M_F \lesssim 10^{16} \,\text{TeV})
\end{aligned}
\end{equation}
for  $\alpha_\lambda \geq 0$ up to the Planck scale (at the Planck scale). The lower limits are stronger than (similar to)  the ones from direct searches. 

Experimentally, flavorful Yukawa portal couplings \eq{FlavorfulYuk} are severely constrained by flavor-changing neutral current (FCNC) processes. 
These are generically induced at tree-level from $Z$-penguins via fermion mixing, and from Higgs-VLF-loops.
These bounds are typically orders of magnitude stronger than the single Yukawa ones, such as \eq{SMEFTBoundL}.
For instance, in Model L combining Higgs stability conditions with  FCNC exclusion limits from $D$-mixing \cite{Ishiwata:2015cga},
 $\alpha_\kappa/M_F \lesssim 1.3 \cdot 10^{-4} \, \text{TeV}^{-1}$,  yields in the flavor universal benchmark $\alpha_\kappa(\mu_0)=\alpha_{\kt}(\mu_0)$ the  limits
\begin{equation}\label{eq:FCNCBoundL}
\begin{aligned}
    480\,\text{TeV} \lesssim &M_F \lesssim 10^7 \,\text{TeV},\\
    (19\,\text{TeV} \lesssim &M_F \lesssim 10^{16} \,\text{TeV})
\end{aligned}
\end{equation}
which can be read off from \fig{FlavorfulModelL}. The lower limits are significantly stronger than those in (\ref{eq:SMEFTBoundL}), and existing  collider search limits.
The stronger (weaker)  limits refer to   $\alpha_\lambda \geq 0$ up to the Planck scale (at the Planck scale).

All models A-M exhibit a very similar pattern of regions (poles, stability, metastability) in the $\alpha_\kappa$ - $M_F$  surface plots. As exemplified by \fig{SingleYukawaSurface}, the main difference is  in the lower left corner, covering a  low scale gauge portal with 
feeble Yukawas and masses in the (multi-)TeV range. Models that do not feature stability here, such as model L, but also VLQs I, J   are subject to BSM mass constraints
from a combination of experimental searches and stability as in \eq{SMEFTBoundL} and (\ref{eq:FCNCBoundL}).
For the other models,  such a model M,  
no lower mass limit can be obtained from our analysis. Furthermore,
if the Yukawa portal is too small or  vanishes,
 collider searches for long lived  particles become important. 
Long lived charged  particles \cite{Altakach:2022hgn}  leave ionization tracks or form resonances that decay to photons.
Long-lived VLQs in addition lead to $R$-hadron or dijet resonance signatures \cite{Bond:2017wut}.

For VLLs, mass bounds are even stronger than  \eq{FCNCBoundL} as charged lepton flavor violation is experimentally more severely constrained than FCNCs in the quark sector. In particular, results from $\mu-e$--conversion experiments \cite{Ishiwata:2015cga} yield bounds as strong as $M_F \gtrsim 1000$ TeV for stability all the way to $M_{\text {Pl}}$, and
one order of magnitude stronger than the quark ones  if one requires only $\alpha_\lambda \geq 0$ at  $M_{\text {Pl}}$. However, these constraints can  be evaded by 
only one sizable Yukawa to one SM species, while assuming feeble ones for the other two SM generations.
In this case, the models essentially resemble the ones in \Sec{3}.

\section{Conclusions \label{sec:con}}

For a decade the Higgs has been fact, and with it came the possibility  of an unstable ``great desert'' (\fig{SM-run}).
We have considered avoiding the SM metastability as a primary model building task and put forward mechanisms for stability at the Planck scale.
If a mechanism offers ``too little  too late'' it will not lift the instability, while ``too much too soon'' can lead to unwanted singularities.
In between, these constraints lead to 
stable, predictive, and testable BSM models including  bounds for masses, couplings and  flavors.

A minimally invasive path to  vacuum stability is given by the
gauge portal mechanism, solely requiring  SM-charged BSM fermions. The underlying interplay of  couplings  leads to a squeeze (\fig{pulldown}) and an uplift \eq{quartic-enhancement} of the BSM Higgs quartic over the SM one. 
We showed that this mechanism  works for all SM gauge interactions, $i.e.$~strong, weak, hypercharge, and combinations thereof, with the electroweak portals  a new addition of this work. In consequence, stability requires more flavors for larger masses, hence the closer the BSM mass is to the TeV-scale, the more perturbative the gauge portal  remains.
All gauge portals are  most efficient for  lighter (TeV-scale) BSM fermions,
 where this minimal fix implies a weakly coupled, yet stable ``great desert''.

Depending on the VLF representation,  vacuum stability  can require a range of flavor multiplicities as shown in Figs.~\ref{fig:StrongPortal}, \ref{fig:WeakPortal}, \ref{fig:HyperchargePortal}, \ref{fig:MixedPortal} for various incarnations of the gauge portal.
If Yukawa portals are unavailable, or 
too feeble, BSM fermions lead to long lived particle signatures,  such as diboson resonances, $R$-hadrons if colored, or long-lived charged particles otherwise \cite{Altakach:2022hgn}. 
However, with already super-feeble Yukawas present, $\alpha_\kappa M_F \gtrsim 10^{-14}$ TeV \cite{Hiller:2020fbu}, the  BSM fermions decay promptly to SM particles, and  can be searched for experimentally with these signatures.

We also find that all thirteen models  with gauge portals and renormalizable Yukawa interactions to SM fermions and the Higgs   (\tab{VLF-models})   can become stable at the Planck scale, without prior Landau poles.
Further, and except for model E which only features singlet BSM fermions, all models  can generically become safe through both a gauge or  a Yukawa portal.
Interestingly, most models (except E, I, J and L) can even be stabilized with a TeV-scale gauge portal (\tab{SM+VLFs}).
``Minimal'' low scale gauge portals, with a single generation of VLFs, feature stability with 
VLLs (models B, D, F) and VLQs (models K, M),  as demonstrated in the upper plot of \fig{SingleYukawaSurface}.
The Yukawa portal stabilizes in a walking regime and allows all models to achieve stability minimally and with TeV-ish BSM masses
  (conditions compiled in \tab{SM+VLF+Yukawa}),  as demonstrated in the lower plot of \fig{SingleYukawaSurface}.

We find that models without a TeV-ish gauge portal  (models E, I, J and L) are subject to novel mass constraints, obtained from combining
conditions for stability at the Planck scale  with experimental search limits. The interplay of constraints is illustrated for model M and L in 
Figs.~\ref{fig:SingleYukawaSurface}, \ref{fig:FlavorfulModelM} and \ref{fig:FlavorfulModelL}.
Lower mass limits can improve  search limits, see \eq{SMEFTBoundL} (model L). If  couplings to more than one SM generation are switched on,
FCNC bounds  can induce  lower limits on $M_F$ as strong as a few hundred TeV, see \eq{FCNCBoundL}.

We also observe upper limits on fermion masses from stability, which are outside the reach of colliders.

We close with two remarks on how the gauge or Yukawa portal mechanism can be exploited for other purposes. The gauge portal lends itself naturally for the search of new grand unified theories (GUTs) or SM extensions with Higgs criticality. In fact, the mild modifications of the RG running of couplings can be used to search systematically for vacuum stability  in combination with gauge coupling unification at or below the Planck scale. Also, having model-building control over the gauge, Yukawa and Higgs  couplings and their beta functions at the Planck scale may  prove useful to find extensions of the SM which connect successfully with quantized gravity~\cite{Shaposhnikov:2009pv}.  Further directions  include the exploration of cosmological consequences of our models.
 We hope to return to some of this in the future.

\begin{acknowledgments}
This work is supported by the \textit{Studienstiftung des Deutschen Volkes} (TH), and by the Science Technology and Facilities Council (STFC)  Consolidated Grant ST/T00102X/1 (DL).
\end{acknowledgments}

\clearpage

\bibliographystyle{jhep}
\bibliography{ref}

\providecommand{\href}[2]{#2}\begingroup\raggedright\begin{thebibliography}{100}

\bibitem{CMS:2012qbp}
{\scshape CMS} collaboration, \emph{{Observation of a New Boson at a Mass of
  125 GeV with the CMS Experiment at the LHC}},
  \href{https://doi.org/10.1016/j.physletb.2012.08.021}{\emph{Phys. Lett. B}
  {\bfseries 716} (2012) 30} [\href{https://arxiv.org/abs/1207.7235}{{\ttfamily
  1207.7235}}].

\bibitem{ATLAS:2012yve}
{\scshape ATLAS} collaboration, \emph{{Observation of a new particle in the
  search for the Standard Model Higgs boson with the ATLAS detector at the
  LHC}}, \href{https://doi.org/10.1016/j.physletb.2012.08.020}{\emph{Phys.
  Lett. B} {\bfseries 716} (2012) 1}
  [\href{https://arxiv.org/abs/1207.7214}{{\ttfamily 1207.7214}}].

\bibitem{Degrassi:2012ry}
G.~Degrassi, S.~Di~Vita, J.~Elias-Miro, J.~R. Espinosa, G.~F. Giudice,
  G.~Isidori et~al., \emph{{Higgs mass and vacuum stability in the Standard
  Model at NNLO}}, \href{https://doi.org/10.1007/JHEP08(2012)098}{\emph{JHEP}
  {\bfseries 08} (2012) 098} [\href{https://arxiv.org/abs/1205.6497}{{\ttfamily
  1205.6497}}].

\bibitem{Buttazzo:2013uya}
D.~Buttazzo, G.~Degrassi, P.~P. Giardino, G.~F. Giudice, F.~Sala, A.~Salvio
  et~al., \emph{{Investigating the near-criticality of the Higgs boson}},
  \href{https://doi.org/10.1007/JHEP12(2013)089}{\emph{JHEP} {\bfseries 12}
  (2013) 089} [\href{https://arxiv.org/abs/1307.3536}{{\ttfamily 1307.3536}}].

\bibitem{Elias-Miro:2012eoi}
J.~Elias-Miro, J.~R. Espinosa, G.~F. Giudice, H.~M. Lee and A.~Strumia,
  \emph{{Stabilization of the Electroweak Vacuum by a Scalar Threshold
  Effect}}, \href{https://doi.org/10.1007/JHEP06(2012)031}{\emph{JHEP}
  {\bfseries 06} (2012) 031} [\href{https://arxiv.org/abs/1203.0237}{{\ttfamily
  1203.0237}}].

\bibitem{Gonderinger:2012rd}
M.~Gonderinger, H.~Lim and M.~J. Ramsey-Musolf, \emph{{Complex Scalar Singlet
  Dark Matter: Vacuum Stability and Phenomenology}},
  \href{https://doi.org/10.1103/PhysRevD.86.043511}{\emph{Phys. Rev. D}
  {\bfseries 86} (2012) 043511}
  [\href{https://arxiv.org/abs/1202.1316}{{\ttfamily 1202.1316}}].

\bibitem{Anchordoqui:2012fq}
L.~A. Anchordoqui, I.~Antoniadis, H.~Goldberg, X.~Huang, D.~Lust, T.~R. Taylor
  et~al., \emph{{Vacuum Stability of Standard Model$^{++}$}},
  \href{https://doi.org/10.1007/JHEP02(2013)074}{\emph{JHEP} {\bfseries 02}
  (2013) 074} [\href{https://arxiv.org/abs/1208.2821}{{\ttfamily 1208.2821}}].

\bibitem{Arkani-Hamed:2012dcq}
N.~Arkani-Hamed, K.~Blum, R.~T. D'Agnolo and J.~Fan, \emph{{2:1 for Naturalness
  at the LHC?}}, \href{https://doi.org/10.1007/JHEP01(2013)149}{\emph{JHEP}
  {\bfseries 01} (2013) 149} [\href{https://arxiv.org/abs/1207.4482}{{\ttfamily
  1207.4482}}].

\bibitem{Joglekar:2012vc}
A.~Joglekar, P.~Schwaller and C.~E.~M. Wagner, \emph{{Dark Matter and Enhanced
  Higgs to Di-photon Rate from Vector-like Leptons}},
  \href{https://doi.org/10.1007/JHEP12(2012)064}{\emph{JHEP} {\bfseries 12}
  (2012) 064} [\href{https://arxiv.org/abs/1207.4235}{{\ttfamily 1207.4235}}].

\bibitem{Fairbairn:2013xaa}
M.~Fairbairn and P.~Grothaus, \emph{{Baryogenesis and Dark Matter with
  Vector-like Fermions}},
  \href{https://doi.org/10.1007/JHEP10(2013)176}{\emph{JHEP} {\bfseries 10}
  (2013) 176} [\href{https://arxiv.org/abs/1307.8011}{{\ttfamily 1307.8011}}].

\bibitem{Altmannshofer:2013zba}
W.~Altmannshofer, M.~Bauer and M.~Carena, \emph{{Exotic Leptons: Higgs, Flavor
  and Collider Phenomenology}},
  \href{https://doi.org/10.1007/JHEP01(2014)060}{\emph{JHEP} {\bfseries 01}
  (2014) 060} [\href{https://arxiv.org/abs/1308.1987}{{\ttfamily 1308.1987}}].

\bibitem{Gabrielli:2013hma}
E.~Gabrielli, M.~Heikinheimo, K.~Kannike, A.~Racioppi, M.~Raidal and
  C.~Spethmann, \emph{{Towards Completing the Standard Model: Vacuum Stability,
  EWSB and Dark Matter}},
  \href{https://doi.org/10.1103/PhysRevD.89.015017}{\emph{Phys. Rev. D}
  {\bfseries 89} (2014) 015017}
  [\href{https://arxiv.org/abs/1309.6632}{{\ttfamily 1309.6632}}].

\bibitem{BhupalDev:2013xol}
P.~S. Bhupal~Dev, D.~K. Ghosh, N.~Okada and I.~Saha, \emph{{125 GeV Higgs Boson
  and the Type-II Seesaw Model}},
  \href{https://doi.org/10.1007/JHEP03(2013)150}{\emph{JHEP} {\bfseries 03}
  (2013) 150} [\href{https://arxiv.org/abs/1301.3453}{{\ttfamily 1301.3453}}].

\bibitem{Datta:2013mta}
A.~Datta, A.~Elsayed, S.~Khalil and A.~Moursy, \emph{{Higgs vacuum stability in
  the $B-L$ extended standard model}},
  \href{https://doi.org/10.1103/PhysRevD.88.053011}{\emph{Phys. Rev. D}
  {\bfseries 88} (2013) 053011}
  [\href{https://arxiv.org/abs/1308.0816}{{\ttfamily 1308.0816}}].

\bibitem{Khan:2014kba}
N.~Khan and S.~Rakshit, \emph{{Study of electroweak vacuum metastability with a
  singlet scalar dark matter}},
  \href{https://doi.org/10.1103/PhysRevD.90.113008}{\emph{Phys. Rev. D}
  {\bfseries 90} (2014) 113008}
  [\href{https://arxiv.org/abs/1407.6015}{{\ttfamily 1407.6015}}].

\bibitem{Costa:2014qga}
R.~Costa, A.~P. Morais, M.~O.~P. Sampaio and R.~Santos, \emph{{Two-loop
  stability of a complex singlet extended Standard Model}},
  \href{https://doi.org/10.1103/PhysRevD.92.025024}{\emph{Phys. Rev. D}
  {\bfseries 92} (2015) 025024}
  [\href{https://arxiv.org/abs/1411.4048}{{\ttfamily 1411.4048}}].

\bibitem{Khoze:2014xha}
V.~V. Khoze, C.~McCabe and G.~Ro, \emph{{Higgs vacuum stability from the dark
  matter portal}}, \href{https://doi.org/10.1007/JHEP08(2014)026}{\emph{JHEP}
  {\bfseries 08} (2014) 026} [\href{https://arxiv.org/abs/1403.4953}{{\ttfamily
  1403.4953}}].

\bibitem{Xiao:2014kba}
M.-L. Xiao and J.-H. Yu, \emph{{Stabilizing electroweak vacuum in a vectorlike
  fermion model}},
  \href{https://doi.org/10.1103/PhysRevD.90.014007}{\emph{Phys. Rev. D}
  {\bfseries 90} (2014) 014007}
  [\href{https://arxiv.org/abs/1404.0681}{{\ttfamily 1404.0681}}].

\bibitem{Coriano:2014mpa}
C.~Coriano, L.~Delle~Rose and C.~Marzo, \emph{{Vacuum Stability in U(1)-Prime
  Extensions of the Standard Model with TeV Scale Right Handed Neutrinos}},
  \href{https://doi.org/10.1016/j.physletb.2014.09.001}{\emph{Phys. Lett. B}
  {\bfseries 738} (2014) 13} [\href{https://arxiv.org/abs/1407.8539}{{\ttfamily
  1407.8539}}].

\bibitem{DiChiara:2014wha}
S.~Di~Chiara, V.~Keus and O.~Lebedev, \emph{{Stabilizing the Higgs potential
  with a Z$'$}},
  \href{https://doi.org/10.1016/j.physletb.2015.03.013}{\emph{Phys. Lett. B}
  {\bfseries 744} (2015) 59} [\href{https://arxiv.org/abs/1412.7036}{{\ttfamily
  1412.7036}}].

\bibitem{Haba:2014oxa}
N.~Haba, H.~Ishida, R.~Takahashi and Y.~Yamaguchi, \emph{{Hierarchy problem,
  gauge coupling unification at the Planck scale, and vacuum stability}},
  \href{https://doi.org/10.1016/j.nuclphysb.2015.09.004}{\emph{Nucl. Phys. B}
  {\bfseries 900} (2015) 244}
  [\href{https://arxiv.org/abs/1412.8230}{{\ttfamily 1412.8230}}].

\bibitem{Lalak:2014qua}
Z.~Lalak, M.~Lewicki and P.~Olszewski, \emph{{Higher-order scalar interactions
  and SM vacuum stability}},
  \href{https://doi.org/10.1007/JHEP05(2014)119}{\emph{JHEP} {\bfseries 05}
  (2014) 119} [\href{https://arxiv.org/abs/1402.3826}{{\ttfamily 1402.3826}}].

\bibitem{Altmannshofer:2014vra}
W.~Altmannshofer, W.~A. Bardeen, M.~Bauer, M.~Carena and J.~D. Lykken,
  \emph{{Light Dark Matter, Naturalness, and the Radiative Origin of the
  Electroweak Scale}},
  \href{https://doi.org/10.1007/JHEP01(2015)032}{\emph{JHEP} {\bfseries 01}
  (2015) 032} [\href{https://arxiv.org/abs/1408.3429}{{\ttfamily 1408.3429}}].

\bibitem{Belanger:2014bga}
G.~B\'elanger, K.~Kannike, A.~Pukhov and M.~Raidal, \emph{{Minimal
  semi-annihilating $\mathbb{Z}_N$ scalar dark matter}},
  \href{https://doi.org/10.1088/1475-7516/2014/06/021}{\emph{JCAP} {\bfseries
  06} (2014) 021} [\href{https://arxiv.org/abs/1403.4960}{{\ttfamily
  1403.4960}}].

\bibitem{Salvio:2015cja}
A.~Salvio, \emph{{A Simple Motivated Completion of the Standard Model below the
  Planck Scale: Axions and Right-Handed Neutrinos}},
  \href{https://doi.org/10.1016/j.physletb.2015.03.015}{\emph{Phys. Lett. B}
  {\bfseries 743} (2015) 428}
  [\href{https://arxiv.org/abs/1501.03781}{{\ttfamily 1501.03781}}].

\bibitem{Chakrabarty:2015kmt}
N.~Chakrabarty, \emph{{High-scale validity of a model with
  Three-Higgs-doublets}},
  \href{https://doi.org/10.1103/PhysRevD.93.075025}{\emph{Phys. Rev. D}
  {\bfseries 93} (2016) 075025}
  [\href{https://arxiv.org/abs/1511.08137}{{\ttfamily 1511.08137}}].

\bibitem{Blum:2015rpa}
K.~Blum, R.~T. D'Agnolo and J.~Fan, \emph{{Vacuum stability bounds on Higgs
  coupling deviations in the absence of new bosons}},
  \href{https://doi.org/10.1007/JHEP03(2015)166}{\emph{JHEP} {\bfseries 03}
  (2015) 166} [\href{https://arxiv.org/abs/1502.01045}{{\ttfamily
  1502.01045}}].

\bibitem{Bagnaschi:2015pwa}
E.~Bagnaschi, F.~Br\"ummer, W.~Buchm\"uller, A.~Voigt and G.~Weiglein,
  \emph{{Vacuum stability and supersymmetry at high scales with two Higgs
  doublets}}, \href{https://doi.org/10.1007/JHEP03(2016)158}{\emph{JHEP}
  {\bfseries 03} (2016) 158}
  [\href{https://arxiv.org/abs/1512.07761}{{\ttfamily 1512.07761}}].

\bibitem{Duch:2015jta}
M.~Duch, B.~Grzadkowski and M.~McGarrie, \emph{{A stable Higgs portal with
  vector dark matter}},
  \href{https://doi.org/10.1007/JHEP09(2015)162}{\emph{JHEP} {\bfseries 09}
  (2015) 162} [\href{https://arxiv.org/abs/1506.08805}{{\ttfamily
  1506.08805}}].

\bibitem{Dhuria:2015ufo}
M.~Dhuria and G.~Goswami, \emph{{Perturbativity, vacuum stability, and
  inflation in the light of 750 GeV diphoton excess}},
  \href{https://doi.org/10.1103/PhysRevD.94.055009}{\emph{Phys. Rev. D}
  {\bfseries 94} (2016) 055009}
  [\href{https://arxiv.org/abs/1512.06782}{{\ttfamily 1512.06782}}].

\bibitem{Salvio:2015jgu}
A.~Salvio and A.~Mazumdar, \emph{{Higgs Stability and the 750 GeV Diphoton
  Excess}}, \href{https://doi.org/10.1016/j.physletb.2016.02.057}{\emph{Phys.
  Lett. B} {\bfseries 755} (2016) 469}
  [\href{https://arxiv.org/abs/1512.08184}{{\ttfamily 1512.08184}}].

\bibitem{Son:2015vfl}
M.~Son and A.~Urbano, \emph{{A new scalar resonance at 750 GeV: Towards a proof
  of concept in favor of strongly interacting theories}},
  \href{https://doi.org/10.1007/JHEP05(2016)181}{\emph{JHEP} {\bfseries 05}
  (2016) 181} [\href{https://arxiv.org/abs/1512.08307}{{\ttfamily
  1512.08307}}].

\bibitem{Hamada:2015bra}
Y.~Hamada, K.~Kawana and K.~Tsumura, \emph{{Landau pole in the Standard Model
  with weakly interacting scalar fields}},
  \href{https://doi.org/10.1016/j.physletb.2015.05.072}{\emph{Phys. Lett. B}
  {\bfseries 747} (2015) 238}
  [\href{https://arxiv.org/abs/1505.01721}{{\ttfamily 1505.01721}}].

\bibitem{Ng:2015eia}
J.~N. Ng and A.~de~la Puente, \emph{{Electroweak Vacuum Stability and the
  Seesaw Mechanism Revisited}},
  \href{https://doi.org/10.1140/epjc/s10052-016-3981-4}{\emph{Eur. Phys. J. C}
  {\bfseries 76} (2016) 122}
  [\href{https://arxiv.org/abs/1510.00742}{{\ttfamily 1510.00742}}].

\bibitem{Falkowski:2015iwa}
A.~Falkowski, C.~Gross and O.~Lebedev, \emph{{A second Higgs from the Higgs
  portal}}, \href{https://doi.org/10.1007/JHEP05(2015)057}{\emph{JHEP}
  {\bfseries 05} (2015) 057}
  [\href{https://arxiv.org/abs/1502.01361}{{\ttfamily 1502.01361}}].

\bibitem{Han:2015hda}
H.~Han and S.~Zheng, \emph{{New Constraints on Higgs-portal Scalar Dark
  Matter}}, \href{https://doi.org/10.1007/JHEP12(2015)044}{\emph{JHEP}
  {\bfseries 12} (2015) 044}
  [\href{https://arxiv.org/abs/1509.01765}{{\ttfamily 1509.01765}}].

\bibitem{Lindner:2015qva}
M.~Lindner, H.~H. Patel and B.~Radov\v{c}i\'c, \emph{{Electroweak Absolute,
  Meta-, and Thermal Stability in Neutrino Mass Models}},
  \href{https://doi.org/10.1103/PhysRevD.93.073005}{\emph{Phys. Rev. D}
  {\bfseries 93} (2016) 073005}
  [\href{https://arxiv.org/abs/1511.06215}{{\ttfamily 1511.06215}}].

\bibitem{DelleRose:2015bms}
L.~Delle~Rose, C.~Marzo and A.~Urbano, \emph{{On the stability of the
  electroweak vacuum in the presence of low-scale seesaw models}},
  \href{https://doi.org/10.1007/JHEP12(2015)050}{\emph{JHEP} {\bfseries 12}
  (2015) 050} [\href{https://arxiv.org/abs/1506.03360}{{\ttfamily
  1506.03360}}].

\bibitem{Latosinski:2015pba}
A.~Latosinski, A.~Lewandowski, K.~A. Meissner and H.~Nicolai, \emph{{Conformal
  Standard Model with an extended scalar sector}},
  \href{https://doi.org/10.1007/JHEP10(2015)170}{\emph{JHEP} {\bfseries 10}
  (2015) 170} [\href{https://arxiv.org/abs/1507.01755}{{\ttfamily
  1507.01755}}].

\bibitem{Chowdhury:2015yja}
D.~Chowdhury and O.~Eberhardt, \emph{{Global fits of the two-loop renormalized
  Two-Higgs-Doublet model with soft Z$_{2}$ breaking}},
  \href{https://doi.org/10.1007/JHEP11(2015)052}{\emph{JHEP} {\bfseries 11}
  (2015) 052} [\href{https://arxiv.org/abs/1503.08216}{{\ttfamily
  1503.08216}}].

\bibitem{Khan:2015ipa}
N.~Khan and S.~Rakshit, \emph{{Constraints on inert dark matter from the
  metastability of the electroweak vacuum}},
  \href{https://doi.org/10.1103/PhysRevD.92.055006}{\emph{Phys. Rev. D}
  {\bfseries 92} (2015) 055006}
  [\href{https://arxiv.org/abs/1503.03085}{{\ttfamily 1503.03085}}].

\bibitem{Ferreira:2015rha}
P.~Ferreira, H.~E. Haber and E.~Santos, \emph{{Preserving the validity of the
  Two-Higgs Doublet Model up to the Planck scale}},
  \href{https://doi.org/10.1103/PhysRevD.92.033003}{\emph{Phys. Rev. D}
  {\bfseries 92} (2015) 033003}
  [\href{https://arxiv.org/abs/1505.04001}{{\ttfamily 1505.04001}}].

\bibitem{Ferreira:2015pfi}
P.~M. Ferreira and B.~Swiezewska, \emph{{One-loop contributions to neutral
  minima in the inert doublet model}},
  \href{https://doi.org/10.1007/JHEP04(2016)099}{\emph{JHEP} {\bfseries 04}
  (2016) 099} [\href{https://arxiv.org/abs/1511.02879}{{\ttfamily
  1511.02879}}].

\bibitem{Swiezewska:2015paa}
B.~Swiezewska, \emph{{Inert scalars and vacuum metastability around the
  electroweak scale}},
  \href{https://doi.org/10.1007/JHEP07(2015)118}{\emph{JHEP} {\bfseries 07}
  (2015) 118} [\href{https://arxiv.org/abs/1503.07078}{{\ttfamily
  1503.07078}}].

\bibitem{Coriano:2015sea}
C.~Coriano, L.~Delle~Rose and C.~Marzo, \emph{{Constraints on abelian
  extensions of the Standard Model from two-loop vacuum stability and
  $U(1)_{B-L}$}}, \href{https://doi.org/10.1007/JHEP02(2016)135}{\emph{JHEP}
  {\bfseries 02} (2016) 135}
  [\href{https://arxiv.org/abs/1510.02379}{{\ttfamily 1510.02379}}].

\bibitem{Oda:2015gna}
S.~Oda, N.~Okada and D.-s. Takahashi, \emph{{Classically conformal U(1)'
  extended standard model and Higgs vacuum stability}},
  \href{https://doi.org/10.1103/PhysRevD.92.015026}{\emph{Phys. Rev. D}
  {\bfseries 92} (2015) 015026}
  [\href{https://arxiv.org/abs/1504.06291}{{\ttfamily 1504.06291}}].

\bibitem{Das:2015nwk}
A.~Das, N.~Okada and N.~Papapietro, \emph{{Electroweak vacuum stability in
  classically conformal B-L extension of the Standard Model}},
  \href{https://doi.org/10.1140/epjc/s10052-017-4683-2}{\emph{Eur. Phys. J. C}
  {\bfseries 77} (2017) 122}
  [\href{https://arxiv.org/abs/1509.01466}{{\ttfamily 1509.01466}}].

\bibitem{Haba:2015rha}
N.~Haba and Y.~Yamaguchi, \emph{{Vacuum stability in the $U(1)_\chi$ extended
  model with vanishing scalar potential at the Planck scale}},
  \href{https://doi.org/10.1093/ptep/ptv121}{\emph{PTEP} {\bfseries 2015}
  (2015) 093B05} [\href{https://arxiv.org/abs/1504.05669}{{\ttfamily
  1504.05669}}].

\bibitem{Haba:2015nwl}
N.~Haba, H.~Ishida, R.~Takahashi and Y.~Yamaguchi, \emph{{Gauge coupling
  unification in a classically scale invariant model}},
  \href{https://doi.org/10.1007/JHEP02(2016)058}{\emph{JHEP} {\bfseries 02}
  (2016) 058} [\href{https://arxiv.org/abs/1511.02107}{{\ttfamily
  1511.02107}}].

\bibitem{Das:2016zue}
A.~Das, S.~Oda, N.~Okada and D.-s. Takahashi, \emph{{Classically conformal
  U(1)' extended standard model, electroweak vacuum stability, and LHC Run-2
  bounds}}, \href{https://doi.org/10.1103/PhysRevD.93.115038}{\emph{Phys. Rev.
  D} {\bfseries 93} (2016) 115038}
  [\href{https://arxiv.org/abs/1605.01157}{{\ttfamily 1605.01157}}].

\bibitem{Chakrabarty:2016smc}
N.~Chakrabarty and B.~Mukhopadhyaya, \emph{{High-scale validity of a two Higgs
  doublet scenario: metastability included}},
  \href{https://doi.org/10.1140/epjc/s10052-017-4705-0}{\emph{Eur. Phys. J. C}
  {\bfseries 77} (2017) 153}
  [\href{https://arxiv.org/abs/1603.05883}{{\ttfamily 1603.05883}}].

\bibitem{Bandyopadhyay:2016oif}
P.~Bandyopadhyay and R.~Mandal, \emph{{Vacuum stability in an extended standard
  model with a leptoquark}},
  \href{https://doi.org/10.1103/PhysRevD.95.035007}{\emph{Phys. Rev. D}
  {\bfseries 95} (2017) 035007}
  [\href{https://arxiv.org/abs/1609.03561}{{\ttfamily 1609.03561}}].

\bibitem{Khan:2016sxm}
N.~Khan, \emph{{Exploring the hyperchargeless Higgs triplet model up to the
  Planck scale}},
  \href{https://doi.org/10.1140/epjc/s10052-018-5766-4}{\emph{Eur. Phys. J. C}
  {\bfseries 78} (2018) 341}
  [\href{https://arxiv.org/abs/1610.03178}{{\ttfamily 1610.03178}}].

\bibitem{Haba:2016zbu}
N.~Haba, H.~Ishida, N.~Okada and Y.~Yamaguchi, \emph{{Vacuum stability and
  naturalness in type-II seesaw}},
  \href{https://doi.org/10.1140/epjc/s10052-016-4180-z}{\emph{Eur. Phys. J. C}
  {\bfseries 76} (2016) 333}
  [\href{https://arxiv.org/abs/1601.05217}{{\ttfamily 1601.05217}}].

\bibitem{Mambrini:2016dca}
Y.~Mambrini, N.~Nagata, K.~A. Olive and J.~Zheng, \emph{{Vacuum Stability and
  Radiative Electroweak Symmetry Breaking in an SO(10) Dark Matter Model}},
  \href{https://doi.org/10.1103/PhysRevD.93.111703}{\emph{Phys. Rev. D}
  {\bfseries 93} (2016) 111703}
  [\href{https://arxiv.org/abs/1602.05583}{{\ttfamily 1602.05583}}].

\bibitem{Chang:2016pya}
W.-F. Chang and J.~N. Ng, \emph{{Renormalization Group Study of the Minimal
  Majoronic Dark Radiation and Dark Matter Model}},
  \href{https://doi.org/10.1088/1475-7516/2016/07/027}{\emph{JCAP} {\bfseries
  07} (2016) 027} [\href{https://arxiv.org/abs/1604.02017}{{\ttfamily
  1604.02017}}].

\bibitem{Hamada:2016vwk}
Y.~Hamada, H.~Kawai, K.~Kawana and K.~Tsumura, \emph{{Models of the LHC
  diphoton excesses valid up to the Planck scale}},
  \href{https://doi.org/10.1103/PhysRevD.94.014007}{\emph{Phys. Rev. D}
  {\bfseries 94} (2016) 014007}
  [\href{https://arxiv.org/abs/1602.04170}{{\ttfamily 1602.04170}}].

\bibitem{Ghosh:2017pxl}
D.~K. Ghosh, N.~Ghosh, I.~Saha and A.~Shaw, \emph{{Revisiting the high-scale
  validity of the type II seesaw model with novel LHC signature}},
  \href{https://doi.org/10.1103/PhysRevD.97.115022}{\emph{Phys. Rev. D}
  {\bfseries 97} (2018) 115022}
  [\href{https://arxiv.org/abs/1711.06062}{{\ttfamily 1711.06062}}].

\bibitem{Oda:2017kwl}
S.~Oda, N.~Okada and D.-s. Takahashi, \emph{{Right-handed neutrino dark matter
  in the classically conformal U(1)' extended standard model}},
  \href{https://doi.org/10.1103/PhysRevD.96.095032}{\emph{Phys. Rev. D}
  {\bfseries 96} (2017) 095032}
  [\href{https://arxiv.org/abs/1704.05023}{{\ttfamily 1704.05023}}].

\bibitem{Garg:2017iva}
I.~Garg, S.~Goswami, K.~N. Vishnudath and N.~Khan, \emph{{Electroweak vacuum
  stability in presence of singlet scalar dark matter in TeV scale seesaw
  models}}, \href{https://doi.org/10.1103/PhysRevD.96.055020}{\emph{Phys. Rev.
  D} {\bfseries 96} (2017) 055020}
  [\href{https://arxiv.org/abs/1706.08851}{{\ttfamily 1706.08851}}].

\bibitem{Ghosh:2017fmr}
P.~Ghosh, A.~K. Saha and A.~Sil, \emph{{Study of Electroweak Vacuum Stability
  from Extended Higgs Portal of Dark Matter and Neutrinos}},
  \href{https://doi.org/10.1103/PhysRevD.97.075034}{\emph{Phys. Rev. D}
  {\bfseries 97} (2018) 075034}
  [\href{https://arxiv.org/abs/1706.04931}{{\ttfamily 1706.04931}}].

\bibitem{Goswami:2018jar}
S.~Goswami, K.~N. Vishnudath and N.~Khan, \emph{{Constraining the minimal
  type-III seesaw model with naturalness, lepton flavor violation, and
  electroweak vacuum stability}},
  \href{https://doi.org/10.1103/PhysRevD.99.075012}{\emph{Phys. Rev. D}
  {\bfseries 99} (2019) 075012}
  [\href{https://arxiv.org/abs/1810.11687}{{\ttfamily 1810.11687}}].

\bibitem{DuttaBanik:2018emv}
A.~Dutta~Banik, A.~K. Saha and A.~Sil, \emph{{Scalar assisted singlet doublet
  fermion dark matter model and electroweak vacuum stability}},
  \href{https://doi.org/10.1103/PhysRevD.98.075013}{\emph{Phys. Rev. D}
  {\bfseries 98} (2018) 075013}
  [\href{https://arxiv.org/abs/1806.08080}{{\ttfamily 1806.08080}}].

\bibitem{Schuh:2018hig}
P.~Schuh, \emph{{Vacuum Stability of Asymptotically Safe Two Higgs Doublet
  Models}}, \href{https://doi.org/10.1140/epjc/s10052-019-7426-8}{\emph{Eur.
  Phys. J. C} {\bfseries 79} (2019) 909}
  [\href{https://arxiv.org/abs/1810.07664}{{\ttfamily 1810.07664}}].

\bibitem{Marzo:2018nov}
C.~Marzo, L.~Marzola and V.~Vaskonen, \emph{{Phase transition and vacuum
  stability in the classically conformal B\textendash{}L model}},
  \href{https://doi.org/10.1140/epjc/s10052-019-7076-x}{\emph{Eur. Phys. J. C}
  {\bfseries 79} (2019) 601}
  [\href{https://arxiv.org/abs/1811.11169}{{\ttfamily 1811.11169}}].

\bibitem{Ellis:2018khn}
S.~A.~R. Ellis, T.~Gherghetta, K.~Kaneta and K.~A. Olive, \emph{{New Weak-Scale
  Physics from SO(10) with High-Scale Supersymmetry}},
  \href{https://doi.org/10.1103/PhysRevD.98.055009}{\emph{Phys. Rev. D}
  {\bfseries 98} (2018) 055009}
  [\href{https://arxiv.org/abs/1807.06488}{{\ttfamily 1807.06488}}].

\bibitem{Boucenna:2018wjc}
S.~M. Boucenna, T.~Ohlsson and M.~Pernow, \emph{{A minimal non-supersymmetric
  SO(10) model with Peccei--Quinn symmetry}},
  \href{https://doi.org/10.1016/j.physletb.2019.03.045}{\emph{Phys. Lett. B}
  {\bfseries 792} (2019) 251}
  [\href{https://arxiv.org/abs/1812.10548}{{\ttfamily 1812.10548}}].

\bibitem{Gopalakrishna:2018uxn}
S.~Gopalakrishna and A.~Velusamy, \emph{{Higgs vacuum stability with vectorlike
  fermions}}, \href{https://doi.org/10.1103/PhysRevD.99.115020}{\emph{Phys.
  Rev. D} {\bfseries 99} (2019) 115020}
  [\href{https://arxiv.org/abs/1812.11303}{{\ttfamily 1812.11303}}].

\bibitem{Wang:2018lhk}
J.-W. Wang, X.-J. Bi, P.-F. Yin and Z.-H. Yu, \emph{{Impact of Fermionic
  Electroweak Multiplet Dark Matter on Vacuum Stability with One-loop
  Matching}}, \href{https://doi.org/10.1103/PhysRevD.99.055009}{\emph{Phys.
  Rev. D} {\bfseries 99} (2019) 055009}
  [\href{https://arxiv.org/abs/1811.08743}{{\ttfamily 1811.08743}}].

\bibitem{Okada:2019bqa}
N.~Okada, D.~Raut and Q.~Shafi, \emph{{Inflation, proton decay, and
  Higgs-portal dark matter in $SO(10) \times U(1)_\psi $}},
  \href{https://doi.org/10.1140/epjc/s10052-019-7550-5}{\emph{Eur. Phys. J. C}
  {\bfseries 79} (2019) 1036}
  [\href{https://arxiv.org/abs/1906.06869}{{\ttfamily 1906.06869}}].

\bibitem{Hiller:2019mou}
G.~Hiller, C.~Hormigos-Feliu, D.~F. Litim and T.~Steudtner, \emph{{Anomalous
  magnetic moments from asymptotic safety}},
  \href{https://doi.org/10.1103/PhysRevD.102.071901}{\emph{Phys. Rev. D}
  {\bfseries 102} (2020) 071901}
  [\href{https://arxiv.org/abs/1910.14062}{{\ttfamily 1910.14062}}].

\bibitem{Bhattacharya:2019fgs}
S.~Bhattacharya, P.~Ghosh, A.~K. Saha and A.~Sil, \emph{{Two component dark
  matter with inert Higgs doublet: neutrino mass, high scale validity and
  collider searches}},
  \href{https://doi.org/10.1007/JHEP03(2020)090}{\emph{JHEP} {\bfseries 03}
  (2020) 090} [\href{https://arxiv.org/abs/1905.12583}{{\ttfamily
  1905.12583}}].

\bibitem{Mandal:2019ndp}
S.~Mandal, R.~Srivastava and J.~W.~F. Valle, \emph{{Consistency of the
  dynamical high-scale type-I seesaw mechanism}},
  \href{https://doi.org/10.1103/PhysRevD.101.115030}{\emph{Phys. Rev. D}
  {\bfseries 101} (2020) 115030}
  [\href{https://arxiv.org/abs/1903.03631}{{\ttfamily 1903.03631}}].

\bibitem{Hiller:2020fbu}
G.~Hiller, C.~Hormigos-Feliu, D.~F. Litim and T.~Steudtner, \emph{{Model
  Building from Asymptotic Safety with Higgs and Flavor Portals}},
  \href{https://doi.org/10.1103/PhysRevD.102.095023}{\emph{Phys. Rev. D}
  {\bfseries 102} (2020) 095023}
  [\href{https://arxiv.org/abs/2008.08606}{{\ttfamily 2008.08606}}].

\bibitem{Borah:2020nsz}
D.~Borah, R.~Roshan and A.~Sil, \emph{{Sub-TeV singlet scalar dark matter and
  electroweak vacuum stability with vectorlike fermions}},
  \href{https://doi.org/10.1103/PhysRevD.102.075034}{\emph{Phys. Rev. D}
  {\bfseries 102} (2020) 075034}
  [\href{https://arxiv.org/abs/2007.14904}{{\ttfamily 2007.14904}}].

\bibitem{Chakrabarty:2020jro}
N.~Chakrabarty, \emph{{Doubly charged scalars and vector-like leptons
  confronting the muon g-2 anomaly and Higgs vacuum stability}},
  \href{https://doi.org/10.1140/epjp/s13360-021-02168-3}{\emph{Eur. Phys. J.
  Plus} {\bfseries 136} (2021) 1183}
  [\href{https://arxiv.org/abs/2010.05215}{{\ttfamily 2010.05215}}].

\bibitem{Mandal:2020lhl}
S.~Mandal, R.~Srivastava and J.~W.~F. Valle, \emph{{Electroweak symmetry
  breaking in the inverse seesaw mechanism}},
  \href{https://doi.org/10.1007/JHEP03(2021)212}{\emph{JHEP} {\bfseries 03}
  (2021) 212} [\href{https://arxiv.org/abs/2009.10116}{{\ttfamily
  2009.10116}}].

\bibitem{Fabbrichesi:2020svm}
M.~Fabbrichesi, C.~M. Nieto, A.~Tonero and A.~Ugolotti, \emph{{Asymptotically
  safe SU(5) GUT}},
  \href{https://doi.org/10.1103/PhysRevD.103.095026}{\emph{Phys. Rev. D}
  {\bfseries 103} (2021) 095026}
  [\href{https://arxiv.org/abs/2012.03987}{{\ttfamily 2012.03987}}].

\bibitem{DuttaBanik:2020jrj}
A.~Dutta~Banik, R.~Roshan and A.~Sil, \emph{{Two component singlet-triplet
  scalar dark matter and electroweak vacuum stability}},
  \href{https://doi.org/10.1103/PhysRevD.103.075001}{\emph{Phys. Rev. D}
  {\bfseries 103} (2021) 075001}
  [\href{https://arxiv.org/abs/2009.01262}{{\ttfamily 2009.01262}}].

\bibitem{Bandyopadhyay:2021kue}
P.~Bandyopadhyay, S.~Jangid and A.~Karan, \emph{{Constraining scalar doublet
  and triplet leptoquarks with vacuum stability and perturbativity}},
  \href{https://doi.org/10.1140/epjc/s10052-022-10418-6}{\emph{Eur. Phys. J. C}
  {\bfseries 82} (2022) 516}
  [\href{https://arxiv.org/abs/2111.03872}{{\ttfamily 2111.03872}}].

\bibitem{Bause:2021prv}
R.~Bause, G.~Hiller, T.~H\"ohne, D.~F. Litim and T.~Steudtner,
  \emph{{B-anomalies from flavorful U(1)$'$ extensions, safely}},
  \href{https://doi.org/10.1140/epjc/s10052-021-09957-1}{\emph{Eur. Phys. J. C}
  {\bfseries 82} (2022) 42} [\href{https://arxiv.org/abs/2109.06201}{{\ttfamily
  2109.06201}}].

\bibitem{Litim:2020jvl}
D.~F. Litim and T.~Steudtner, \emph{{ARGES \textendash{} Advanced
  Renormalisation Group Equation Simplifier}},
  \href{https://doi.org/10.1016/j.cpc.2021.108021}{\emph{Comput. Phys. Commun.}
  {\bfseries 265} (2021) 108021}
  [\href{https://arxiv.org/abs/2012.12955}{{\ttfamily 2012.12955}}].

\bibitem{Zyla:2020zbs}
{\scshape Particle Data Group} collaboration, \emph{{Review of Particle
  Physics}}, \href{https://doi.org/10.1093/ptep/ptaa104}{\emph{PTEP} {\bfseries
  2020} (2020) 083C01}.

\bibitem{Branchina:2014rva}
V.~Branchina, E.~Messina and M.~Sher, \emph{{Lifetime of the electroweak vacuum
  and sensitivity to Planck scale physics}},
  \href{https://doi.org/10.1103/PhysRevD.91.013003}{\emph{Phys. Rev. D}
  {\bfseries 91} (2015) 013003}
  [\href{https://arxiv.org/abs/1408.5302}{{\ttfamily 1408.5302}}].

\bibitem{Machacek:1984zw}
M.~E. Machacek and M.~T. Vaughn, \emph{{Two Loop Renormalization Group
  Equations in a General Quantum Field Theory. 3. Scalar Quartic Couplings}},
  \href{https://doi.org/10.1016/0550-3213(85)90040-9}{\emph{Nucl. Phys. B}
  {\bfseries 249} (1985) 70}.

\bibitem{Cohen:1988sq}
A.~G. Cohen and H.~Georgi, \emph{{Walking Beyond the Rainbow}},
  \href{https://doi.org/10.1016/0550-3213(89)90109-0}{\emph{Nucl. Phys. B}
  {\bfseries 314} (1989) 7}.

\bibitem{Cacciapaglia:2020kgq}
G.~Cacciapaglia, C.~Pica and F.~Sannino, \emph{{Fundamental Composite Dynamics:
  A Review}}, \href{https://doi.org/10.1016/j.physrep.2020.07.002}{\emph{Phys.
  Rept.} {\bfseries 877} (2020) 1}
  [\href{https://arxiv.org/abs/2002.04914}{{\ttfamily 2002.04914}}].

\bibitem{Manohar:2018aog}
A.~V. Manohar, \emph{{Introduction to Effective Field Theories}},
  \href{https://arxiv.org/abs/1804.05863}{{\ttfamily 1804.05863}}.

\bibitem{Weinberg:1980wa}
S.~Weinberg, \emph{{Effective Gauge Theories}},
  \href{https://doi.org/10.1016/0370-2693(80)90660-7}{\emph{Phys. Lett. B}
  {\bfseries 91} (1980) 51}.

\bibitem{ATLAS:2020mee}
{ATLAS collaboration}, \emph{{Determination of the strong coupling constant and
  test of asymptotic freedom from Transverse Energy-Energy Correlations in
  multijet events at $\sqrt{s} = 13$ TeV with the ATLAS detector}},
  {\emph{{\href{https://cdsweb.cern.ch/record/2725553}{ATLAS-CONF-2020-025}}}
  (2020) }.

\bibitem{Bond:2017wut}
A.~D. Bond, G.~Hiller, K.~Kowalska and D.~F. Litim, \emph{{Directions for model
  building from asymptotic safety}},
  \href{https://doi.org/10.1007/JHEP08(2017)004}{\emph{JHEP} {\bfseries 08}
  (2017) 004} [\href{https://arxiv.org/abs/1702.01727}{{\ttfamily
  1702.01727}}].

\bibitem{Farina:2016rws}
M.~Farina, G.~Panico, D.~Pappadopulo, J.~T. Ruderman, R.~Torre and A.~Wulzer,
  \emph{{Energy helps accuracy: electroweak precision tests at hadron
  colliders}},
  \href{https://doi.org/10.1016/j.physletb.2017.06.043}{\emph{Phys. Lett. B}
  {\bfseries 772} (2017) 210}
  [\href{https://arxiv.org/abs/1609.08157}{{\ttfamily 1609.08157}}].

\bibitem{Alves:2014cda}
D.~S.~M. Alves, J.~Galloway, J.~T. Ruderman and J.~R. Walsh, \emph{{Running
  Electroweak Couplings as a Probe of New Physics}},
  \href{https://doi.org/10.1007/JHEP02(2015)007}{\emph{JHEP} {\bfseries 02}
  (2015) 007} [\href{https://arxiv.org/abs/1410.6810}{{\ttfamily 1410.6810}}].

\bibitem{Altakach:2022hgn}
M.~M. Altakach, P.~Lamba, R.~Mase\l{}ek, V.~A. Mitsou and K.~Sakurai,
  \emph{{Discovery prospects for long-lived multiply charged particles at the
  LHC}}, \href{https://doi.org/10.1140/epjc/s10052-022-10805-z}{\emph{Eur.
  Phys. J. C} {\bfseries 82} (2022) 848}
  [\href{https://arxiv.org/abs/2204.03667}{{\ttfamily 2204.03667}}].

\bibitem{Falkowski:2019hvp}
A.~Falkowski and D.~Straub, \emph{{Flavourful SMEFT likelihood for Higgs and
  electroweak data}},
  \href{https://doi.org/10.1007/JHEP04(2020)066}{\emph{JHEP} {\bfseries 04}
  (2020) 066} [\href{https://arxiv.org/abs/1911.07866}{{\ttfamily
  1911.07866}}].

\bibitem{Ellis:2020unq}
J.~Ellis, M.~Madigan, K.~Mimasu, V.~Sanz and T.~You, \emph{{Top, Higgs, Diboson
  and Electroweak Fit to the Standard Model Effective Field Theory}},
  \href{https://doi.org/10.1007/JHEP04(2021)279}{\emph{JHEP} {\bfseries 04}
  (2021) 279} [\href{https://arxiv.org/abs/2012.02779}{{\ttfamily
  2012.02779}}].

\bibitem{Ishiwata:2015cga}
K.~Ishiwata, Z.~Ligeti and M.~B. Wise, \emph{{New Vector-Like Fermions and
  Flavor Physics}}, \href{https://doi.org/10.1007/JHEP10(2015)027}{\emph{JHEP}
  {\bfseries 10} (2015) 027}
  [\href{https://arxiv.org/abs/1506.03484}{{\ttfamily 1506.03484}}].

\bibitem{ATLAS:2018ziw}
{\scshape ATLAS} collaboration, \emph{{Combination of the searches for
  pair-produced vector-like partners of the third-generation quarks at
  $\sqrt{s} =$ 13 TeV with the ATLAS detector}},
  \href{https://doi.org/10.1103/PhysRevLett.121.211801}{\emph{Phys. Rev. Lett.}
  {\bfseries 121} (2018) 211801}
  [\href{https://arxiv.org/abs/1808.02343}{{\ttfamily 1808.02343}}].

\bibitem{Erdmann:2018axt}
{\scshape ATLAS} collaboration, \emph{{Overview of searches for single
  production of vector-like top and bottom quarks with the ATLAS experiment at
  13 TeV}},  in \emph{{Proceedings of the 11th International Workshop on Top
  Quark Physics}}, 11, 2018, \href{https://arxiv.org/abs/1811.11496}{{\ttfamily
  1811.11496}}.

\bibitem{HHLS22plus}
{G. Hiller, T. H\"ohne, D. F. Litim and T. Steudtner}, \emph{{in preparation}},
  {\emph{preprint DO-TH 22/17} (2022) }.

\bibitem{Aguilar-Saavedra:2013qpa}
J.~A. Aguilar-Saavedra, R.~Benbrik, S.~Heinemeyer and M.~P\'erez-Victoria,
  \emph{{Handbook of vectorlike quarks: Mixing and single production}},
  \href{https://doi.org/10.1103/PhysRevD.88.094010}{\emph{Phys. Rev. D}
  {\bfseries 88} (2013) 094010}
  [\href{https://arxiv.org/abs/1306.0572}{{\ttfamily 1306.0572}}].

\bibitem{Shaposhnikov:2009pv}
M.~Shaposhnikov and C.~Wetterich, \emph{{Asymptotic safety of gravity and the
  Higgs boson mass}},
  \href{https://doi.org/10.1016/j.physletb.2009.12.022}{\emph{Phys. Lett. B}
  {\bfseries 683} (2010) 196}
  [\href{https://arxiv.org/abs/0912.0208}{{\ttfamily 0912.0208}}].

\end{thebibliography}\endgroup

\end{document}